\documentclass[english]{article}
\usepackage[T1]{fontenc}
\usepackage[utf8]{inputenc}
\usepackage{babel}
\usepackage{refstyle}
\usepackage{float}
\usepackage{mathtools}
\usepackage{bm}
\usepackage{amsmath}
\usepackage{amsthm}
\usepackage{amssymb}
\usepackage{graphicx}
\usepackage[unicode=true,pdfusetitle,
 bookmarks=true,bookmarksnumbered=false,bookmarksopen=false,
 breaklinks=false,pdfborder={0 0 1},backref=false,colorlinks=false]
 {hyperref}

\makeatletter

\AtBeginDocument{\providecommand\secref[1]{\ref{sec:#1}}}
\AtBeginDocument{\providecommand\thmref[1]{\ref{thm:#1}}}
\RS@ifundefined{subsecref}
  {\newref{subsec}{name = \RSsectxt}}
  {}
\RS@ifundefined{thmref}
  {\def\RSthmtxt{theorem~}\newref{thm}{name = \RSthmtxt}}
  {}
\RS@ifundefined{lemref}
  {\def\RSlemtxt{lemma~}\newref{lem}{name = \RSlemtxt}}
  {}

\numberwithin{equation}{section}
\numberwithin{figure}{section}
\theoremstyle{remark}
\newtheorem{rem}{\protect\remarkname}
\theoremstyle{plain}
\newtheorem{thm}{\protect\theoremname}
\theoremstyle{plain}
\newtheorem{prop}{\protect\propositionname}

\usepackage{pgfplots}
\pgfplotsset{compat=newest}
\makeatother

\providecommand{\propositionname}{Proposition}
\providecommand{\remarkname}{Remark}
\providecommand{\theoremname}{Theorem}

\begin{document}
\title{Barriers to the Transport of Diffusive Scalars in Compressible Flows}
\author{George Haller$^{a,}$\thanks{Corresponding author. Email: georgehaller@ethz.ch},
Daniel Karrasch$^{b}$ and Florian Kogelbauer$^{a}$}

\maketitle
(a) Institute for Mechanical Systems, ETH Zürich, Leonhardstrasse
21, 8092 Zürich, Switzerland

(b) Zentrum Mathematik M3, Technische Universität München, Boltzmannstrasse
3, 85748 Garching, Germany
\begin{abstract}
Our recent work identifies material surfaces in incompressible flows
that extremize the transport of an arbitrary, weakly diffusive scalar
field relative to neighboring surfaces. Such barriers and enhancers
of transport can be located directly from the deterministic component
of the velocity field without diffusive or stochastic simulations.
Here we extend these results to compressible flows and to diffusive
concentration fields affected by sources or sinks, as well as by spontaneous
decay. We construct diffusive transport extremizers with and without
constraining them on a specific initial concentration distribution.
For two-dimensional flows, we obtain explicit differential equations
and a diagnostic scalar field that identify the most observable extremizers
with pointwise uniform transport density. We illustrate our results
by uncovering diffusion barriers and enhancers in analytic, numerical,
and observational velocity fields.
\end{abstract}

\section{Introduction}

\textit{\emph{Transport barriers}} can informally be defined as observed
inhibitors of the spread of substances in a flow. They are well documented
in geophysics \cite{Weiss2008}, classical fluid dynamics \cite{Ottino1989},
plasma fusion \cite{Dinklage2005}, reactive flows \cite{Rosner2000}
and molecular dynamics \cite{Toda2005}, yet no general theory for
them has been available until recently. In \cite{Haller2018}, we
have put forward such a theory for incompressible flows and weakly
diffusing substances by defining and solving an extremum problem for
material surfaces that block the diffusive transport of passive scalars
more than neighboring surfaces do. 

The a priori restriction of this optimization problem to material
surfaces (codimension-one invariant manifolds of the flow in the extended
phase space of positions and time) is justified by the complete lack
of advective transport across such invariant surfaces. Indeed, for
small enough diffusivities, pointwise, finite-time, advective transport
through any non-material (i.e., non-invariant) surface is always larger
than diffusive transport. As a consequence, one should seek universal,
diffusion-independent transport barriers among material surfaces. 

Such barriers turn out to be computable and depend on the structure
of the diffusion tensor but not on the actual value of the diffusivities
\cite{Haller2018}. Thus, diffusion barriers remain well defined in
the limit of non-diffusive, purely advective transport. In that limit,
they form surfaces that will prevail as transport inhibitors or enhancers
under the presence of the slightest diffusion or uncertainty modeled
by Brownian motion. This gives a transport-barrier definition in the
advective limit, independent of any preferred coherence principle
and based solely on the physically well-defined and quantitative notion
of diffusive transport through a surface. This limiting property of
diffusion barriers eliminates the current ambiguity in locating coherent
structures in finite-time, advective transport where different coherence
principles give different results on the same flow \cite{Hadjighasem2017}.
Unlike set-based approaches to coherent advective transport, the approach
in \cite{Haller2018} does not require diffusion barriers to be closed,
and hence also finds open bottlenecks to transport such as fronts
and jets. This feature of the method is also important for closed
diffusion barriers to remain detectable even if they do not lie entirely
in the domain where velocity data is available.

While valid in arbitrary dimension, the results in \cite{Haller2018}
rely explicitly on the assumption that the flow carrying the concentration
field of interest is incompressible. Fluid flows arising in applications
are indeed practically incompressible, but air flows are relatively
easy to compress. This precludes the application of \cite{Haller2018}
to atmospheric transport problems, such as the identification of temperature
barriers surrounding the polar vortices (cf.~\cite{Bowman1993,Chen1994,Lekien2010,Serra2017b}).
Notable compressibility also arises in two-dimensional velocity fields
representing horizontal slices of planetary atmospheres, obtained
from observations \cite{Hadjighasem2016b} or from numerical models
\cite{Beron-Vera2008}. The dramatic accumulation of oil and flotsam
on the ocean surface \cite{DAsaro2018}, as well as the characteristically
non-conservative surface patterns formed by algae \cite{Zhong2012},
also necessitate the use of two-dimensional numerical models with
significantly compressible, two-dimensional velocity fields.

These examples of compressible velocity fields nevertheless invariably
conserve mass. For instance, oil remains buoyant and hence confined
to the ocean surface, thus there is no significant loss in the total
oil mass in the absence of other processes eroding it. Accordingly,
a velocity field model for surface oil transport should be mass conserving.
Inspired by such examples, we consider here diffusive transport in
the presence of a carrier flow that may be compressible but conserves
mass. At the same time, we also allow for variations of the transported
concentration field due to effects beyond diffusion. These effects
include contribution from distributed sources and sinks, as well as
spontaneous concentration decay in time governed by a potentially
time-dependent decay rate.

A number of prior approaches exist to weakly diffusive transport (see,
e.g., \cite{Weiss2008} for a survey), but only a handful of these
target structures in the compressible advection-diffusion equation.
Among these, \cite{Tang1996,Thiffeault2003} recast the advection-diffusion
equation in Lagrangian coordinates and suggest a quasi-reduction to
a one-dimensional diffusion PDE along the most contracting direction.
While this approach yields formal asymptotic scaling laws for stretching
and folding statistics along chaotic trajectories, such trajectories
become undefined for finite-time data sets that we seek to analyze
here. The residual velocity field of \cite{Pratt2016} offers an attractive
visualization tool for regions of enhanced or suppressed transport,
but requires already performed diffusive simulations as input, rather
than providing predictions for them from velocity data. The popular
effective diffusivity approach of \cite{Nakamura2008} is based on
the assumption of incompressibility (conservation of area), and hence
becomes inapplicable to compressible flows. We finally mention recent
work in \cite{Karrasch2016b} which provides an advection--diffusion
interpretation for the compressible dynamic isoperimetry methodology
developed in \cite{Froyland2017}. This latter, set-based approach
targets metastable or almost-invariant sets in a purely advective
transport context.

Our analysis here considers a mass-based (rather than volume-based)
concentration field $c(\mathbf{x},t)$. In the absence of diffusion,
spontaneous concentration-decay and concentration sources, the transport
of $c(\mathbf{x},t)$ in and out of an evolving material volume $V(t)$
would be pointwise zero due to the conservation of the mass of $V(t)$
by the flow. Source terms and spontaneous concentration decay add
a deterministic evolution to the concentration along particle trajectories,
and hence the initial concentration remains deterministically reproducible,
i.e., a conserved quantity along all particle motions in the absence
of diffusion. 

The presence of diffusion, however, erodes this conservation law,
as if trajectories were stochastic and hence the value of the initial
concentration along them could not be immediately reproduced just
from the knowledge of the present concentration, the initial time
and initial location along a fluid particle trajectory. Here, we will
seek transport barriers as material surfaces across which this diffusive
erosion of initial concentrations is stationary when compared to nearby
material surfaces. For incompressible flows, this barrier concept
will turn out to simplify exactly to the concept of most impermeable
material barriers to diffusion, as developed in \cite{Haller2018}.
In the present work, we will collectively refer to stationary surfaces
(minimizers, maximizers and saddle-type surfaces) of diffusive transport
as transport barriers without performing a second-order calculation
to identify their types. 

Beyond adding the effects of compressibility, sources, sinks and spontaneous
decay, our present analysis performs the transport extremization both
with and without conditioning it on a known initial concentration
field. In this context, unconstrained barriers are material surfaces
that prevail as stationary surfaces of transport even under concentration
gradients initially normal to them. Constrained barriers, in contrast,
are stationary surfaces of transport under a fixed initial diffusion-gradient
configuration. We derive mathematical criteria for both types of barriers
and illustrate these criteria first on explicitly known Navier--Stokes
flows, then on observational and numerical ocean data.

Our analysis and examples show that several well documented features
in a diffusing scalar field, such as jets and fronts, are technically
not minimizers of diffusive transport when constrained on a given
initial scalar field. This is at odds with the usual terminology by
which surfaces of large concentration gradients are generally referred
to as transport barriers, even though the actual transport through
them appears large precisely because of those large gradients. This
paradox has already been pointed out in the Eulerian frame but has
remained unresolved \cite{Nakamura2008}. Here we recover the same
effect in rigorous terms in the Lagrangian frame, and find that barriers
are transport maximizers with respect to all localized perturbations

\section{Set-up}

We consider the compressible advection-diffusion equation for a mass-unit-based
concentration field $c(\mathbf{x},t)$ in the general form 
\begin{align}
\partial_{t}\left(\rho c\right)+\mathbf{\bm{\nabla\cdot}}\left(\rho c\mathbf{v}\right) & =\nu\mathbf{\bm{\nabla}}\cdot\left(\rho\mathbf{D}\mathbf{\bm{\nabla}}c\right)-k(t)\rho c+f(\mathbf{x},t)\rho,\label{eq:adv-diff}\\
c(\mathbf{x},t_{0}) & =c_{0}(\mathbf{x}),\quad\rho(\mathbf{x},t_{0})=\rho_{0}(\mathbf{x}).\nonumber 
\end{align}
Here $\bm{\mathbf{\nabla}}$ denotes the gradient operation with respect
to the spatial variable $\mathbf{x}\in U\subset\mathbb{R}^{n}$ on
a compact domain $U$ with $n\geq1$; $\mathbf{v}(\mathbf{x},t)$
is an $n$-dimensional smooth, mass-conserving velocity field generating
the advective transport of $c(\mathbf{x},t)$ whose initial distribution
is $c_{0}(\mathbf{x})$; $\mathbf{D}(\mathbf{x},t)=\mathbf{D}^{T}(\mathbf{x},t)\in\mathbb{R}^{n\times n}$
is a dimensionless, positive definite diffusion-structure tensor describing
possible inhomogeneity, anisotropy and temporal variation in the diffusive
transport of $c$; $\rho(\mathbf{x},t)>0$ is the mass-density of
the carrier medium; $\nu\geq0$ is a small diffusivity parameter rendering
the full diffusion tensor $\nu\mathbf{D}$ small in norm; $k(\mathbf{x},t)$
is a space- and time-dependent coefficient governing spontaneous concentration
decay in the absence of diffusion; and $f(\mathbf{x},t)$ describes
the spatiotemporal sink- and source-distribution for the concentration.
We assume that the initial concentration $c(\mathbf{x},t_{0})=c_{0}(\mathbf{x})$
is of class $C^{2}$, and $\mathbf{D}(\mathbf{x},t)$, $k(t)$ and
$f(\mathbf{x},t)$ are at least Hölder-continuous, which certainly
holds if they are continuously differentiable. 

Without the spontaneous decay and source terms, eq.~(\ref{eq:adv-diff})
was apparently first obtained by Landau and Lifschitz \cite{Landau1966}
as a compressible, non-Fickian advection-diffusion equation for $\rho c$
(see also Thiffeault \cite{Thiffeault2003}). We note, however, that
with the modified velocity field
\begin{equation}
\mathbf{w}=\mathbf{v}+\frac{\nu}{\rho}\mathbf{D}\mathbf{\bm{\nabla}}\rho,\label{eq:wdef}
\end{equation}
eq.~(\ref{eq:adv-diff}) can also be recast as
\begin{equation}
\partial_{t}\left(\rho c\right)+\mathbf{\bm{\nabla\cdot}}\left(\rho c\mathbf{w}\right)=\nu\mathbf{\bm{\nabla}}\cdot\left(\mathbf{D}\mathbf{\bm{\nabla}}\left(\rho c\right)\right)-k(t)\rho c+f(\mathbf{x},t)\rho,\label{eq:adv-diff-3}
\end{equation}
an advection-diffusion equation with classic Fickian diffusion for
the scalar field $\rho c$ under the modified velocity field $\mathbf{w}$. 

Given a carrier velocity field \textbf{$\mathbf{v}(\mathbf{x},t)$
}of general divergence $\nabla\cdot\mathbf{v}(\mathbf{x},t)$, the
density $\rho$ featured in eqs.~(\ref{eq:adv-diff})-(\ref{eq:adv-diff-3})
must satisfy the equation of continuity 
\begin{equation}
\partial_{t}\rho+\mathbf{\bm{\nabla\cdot}}\left(\rho\mathbf{v}\right)=0.\label{eq:continuity}
\end{equation}

Combining the continuity equation (\ref{eq:continuity}) with (\ref{eq:adv-diff})
gives an alternative form of the compressible advection-diffusion
equation as 
\begin{equation}
\frac{Dc}{Dt}=\frac{1}{\rho}\nu\mathbf{\bm{\nabla}}\cdot\left(\rho\mathbf{D}\mathbf{\bm{\nabla}}c\right)-kc+f.\label{eq:adv-diff-1}
\end{equation}

The flow map induced by the velocity field $\mathbf{v}$ is $\mathbf{F}_{t_{0}}^{t}\colon\mathbf{x}_{0}\mapsto\mathbf{x}(t;t_{0},\mathbf{x}_{0})$,
mapping initial positions $\mathbf{x}_{0}\in U$ to their later positions
at time $t$. We assume that all trajectories stay in the domain $U$
of known velocities, i.e., $\mathbf{F}_{t_{0}}^{t}(U)\subset U$ holds
for all times $t$ of interest. We will be studying diffusive transport
through \emph{material surfaces} which are time-dependent families
of codimension-one differentiable manifolds satisfying
\[
\mathcal{M}(t)=\mathbf{F}_{t_{0}}^{t}\left(\mathcal{M}_{0}\right)\subset U,
\]
with $\mathcal{M}(t_{0})=\mathcal{M}_{0}$ denoting the initial position
of the material surface. Note that $\left(\mathcal{M}(t),t\right)$
is an $n$-dimensional invariant manifold in the extended phase space
of the non-autonomous ODE $\dot{\mathbf{x}}=\mathbf{v}(\mathbf{x},t).$ 

We denote by $\mathbf{\bm{\nabla}}_{0}\mathbf{F}_{t_{0}}^{t}$ the
gradient of $\mathbf{F}_{t_{0}}^{t}$ with respect to initial positions
$\mathbf{x}_{0}$, satisfying 
\begin{equation}
\mathbf{\det\bm{\nabla}}_{0}\mathbf{F}_{t_{0}}^{t}(\mathbf{x}_{0})=\exp\int_{t_{0}}^{t}\mathbf{\bm{\nabla}}\cdot\mathbf{v}\left(\mathbf{F}_{t_{0}}^{s}(\mathbf{x}_{0}),s\right)\,ds.\label{eq:detform}
\end{equation}
The equation of continuity (\ref{eq:continuity}) together with (\ref{eq:detform})
then yields the relation
\[
\rho\left(\mathbf{F}_{t_{0}}^{t}(\mathbf{x}_{0}),t\right)=\rho_{0}(\mathbf{x}_{0})\exp\left[-\int_{t_{0}}^{t}\mathbf{\bm{\nabla}}\cdot\mathbf{v}\left(\mathbf{F}_{t_{0}}^{s}(\mathbf{x}_{0}),s\right)\,ds\right]=\frac{\rho_{0}(\mathbf{x}_{0})}{\mathbf{\det\bm{\nabla}}_{0}\mathbf{F}_{t_{0}}^{t}(\mathbf{x}_{0})}.
\]

The smallness of the diffusivity parameter $\nu$ is not just a convenient
mathematical assumption: most diffusive processes in nature have very
small $\nu$ values (i.e., large Péclet numbers) associated with them
(see, e.g., Weiss and Provenzale \cite{Weiss2008}). The smallness
of $\nu$, however, does not automatically allow for simple perturbation
approaches, because (\ref{eq:adv-diff}) is a singularly perturbed
PDE for such $\nu$ values. 

\section{The compressible diffusion barrier problem}

To formulate the compressible diffusion barrier problem outlined in
the Introduction in precise terms, we first observe that for $\nu=0$,
eq.~(\ref{eq:adv-diff-1}) is solved by 
\[
c(\mathbf{x},t)=e^{-\int_{t_{0}}^{t}k(s)ds}c_{0}(\mathbf{F}_{t}^{t_{0}}(\mathbf{x}))+\int_{t_{0}}^{t}e^{-\int_{s}^{t}k(\sigma)d\sigma}f(\mathbf{F}_{t}^{s}(\mathbf{x}),s)\,ds.
\]
Therefore, the function 
\begin{equation}
\mu(\mathbf{x},t)\coloneqq e^{\int_{t_{0}}^{t}k(s)ds}c(\mathbf{x},t)-\int_{t_{0}}^{t}e^{\int_{t_{0}}^{s}k(\sigma)d\sigma}f(\mathbf{F}_{t}^{s}(\mathbf{x}),s)\,ds,\label{eq:mudef}
\end{equation}
returning the initial concentration at time $t_{0}$ along characteristics
of (\ref{eq:adv-diff-1}), is conserved along trajectories, given
that $\mu(\mathbf{x}(t),t)=c_{0}(\mathbf{F}_{t}^{t_{0}}(\mathbf{x}))\equiv c_{0}(\mathbf{x}_{0})$. 

For nonzero $\nu$ values, $\mu$ is no longer conserved along trajectories
of $\mathbf{v}(\mathbf{x},t)$. In that case, $\frac{D}{Dt}\mu(\mathbf{x}(t),t)$
measures the irreversibility in the evolution of $c(\mathbf{x},t)$
along trajectories. Specifically, we have 
\begin{align}
\frac{D}{Dt}\mu(\mathbf{x},t) & =\frac{D}{Dt}\left[e^{\int_{t_{0}}^{t}k(s)ds}c(\mathbf{F}_{t_{0}}^{t}(\mathbf{x}_{0}),t)-\int_{t_{0}}^{t}e^{\int_{t_{0}}^{s}k(\sigma)d\sigma}f(\mathbf{F}_{t_{0}}^{s}(\mathbf{x}_{0}),s)\,ds\right]\nonumber \\
 & =k(t)e^{\int_{t_{0}}^{t}k(s)ds}c\left(\mathbf{F}_{t_{0}}^{t}(\mathbf{x}_{0}),t\right)+e^{\int_{t_{0}}^{t}k(s)ds}\frac{D}{Dt}c\left(\mathbf{F}_{t_{0}}^{t}(\mathbf{x}_{0}),t\right)-e^{\int_{t_{0}}^{t}k(s)ds}f\left(\mathbf{F}_{t_{0}}^{t}(\mathbf{x}_{0}),t\right)\nonumber \\
 & =\nu\frac{1}{\rho(\mathbf{x},t)}\mathbf{\bm{\nabla}}\cdot\left(\rho(\mathbf{x},t)\mathbf{D}(\mathbf{x},t)\mathbf{\bm{\nabla}}\left[\mu(\mathbf{x},t)+\int_{t_{0}}^{t}e^{\int_{t_{0}}^{s}k(\sigma)d\sigma}f(\mathbf{F}_{t}^{s}(\mathbf{x}),s)\,ds\right]\right).\label{eq:Dmu/Dt}
\end{align}
This latter PDE for the evolution of $\mu(\mathbf{x},t)$ can also
be rewritten in Lagrangian coordinates applying the formulas of Tang
and Boozer \cite{Tang1996} and Thiffeault \cite{Thiffeault2003}
that transform the classic advection-diffusion equation to Lagrangian
coordinates. When applied to $\hat{\mu}(\mathbf{x}_{0},t):=\mu(\mathbf{F}_{t_{0}}^{t}(\mathbf{x}_{0}),t)$,
those formulas directly give
\begin{align}
\partial_{t}\hat{\mu}(\mathbf{x}_{0},t) & =\nu\frac{1}{\rho_{0}(\mathbf{x}_{0})}\mathbf{\bm{\nabla}}_{0}\cdot\left(\mathbf{T}_{t_{0}}^{t}(\mathbf{x}_{0})\mathbf{\bm{\nabla}}_{0}\left[\hat{\mu}(\mathbf{x}_{0},t)+b(\mathbf{x}_{0},t)\right]\right),\label{eq:mu-adv-diff}
\end{align}
with the notation 
\begin{align}
\mathbf{T}_{t_{0}}^{t}(\mathbf{x}_{0}) & \coloneqq\rho_{0}(\mathbf{x}_{0})\left[\mathbf{\bm{\nabla}}_{0}\mathbf{F}_{t_{0}}^{t}(\mathbf{x}_{0})\right]^{-1}\mathbf{D}(\mathbf{F}_{t_{0}}^{t}(\mathbf{x}_{0}),t)\left[\mathbf{\bm{\nabla}}_{0}\mathbf{F}_{t_{0}}^{t}(\mathbf{x}_{0})\right]^{-T},\label{eq:Tdef}\\
b(\mathbf{x}_{0},t) & \coloneqq\int_{t_{0}}^{t}e^{\int_{t_{0}}^{s}k(\sigma)d\sigma}f(\mathbf{F}_{t_{0}}^{s}(\mathbf{x}_{0}),s)\,ds.\nonumber 
\end{align}
The definition of the\emph{ transport tensor} $\mathbf{T}_{t_{0}}^{t}(\mathbf{x}_{0})$
in (\ref{eq:Tdef}) is similar to that in \cite{Haller2018} but the
present definition also contains the initial density $\rho_{0}(\mathbf{x}_{0})$
as a factor and no longer assumes the flow map to be volume-preserving.
\begin{rem}
It will be crucial in our present derivation that no spatially dependent
terms beyond the initial density remain in front of the divergence
operation in eq.~(\ref{eq:Dmu/Dt}). That is only the case if the
flow map of the characteristics of eq.~(\ref{eq:adv-diff}) is linear
in $\mathbf{x}_{0}$ for $\nu=0$. This, in turn, only holds if the
right-hand side of (\ref{eq:adv-diff}) is a linear function of $c(\mathbf{x},t)$,
as we have assumed. Consequently, (\ref{eq:adv-diff}) is the broadest
class of PDEs to which our present approach is applicable. 

The following result is critical to our analysis, establishing a leading-order
formula for the total transport of the scalar field $\mu(x,t)$ field
through an evolving material surface.
\end{rem}
\begin{thm}
\label{thm:leading_order_transport}The total transport of $\mu$
through an arbitrary, evolving material surface $\mathcal{M}(t)=\mathbf{F}_{t_{0}}^{t}\left(\mathcal{M}_{0}\right)$
over the time interval $[t_{0},t_{1}]$ is given by 
\begin{align}
\Sigma_{t_{0}}^{t_{1}}(\mathcal{M}_{0}) & =\nu\int_{t_{0}}^{t_{0}}\int_{\mathcal{M}_{0}}\left\langle \mathbf{T}_{t_{0}}^{t}\left(\mathbf{\bm{\nabla}}_{0}c_{0}(\mathbf{x}_{0})+\mathbf{\bm{\nabla}}_{0}b(\mathbf{x}_{0},)\right)t,\mathbf{n}_{0}\right\rangle \,dA_{0}dt+o(\nu),\label{eq:Sigmadef}
\end{align}
with $o(\nu)$ denoting a quantity that, for $\nu\to0,$ tends to
zero pointwise at any $\mathbf{x}_{0}\in\mathcal{M}_{0}$ even after
division by $\nu$. 
\end{thm}
\begin{proof}
See \secref{Appendix-A}.
\end{proof}
Next, we will seek diffusion barriers as codimension-one stationary
surfaces of the leading-order term in the expression of $\Sigma_{t_{0}}^{t_{1}}(\mathcal{M}_{0})$
in two different settings. First, we consider the initial tracer concentration
unknown or uncertain, and assume the most diffusion-prone initial
concentration distribution for $c$ along each material surface. Next,
we consider an arbitrary but fixed initial concentration and seek
material surfaces that render diffusive transport stationary under
this initial concentration.

\section{Unconstrained diffusion barriers\label{sec:Unconstrained-diffusion-barriers}}

To compare the intrinsic ability of different material surfaces to
withstand diffusion, we now subject each material surface to the same,
locally customized initial concentration gradient setting that makes
the surface a priori maximally conducive to diffusive transport. Specifically,
we initialize the initial concentration along the initial position
$\mathcal{M}_{0}$ of any material surface in such a way that 
\begin{equation}
\mathbf{\bm{\nabla}}_{0}c_{0}(\mathbf{x}_{0})=\frac{K_{0}}{\nu^{\alpha}}\mathbf{n}_{0}(\mathbf{x}_{0}),\quad\nu>0,\quad\mathbf{x}_{0}\in\mathcal{M}_{0},\label{eq:uniformity assumption}
\end{equation}
for some constant $\alpha\in(0,1)$. In other words, we prescribe
uniformly high concentration gradients along $\mathcal{M}_{0}$ that
are perfectly aligned with the normals of $\mathcal{M}_{0}$ and grow
as $\nu^{-\alpha}$ as $\nu\to0$. We will refer to (\ref{eq:uniformity assumption})
as the \emph{uniformity assumption. }This assumption focuses our analysis
on the intrinsic ability of a material surface to block diffusion,
rather than on its position relative to features present in a specific
initial concentration field. 
\begin{rem}
The uniformity assumption in \cite{Haller2018} is a specific case
of (\ref{eq:uniformity assumption}) with $\alpha=0.$ If sinks and
sources are absent ($f(\mathbf{x},t)\equiv0$), as is the case in
\cite{Haller2018}, we can also select $\alpha=0$ in (\ref{eq:uniformity assumption})
and still obtain the upcoming results of this section.
\end{rem}
By the compactness of $U$ and the time interval $[t_{0},t_{1}],$
we can also select a constant bound $M_{0}>0$ so that 
\[
(t_{1}-t_{0})\left|\int_{t_{0}}^{t_{1}}\mathbf{\bm{\nabla}}_{0}\left[\frac{1}{\rho_{0}(\mathbf{x}_{0})}\mathbf{\bm{\nabla}}_{0}\cdot\left(\mathbf{T}_{t_{0}}^{s}(\mathbf{x}_{0})\mathbf{\bm{\nabla}}_{0}b(\mathbf{x}_{0},s)\right)\right]ds\right|\leq M_{0}
\]
for all $\mathbf{x}_{0}\in U$, given that $\mathbf{T}_{t_{0}}^{s}(\mathbf{x}_{0})\mathbf{\bm{\nabla}}_{0}b(\mathbf{x}_{0},s)$
is assumed $C^{1}$ for all $s$ values. We can then rewrite $\Sigma_{t_{0}}^{t_{1}}(\mathcal{M}_{0})$
in (\ref{eq:Sigmadef}) as 
\[
\Sigma_{t_{0}}^{t_{1}}(\mathcal{M}_{0})=\nu(t_{1}-t_{0})K\int_{\mathcal{M}_{0}}\left\langle \mathbf{\bar{T}}_{t_{0}}^{t_{1}}\mathbf{n}_{0},\mathbf{n}_{0}\right\rangle \,dA_{0}+o(\nu)+\mathcal{O}\text{\ensuremath{\left(\nu\frac{M_{0}\rho_{0}\nu^{\alpha}}{K_{0}}\right)}}.
\]
This leads to the normalized total transport of $\mu(\mathbf{x},t)$
in the form

\begin{equation}
\tilde{\Sigma}_{t_{0}}^{t_{1}}(\mathcal{M}_{0}):=\frac{\Sigma_{t_{0}}^{t_{1}}(\mathcal{M}_{0})}{\nu K\left(t_{1}-t_{0}\right)A_{0}(\mathcal{M}_{0})}=\mathcal{T}{}_{t_{0}}^{t_{1}}(\mathcal{M}_{0})+o(\nu^{\alpha}),\quad\alpha\in(0,1],\label{eq:total transport}
\end{equation}
where the \emph{transport functional, }
\begin{equation}
\mathcal{T}{}_{t_{0}}^{t_{1}}(\mathcal{M}_{0}):=\frac{\int_{\mathcal{M}_{0}}\left\langle \mathbf{n}_{0},\mathbf{\bar{T}}_{t_{0}}^{t_{1}}\mathbf{n}_{0}\right\rangle dA_{0}}{\int_{\mathcal{M}_{0}}dA_{0}},\label{eq:transport functional}
\end{equation}
measures the leading-order diffusive transport of $c(\mathbf{x},t)$
through the material surface $\mathcal{M}(t)$ over the period $[t_{0},t_{1}]$.
This functional formally coincides with the transport functional obtained
in \cite{Haller2018} for incompressible flows with $k(t)\equiv f(\mathbf{x},t)\equiv0$
in the PDE (\ref{eq:adv-diff}). The only, minor difference here is
in the definition (\ref{eq:Tdef}) of $\mathbf{\bar{T}}_{t_{0}}^{t_{1}}$,
which now features the initial density field $\rho_{0}(\mathbf{x}_{0})$.
Importantly, the present theory returns the results of the incompressible
theory when applied to incompressible flows with uniform density.
We stress that $\mathcal{T}{}_{t_{0}}^{t_{1}}(\mathcal{M}_{0})$ can
be computed for any initial surface $\mathcal{M}_{0}$ directly from
the trajectories of \textbf{$\mathbf{v}$} without solving the PDE
(\ref{eq:adv-diff}).

We propose that what makes a barrier most observable is a near-uniform
concentration jump along it. By continuity with respect to all quantities
involved over finite time intervals, however, surfaces delineating
near-uniform concentration jumps form continuous families and hence
cannot be uniquely identified. The theoretical centerpiece of such
a family is still well defined by \emph{uniform barriers} which are
characterized by constant pointwise transport density at leading order.
By formula (\ref{eq:transport functional}), these surfaces satisfy
\begin{equation}
\left\langle \mathbf{n}_{0},\mathbf{\bar{T}}_{t_{0}}^{t_{1}}\mathbf{n}_{0}\right\rangle =\mathcal{T}_{0}=const.\label{eq:uniform barrier construntion}
\end{equation}
Because of the formal coincidence between the transport functional
$\mathcal{T}{}_{t_{0}}^{t_{1}}(\mathcal{M}_{0})$ defined in (\ref{eq:transport functional})
and that arising in the incompressible case, the general necessary
criterion of \cite{Haller2018} for uniform barriers remains valid
here and can be stated with the help of the tensor family
\begin{equation}
\mathbf{E}_{\mathcal{T}_{0}}(\mathbf{x}_{0}):=\mathbf{\bar{T}}_{t_{0}}^{t_{1}}-\mathcal{T}_{0}\mathbf{I}\label{eq:nullsurface}
\end{equation}
as follows. 
\begin{thm}
\label{thm:uncnstrained barriers general}Under assumption (\ref{eq:uniformity assumption}),
a uniform minimizer $\mathcal{M}_{0}$ of the transport functional
$\mathcal{T}{}_{t_{0}}^{t_{1}}$ is necessarily a non-negatively traced
null-surface of the tensor field $\mathbf{E}_{\mathcal{T}}$, i.e,
\begin{equation}
\left\langle \mathbf{n}_{0}(\mathbf{x}_{0}),\mathbf{E}_{\mathcal{T}_{0}}(\mathbf{x}_{0})\mathbf{n}_{0}(\mathbf{x}_{0})\right\rangle =0,\qquad\mathrm{trace\,\mathbf{E}_{\mathcal{T}_{0}}(\mathbf{x}_{0})\geq0},\label{eq:barriercrit}
\end{equation}
holds at every point $\mathbf{x}_{0}\in\mathcal{M}_{0}$ with unit
normal $\mathbf{n}_{0}(\mathbf{x}_{0})$ to $\mathcal{M}_{0}$. Similarly,
a uniform maximizer $\mathcal{M}_{0}$ of $\mathcal{T}{}_{t_{0}}^{t_{1}}$
is necessarily a non-positively traced null surface of the tensor
field $\mathbf{E}_{\mathcal{T}}$, i.e, 
\begin{equation}
\left\langle \mathbf{n}_{0}(\mathbf{x}_{0}),\mathbf{E}_{\mathcal{T}_{0}}(\mathbf{x}_{0})\mathbf{n}_{0}(\mathbf{x}_{0})\right\rangle =0,\qquad\mathrm{trace\,\mathbf{E}_{\mathcal{T}_{0}}(\mathbf{x}_{0})\leq0},\label{eq:enhancercrit}
\end{equation}
holds at every point $\mathbf{x}_{0}\in\mathcal{M}_{0}$.
\end{thm}
Finally, by direct analogy with the incompressible case treated in
\cite{Haller2018}, a diagnostic field measuring the strength of unconstrained
diffusion barriers is given by the Diffusion Barriers Strength (DBS)
field, defined as
\begin{equation}
DBS_{t_{0}}^{t_{1}}(\mathbf{x}_{0})=\mathrm{trace}\,\mathbf{\bar{T}}_{t_{0}}^{t_{1}}(\mathbf{x}_{0}).\label{eq:DBS unconstrained}
\end{equation}
This follows because, as shown in \cite{Haller2018}, the leading-order
change in the non-dimensionalized transport functional $\mathcal{T}{}_{t_{0}}^{t_{1}}$
under a small, surface-area-preserving perturbation localized in an
$\mathcal{O}(\epsilon)$ neighborhood of a point $\mathbf{x}_{0}\in\mathcal{M}_{0}$
is given by $\epsilon DBS_{t_{0}}^{t_{1}}(\mathbf{x}_{0})$. Of all
uniform extremizers, therefore, those along the ridges of $\left|DBS_{t_{0}}^{t_{1}}(\mathbf{x}_{0})\right|$
will prevail as the strongest inhibitors or enhancers of diffusive
transport. 

\subsection{Relationship between diffusion barriers and classic invariant manifolds}

Assume that the diffusion structure tensor is $\mathbf{D}\equiv\mathbf{I}$
(homogeneous and isotropic diffusion), $\rho_{0}(\mathbf{x}_{0})=\rho_{0}=const.$
(homogeneous initial density) and the conservation law (\ref{eq:uniform barriers})
holds for $t_{1}\in(t_{0},\infty)$ over the material surface evolving
from $\mathcal{M}_{0}$ under the flow map. This implies 
\[
\left\langle \mathbf{n}_{0},\left(\frac{1}{t_{1}-t_{0}}\int_{t_{0}}^{t_{1}}\rho_{0}\left[\mathbf{\bm{\nabla}}_{0}\mathbf{F}_{t_{0}}^{t}(\mathbf{x}_{0})\right]^{-1}\left[\mathbf{\bm{\nabla}}_{0}\mathbf{F}_{t_{0}}^{t}(\mathbf{x}_{0})\right]^{-T}dt\right)\mathbf{n}_{0}\right\rangle =\mathcal{T}_{0},
\]
or, equivalently, 
\[
\frac{1}{t_{1}-t_{0}}\int_{t_{0}}^{t_{1}}\left|\left[\mathbf{\bm{\nabla}}_{0}\mathbf{F}_{t_{0}}^{t}(\mathbf{x}_{0})\right]^{-T}\mathbf{n}_{0}(\mathbf{x}_{0})\right|^{2}dt=\frac{\mathcal{T}_{0}}{\rho_{0}}=const.,\qquad t_{1}\in(t_{0},\infty).
\]

At the same time, the normal component of an initially normal unit
perturbation to $\mathcal{M}_{0}$, represented by $\mathbf{n}_{0}(\mathbf{x}_{0})$,
is given by the orthogonal projection of the advected normal $\mathbf{\bm{\nabla}}_{0}\mathbf{F}_{t_{0}}^{t}(\mathbf{x}_{0})\mathbf{n}_{0}(\mathbf{x}_{0})$
onto the unit normal
\[
\mathbf{n}\left(\mathbf{F}_{t_{0}}^{t}(\mathbf{x}_{0})\right)=\frac{\left[\mathbf{\bm{\nabla}}_{0}\mathbf{F}_{t_{0}}^{t}(\mathbf{x}_{0})\right]^{-T}\mathbf{n}_{0}(\mathbf{x}_{0})}{\left|\left[\mathbf{\bm{\nabla}}_{0}\mathbf{F}_{t_{0}}^{t}(\mathbf{x}_{0})\right]^{-T}\mathbf{n}_{0}(\mathbf{x}_{0})\right|},
\]
i.e., by the the normal repulsion rate $\mathbf{\sigma}_{t_{0}}^{t}(\mathbf{x}_{0})$,
computed as
\begin{align*}
\mathbf{\sigma}_{t_{0}}^{t}(\mathbf{x}_{0}) & =\left\langle \mathbf{\bm{\nabla}}_{0}\mathbf{F}_{t_{0}}^{t}(\mathbf{x}_{0})\mathbf{n}_{0}(\mathbf{x}_{0}),\frac{\left[\mathbf{\bm{\nabla}}_{0}\mathbf{F}_{t_{0}}^{t}(\mathbf{x}_{0})\right]^{-T}\mathbf{n}_{0}(\mathbf{x}_{0})}{\left|\left[\mathbf{\bm{\nabla}}_{0}\mathbf{F}_{t_{0}}^{t}(\mathbf{x}_{0})\right]^{-T}\mathbf{n}_{0}(\mathbf{x}_{0})\right|}\right\rangle \\
 & =\frac{1}{\left|\left[\mathbf{\bm{\nabla}}_{0}\mathbf{F}_{t_{0}}^{t}(\mathbf{x}_{0})\right]^{-T}\mathbf{n}_{0}(\mathbf{x}_{0})\right|}.
\end{align*}
Therefore, we have
\[
\frac{1}{t_{1}-t_{0}}\int_{t_{0}}^{t_{1}}\frac{1}{\left[\mathbf{\sigma}_{t_{0}}^{t}(\mathbf{x}_{0})\right]^{2}}dt=\frac{\mathcal{T}_{0}}{\rho_{0}}=const.,
\]
or, equivalently,
\[
\left\Vert \mathbf{\sigma}_{t_{0}}^{t}(\mathbf{x}_{0})^{-1}\right\Vert _{L^{2}\left(t_{0},t_{1}\right)}=\frac{\mathcal{T}_{0}}{\rho_{0}}=const.
\]
This shows that the temporally $L^{2}$-normed reciprocal of the normal
stretching rate along the manifold $\mathcal{M}_{0}$ should be spatially
constant along diffusion barriers. This is certainly satisfied for
$t_{1}\to\infty$ along stable manifolds of fixed points and periodic
orbits, as well as along quasiperiodic invariant tori.

\subsection{Unconstrained diffusion barriers in two-dimensional flows}

In arbitrary dimensions, the first equation defining null surfaces
in (\ref{eq:barriercrit}) and (\ref{eq:enhancercrit}) are partial
differential equations with a priori unknown solvability properties.
For two-dimensional flows, however, the null-surfaces become curves
satisfying ordinary differential equations that turn out to be explicitly
solvable (cf.~\cite{Haller2018}). These differential equations can
be expressed in terms of the invariants of the time-averaged, diffusion-structure-weighted
Cauchy--Green strain tensor 
\[
\mathbf{\bar{C}}_{\mathbf{D}}(\mathbf{x}_{0})\coloneqq\frac{1}{t_{1}-t_{0}}\int_{t_{0}}^{t_{1}}\det\left[\mathbf{D}\left(\mathbf{F}_{t_{0}}^{t}(\mathbf{x}_{0}),t\right)\right]\left[\mathbf{T}_{t_{0}}^{t}(\mathbf{x}_{0})\right]^{-1}dt
\]
as follows.
\begin{thm}
\label{thm:unconstrained barriers 2D}For two-dimensional flows ($n=2$),
let $\bm{\xi}_{i}(\mathbf{x}_{0})\in\mathbb{R}^{2}$ denote the unit
eigenvectors corresponding to the eigenvalues $0<\lambda_{1}(\mathbf{x}_{0})\leq\lambda_{2}(\mathbf{x}_{0})$
of the positive definite tensor $\mathbf{\bar{C}}_{\mathbf{D}}(\mathbf{x}_{0})$.
A uniform extremizer $\mathcal{M}_{0}$ of the transport functional
$\mathcal{T}{}_{t_{0}}^{t_{1}}$ is then necessarily a trajectory
of the direction field family
\begin{equation}
\mathbf{x}_{0}^{\prime}=\bm{\eta}_{\mathcal{T}_{0}}^{\pm}(\mathbf{x}_{0})\coloneqq\sqrt{\frac{\lambda_{2}(\mathbf{x}_{0})-\mathcal{T}_{0}}{\lambda_{2}(\mathbf{x}_{0})-\lambda_{1}(\mathbf{x}_{0})}}\mathbf{\bm{\xi}}_{1}(\mathbf{x}_{0})\pm\sqrt{\frac{\mathcal{T}_{0}-\lambda_{1}(\mathbf{x}_{0})}{\lambda_{2}(\mathbf{x}_{0})-\lambda_{1}(\mathbf{x}_{0})}}\mathbf{\bm{\xi}}_{2}(\mathbf{x}_{0}),\qquad\mathbf{x}_{0}\in U_{\mathcal{T}_{0}},\label{eq:eta field}
\end{equation}
 defined on the spatial domain $U_{\mathcal{T}_{0}}=\left\{ \mathbf{x}_{0}\in U:\,\lambda_{1}(\mathbf{x}_{0})\leq\mathcal{T}_{0}\leq\lambda_{2}(\mathbf{x}_{0})\right\} $.
\end{thm}
The domain of definition $U_{\mathcal{T}_{0}}$ of the direction field
family $\bm{\eta}_{\mathcal{T}_{0}}^{\pm}(\mathbf{x}_{0})$ is precisely
the spatial domain where the tensor $\mathbf{E}_{\mathcal{T}_{0}}(\mathbf{x}_{0})$
defined in (\ref{eq:nullsurface}) is indefinite (Lorentzian) and
hence indeed have well-defined null-surfaces. Formula (\ref{eq:eta field})
enables a detailed computation of diffusion barriers in two-dimensions
based on a numerical approximation of the flow map $\mathbf{F}_{t_{0}}^{t}(\mathbf{x}_{0})$
obtained on a grid of initial conditions (see \cite{Haller2018} for
details). 

As shown in \cite{Haller2018} for two-dimensional flows, we have
$\mathbf{\bar{T}}_{t_{0}}^{t_{1}}=\det\mathbf{\bar{C}}_{\mathbf{D}}\mathbf{\bar{C}}_{\mathbf{D}}^{-1}.$
This implies that the DBS field defined in (\ref{eq:DBS unconstrained})
in the present, two-dimensional case, can be evaluated as
\begin{equation}
DBS_{t_{0}}^{t_{1}}(\mathbf{x}_{0})=\mathrm{trace}\,\mathbf{\bar{T}}_{t_{0}}^{t_{1}}(\mathbf{x}_{0})=\det\mathbf{\bar{C}}_{\mathbf{D}}(\mathbf{x}_{0})\mathbf{\mathrm{trace}\,\bar{C}}_{\mathbf{D}}^{-1}(\mathbf{x}_{0})=\mathbf{\mathrm{trace}\,\bar{C}}_{\mathbf{D}}(\mathbf{x}_{0}).\label{eq:unconstrained DBS}
\end{equation}

The results in Theorems \ref{thm:leading_order_transport} and \ref{thm:uncnstrained barriers general}
are directly applicable to incompressible flows as well, in which
case they agree with the results in \cite{Haller2018}. The derivation
in \cite{Haller2018}, however, does not apply to the present, compressible
case, while our argument leading to the normalized total transport
$\tilde{\Sigma}_{t_{0}}^{t_{1}}(\mathcal{M}_{0})$ in (\ref{eq:total transport})
is general enough to cover the incompressible case as well.

\section{Constrained diffusion extremizers\label{sec:Constrained-diffusion-barriers}}

The above formulation is independent of the initial concentration
$c_{0}(\mathbf{x})$ and assumes large enough initial gradients along
the initial surface $\mathcal{M}_{0}$ that continue to dominate contributions
from concentration decay and source terms (cf. the uniformity assumption
(\ref{eq:uniformity assumption})). If, however, we wish to extremize
diffusive transport with respect to a specific initial concentration
field $c_{0}(\mathbf{x}_{0})$, as opposed to unspecified or uncertain
initial concentrations, then we can no longer prescribe (\ref{eq:uniformity assumption})
along an arbitrary material surface $\mathcal{M}_{0}$. Indeed, some
material surfaces will experience $\mathbf{\bm{\nabla}}_{0}c_{0}$
vectors favorable to cross-surface diffusion while others will not. 

Considering $\mathbf{\bm{\nabla}}_{0}c_{0}$ as a given quantity and
using formula (\ref{eq:Sigmadef}), we rewrite the normed, total transport
of $\mu(\mathbf{x},t)$ as 
\begin{align}
\tilde{\Sigma}_{t_{0}}^{t_{1}}(\mathcal{M}_{0}): & =\frac{1}{\int_{\mathcal{M}_{0}}dA_{0}}\int_{t_{0}}^{t_{1}}\int_{\partial V(t_{0})}\left\langle \mathbf{T}_{t_{0}}^{s}\left(\mathbf{\bm{\nabla}}_{0}c(\mathbf{x}_{0})+\mathbf{\bm{\nabla}}_{0}b(\mathbf{x}_{0},s)\right),\mathbf{n}_{0}\right\rangle \,dA_{0}ds+o(\nu^{\alpha})\nonumber \\
 & =\frac{\int_{\mathcal{M}_{0}}\left\langle \mathbf{\bar{q}}_{t_{0}}^{t_{1}},\mathbf{n}_{0}\right\rangle \,dA_{0}}{\int_{\mathcal{M}_{0}}dA_{0}}+o(\nu^{\alpha}),\label{eq:sigma tilde}
\end{align}
with the help of the \emph{transport vector field}
\begin{equation}
\mathbf{\bar{q}}_{t_{0}}^{t_{1}}(\mathbf{x}_{0}):=\int_{t_{0}}^{t_{1}}\left[\mathbf{T}_{t_{0}}^{t}(\mathbf{x}_{0})\left(\mathbf{\bm{\nabla}}_{0}c_{0}(\mathbf{x}_{0})+\mathbf{\bm{\nabla}}_{0}b(\mathbf{x}_{0},t)\right)\right]dt.\label{eq:qdef}
\end{equation}
The sign of the net total transport $\tilde{\Sigma}_{t_{0}}^{t_{1}}(\mathcal{M}_{0})$
through $\mathcal{M}_{0}$ is now not necessarily positive, which
necessitates extremizing the normed transport (the time-integral of
the geometric flux in the terminology of MacKay \cite{MacKay1994}).
To this end, we seek to extremize the area-normalized, leading-order,
normed diffusive transport 
\[
\tilde{\mathcal{E}}(\mathcal{M}_{0}):=\frac{\int_{\mathcal{M}_{0}}\left|\left\langle \mathbf{\bar{q}}_{t_{0}}^{s},\mathbf{n}_{0}\right\rangle \right|\,dA_{0}}{\int_{\mathcal{M}_{0}}dA_{0}}
\]
with respect to the initial surface $\mathcal{M}_{0}$, for which
a necessary condition is the vanishing Gâteaux derivative
\begin{equation}
\delta\mathcal{\tilde{E}}(\mathcal{M}_{0})=0.\label{eq:variproblem_known_IC}
\end{equation}
For such stationary surfaces of $\tilde{\mathcal{E}}$, we have the
following result.
\begin{thm}
\label{thm:first_integral}Along any solution $\mathcal{M}_{0}$ of
the variational problem (\ref{eq:variproblem_known_IC}), there exists
a constant $C\in\mathbb{R}$ such that 
\[
\left[\left|\left\langle \mathbf{\bar{q}}_{t_{0}}^{t_{1}}\left(\mathbf{x}_{0}\right),\mathbf{n}_{0}\left(\partial_{\mathbf{s}}\mathbf{x}_{0}\right)\right\rangle \right|-\mathcal{T}_{0}\right]\sqrt{\det\mathbf{G}\left(\partial_{\mathbf{s}}\mathbf{x}_{0}\right)}=C,\qquad\mathbf{x}_{0}\in\mathcal{M}_{0},
\]
with $\mathcal{T}_{0}:=\tilde{\mathcal{E}}(\mathcal{M}_{0})$ denoting
the value of the normed transport functional on $\mathcal{M}_{0}$. 
\end{thm}
\begin{proof}
See \secref{Appendix-B}.
\end{proof}
As argued in \cite{Haller2018}, the most observable stationary surfaces
of diffusive transport are those with nearly uniformly high gradients
along them, associated with a nearly uniform pointwise transport density.
The theoretical centerpieces of such regions are provided by surfaces
with perfectly uniform transport density, which, by the statement
of Theorem (\ref{thm:first_integral}) with $C=0$, satisfy the implicit
equation
\begin{equation}
\left|\left\langle \mathbf{\bar{q}}_{t_{0}}^{t_{1}}(\mathbf{x}_{0}(\mathbf{s})),\mathbf{n}_{0}(\partial_{\mathbf{s}}\mathbf{x}_{0}(\mathbf{s}))\right\rangle \right|=\mathcal{T}_{0}\label{eq:uniform barriers}
\end{equation}
for any selected pointwise transport density $\mathcal{T}_{0}\geq0.$ 
\begin{rem}
\label{rem:DBSremark} By analogy with the DBS field defined in (\ref{eq:DBS unconstrained}),
a direct measure of the local strength of a uniform constrained barrier
is 
\begin{equation}
DBS_{t_{0}}^{t_{1}}(\mathbf{x}_{0})=\left|\mathbf{\bar{q}}_{t_{0}}^{t_{1}}\left(\mathbf{x}_{0}\right)\right|,\label{eq:DBSdef}
\end{equation}
a predictive diagnostic field applicable in any dimension. This diagnostic
emerges as the normed, leading-order change in the functional $\mathcal{E}(\mathcal{M}_{0})$
under small, localized perturbations to a stationary surface $\mathcal{M}_{0}$.
Ridges of $DBS_{t_{0}}^{t_{1}}(\mathbf{x}_{0})$ are expected to highlight
the strongest diffusive transport barriers over the time interval
$[t_{0},t_{1}].$
\end{rem}
\begin{rem}
\label{rem:simplified q}When $k(t)\equiv0$ and $f(\mathbf{x},t)\equiv0$
holds (no sources or sinks), then we have $\hat{\mu}(\mathbf{x}_{0},t)\equiv c(\mathbf{F}_{t_{0}}^{t}(\mathbf{x}_{0}),t)$
and $\tilde{\Sigma}_{t_{0}}^{t_{1}}(\mathcal{M}_{0})$ in (\ref{eq:sigma tilde})
can exactly---i.e., without the $o(\nu)$ error---be represented
as 
\begin{equation}
\tilde{\Sigma}_{t_{0}}^{t_{1}}(\mathcal{M}_{0})=\nu\frac{1}{\int_{\mathcal{M}_{0}}dA_{0}}\int_{t_{0}}^{t_{1}}\int_{\mathcal{M}_{0}}\left\langle \mathbf{T}_{t_{0}}^{t}(\mathbf{x}_{0})\mathbf{\bm{\nabla}}_{0}c(\mathbf{F}_{t_{0}}^{t}(\mathbf{x}_{0}),t),\mathbf{n}_{0}(\mathbf{x}_{0})\right\rangle \,dA_{0}\,dt,\label{eq:sigma tilde simp1}
\end{equation}
as ones verifies from eq.~(\ref{eq:sigma first eq}) of \secref{Appendix-A}.
Furthermore, noting that in this case, $\hat{\mu}(\mathbf{x}_{0},t)\equiv c(\mathbf{F}_{t_{0}}^{t}(\mathbf{x}_{0}),t),$
we obtain from the chain rule that
\begin{align}
\mathbf{T}_{t_{0}}^{t}(\mathbf{x}_{0})\mathbf{\bm{\nabla}}_{0}c(\mathbf{F}_{t_{0}}^{t}(\mathbf{x}_{0}),t) & =\rho_{0}(\mathbf{x}_{0})\left[\mathbf{\bm{\nabla}}_{0}\mathbf{F}_{t_{0}}^{t}(\mathbf{x}_{0})\right]^{-1}\mathbf{D}(\mathbf{F}_{t_{0}}^{t}(\mathbf{x}_{0}),t)\left[\mathbf{\bm{\nabla}}_{0}\mathbf{F}_{t_{0}}^{t}(\mathbf{x}_{0})\right]^{-T}\mathbf{\bm{\nabla}}_{0}c(\mathbf{F}_{t_{0}}^{t}(\mathbf{x}_{0}),t)\nonumber \\
 & =\rho_{0}(\mathbf{x}_{0})\left[\mathbf{\bm{\nabla}}_{0}\mathbf{F}_{t_{0}}^{t}(\mathbf{x}_{0})\right]^{-1}\mathbf{D}(\mathbf{F}_{t_{0}}^{t}(\mathbf{x}_{0}),t)\mathbf{\bm{\nabla}}c(\mathbf{F}_{t_{0}}^{t}(\mathbf{x}_{0}),t).\label{eq:sigma tilde simp2}
\end{align}
Therefore, formulas (\ref{eq:sigma tilde simp1})-(\ref{eq:sigma tilde simp2})
give an exact expression for $\tilde{\Sigma}_{t_{0}}^{t_{1}}(\mathcal{M}_{0})$
with a redefined form of $\mathbf{\bar{q}}_{t_{0}}^{t_{1}}$ as 
\begin{align}
\tilde{\Sigma}_{t_{0}}^{t_{1}}(\mathcal{M}_{0}) & \coloneqq\frac{\int_{\mathcal{M}_{0}}\left\langle \mathbf{\bar{q}}_{t_{0}}^{t_{1}},\mathbf{n}_{0}\right\rangle \,dA_{0}}{\int_{\mathcal{M}_{0}}dA_{0}},\nonumber \\
\mathbf{\bar{q}}_{t_{0}}^{t_{1}}(\mathbf{x}_{0}) & \coloneqq\int_{t_{0}}^{t_{1}}\rho_{0}(\mathbf{x}_{0})\left[\mathbf{\bm{\nabla}}_{0}\mathbf{F}_{t_{0}}^{t}(\mathbf{x}_{0})\right]^{-1}\mathbf{D}(\mathbf{F}_{t_{0}}^{t}(\mathbf{x}_{0}),t)\mathbf{\bm{\nabla}}c(\mathbf{F}_{t_{0}}^{t}(\mathbf{x}_{0}),t)\,dt.\label{eq:q redefined}
\end{align}
Using the form (\ref{eq:q redefined}) of $\mathbf{\bar{q}}_{t_{0}}^{t_{1}}(\mathbf{x}_{0})$
in all our results below increases the accuracy of transport extremizer
detection. At the same time, formula (\ref{eq:q redefined}) requires
explicit knowledge of the current concentration $c(\mathbf{x},t),$
which generally necessitates the numerical solution of the advection-diffusion
equation (\ref{eq:adv-diff}). A notable case in which (\ref{eq:q redefined})
is useful is when $c(\mathbf{x},t)=\omega(\mathbf{x},t)$ is the scalar
vorticity associated with a two-dimensional velocity field $\mathbf{v}(\mathbf{x},t).$
In that case, once $\mathbf{v}(\mathbf{x},t)$ is known, $\omega(\mathbf{x},t)$
is readily obtained as the plane-normal component of $\nabla\omega(\mathbf{x},t)$
without the need to solve the vorticity-transport equation. In this
latter case, no assumption is needed on the smallness of the viscosity
$\nu$.
\end{rem}

\subsection{Perfect constrained barriers and enhancers to diffusion}

In any dimension, a distinguished subset of uniform constrained barriers,
the \emph{perfect barriers,} inhibit transport completely pointwise
at leading order, i.e., are characterized by $\mathcal{T}_{0}=0$.
By eq.~(\ref{eq:uniform barriers}), the time-$t_{0}$ positions
of perfect constrained barriers satisfy
\begin{equation}
\left\langle \mathbf{\bar{q}}_{t_{0}}^{t_{1}},\mathbf{n}_{0}\right\rangle =0.\label{eq:zero inner product}
\end{equation}
Therefore, the vector $\mathbf{\bar{q}}_{t_{0}}^{t_{1}}(\mathbf{x}_{0})$
is necessarily tangent to a perfect barrier at every point $\mathbf{x}_{0}$.
In other words, time-$t_{0}$ positions of material surfaces acting
as perfect constrained material barriers to diffusive transport are
necessarily codimension-one invariant manifolds of the autonomous
dynamical system
\begin{equation}
\mathbf{x}_{0}^{\prime}=\mathbf{\bar{q}}_{t_{0}}^{t_{1}}\left(\mathbf{x}_{0}\right)=\mathbf{\bar{T}}_{t_{0}}^{t_{1}}\left(\mathbf{x}_{0}\right)\mathbf{\bm{\nabla}}_{0}c_{0}\left(\mathbf{x}_{0}\right)+\mathbf{\bar{B}}_{t_{0}}^{t_{1}}\left(\mathbf{x}_{0}\right),\label{eq:local_general-1}
\end{equation}
with
\[
\mathbf{\bar{B}}_{t_{0}}^{t_{1}}\left(\mathbf{x}_{0}\right)\coloneqq\int_{t_{0}}^{t_{1}}\mathbf{T}_{t_{0}}^{t}\mathbf{\bm{\nabla}}_{0}b(\mathbf{x}_{0},t)\,dt.
\]
In the absence of sources and sinks ($b(\mathbf{x}_{0},t)\equiv0)$,
eq.~(\ref{eq:local_general-1}) simplifies to 
\begin{equation}
\mathbf{x}_{0}^{\prime}=\mathbf{\bar{T}}_{t_{0}}^{t_{1}}\left(\mathbf{x}_{0}\right)\mathbf{\bm{\nabla}}_{0}c_{0}\left(\mathbf{x}_{0}\right),\label{eq:local_nosource}
\end{equation}
which leads to the following result.
\begin{prop}
\label{prop: nonexistence2} Consider the time $t_{0}$ position of
a perfect constrained diffusion barrier along which $\mathbf{\bm{\nabla}}_{0}c_{0}$
is not identically zero. Then the barrier can contain no homoclinic,
periodic, quasiperiodic or almost periodic orbit. 
\end{prop}
\begin{proof}
We use the function $V(\mathbf{x}_{0})=c_{0}(\mathbf{x}_{0})$ to
note that 
\[
\frac{d}{ds}V(\mathbf{x}_{0}(s))=\mathbf{\bm{\nabla}}_{0}c_{0}\cdot\mathbf{x}_{0}^{\prime}=\left\langle \mathbf{\bm{\nabla}}_{0}c_{0},\mathbf{\bar{T}}_{t_{0}}^{t_{1}}\mathbf{\bm{\nabla}}_{0}c_{0}\right\rangle 
\]
along trajectories of (\ref{eq:local_nosource}). Since $\mathbf{\bar{T}}_{t_{0}}^{t_{1}}$
is positive definite, $V(\mathbf{x}_{0}(s))$ strictly increases at
points where $\mathbf{\bm{\nabla}}_{0}c_{0}$ does not vanish. This
excludes the existence of any recurrent motion that contains at least
one point where $\mathbf{\bm{\nabla}}_{0}c_{0}$ does not vanish. 
\end{proof}
A consequence of Proposition \ref{prop: nonexistence2} in two-dimensions:
no closed perfect constrained barriers can exist apart from closed
ridges and trenches of the initial concentration field. In three dimensions,
Proposition \ref{prop: nonexistence2} implies that no two-dimensional,
quasiperiodic invariant tori can arise as perfect constrained diffusion
barriers, apart from toroidal ridges or trenches of the initial concentration
field. Finally, for perfect constrained barriers, the diffusion barrier
strength field defined in (\ref{eq:DBSdef}) simplifies to 
\begin{equation}
DBS_{t_{1}}^{t_{0}}(\mathbf{x}_{0})=\left|\mathbf{\bar{T}}_{t_{0}}^{t_{1}}\left(\mathbf{x}_{0}\right)\mathbf{\bm{\nabla}}_{0}c_{0}\left(\mathbf{x}_{0}\right)\right|.\label{eq:DBS for perfect barriers}
\end{equation}

In contrast to perfect barriers to diffusion, \emph{perfect enhancers}
to diffusive transport can be defined as material surfaces that pointwise
maximize diffusive transport. By eq.~(\ref{eq:uniform barriers}),
the time-$t_{0}$ positions of such perfect constrained enhancers
must have unit normals $\mathbf{n}_{0}$ satisfying
\begin{equation}
\left|\left\langle \mathbf{\bar{q}}_{t_{0}}^{t_{1}}\left(\mathbf{x}_{0}\right),\mathbf{n}_{0}\left(\mathbf{x}_{0}\right)\right\rangle \right|=\left|\mathbf{\bar{q}}_{t_{0}}^{t_{1}}\left(\mathbf{x}_{0}\right)\right|,\label{eq:zero inner product-1}
\end{equation}
Note that the norm of $\mathbf{\bar{q}}_{t_{0}}^{t_{1}}$ is not necessarily
constant along such surfaces, and hence perfect transport enhancers
are generally not solutions of the constrained variational problem
(\ref{eq:variproblem_known_IC}). Instead, perfect enhancers to transport
are simply  surfaces that are pointwise normal to $\mathbf{\bar{q}}_{t_{0}}^{t_{1}}$,
thus experiencing the locally strongest transport possible at each
of their points.

In two dimensions, perfect transport enhancers are curves satisfying
the ODE
\begin{equation}
\mathbf{x}_{0}^{\prime}=\mathbf{\bm{\Omega}}\mathbf{\bar{q}}_{t_{0}}^{t_{1}}\left(\mathbf{x}_{0}\right)=\mathbf{\bm{\Omega}}\left[\mathbf{\bar{T}}_{t_{0}}^{t_{1}}\left(\mathbf{x}_{0}\right)\mathbf{\bm{\nabla}}_{0}c_{0}\left(\mathbf{x}_{0}\right)+\mathbf{\bar{B}}_{t_{0}}^{t_{1}}\left(\mathbf{x}_{0}\right)\right],\label{eq:local_general-1-1}
\end{equation}
along which $\mathbf{\bar{q}}_{t_{0}}^{t_{1}}$ has constant norm.
Here we have used the notation
\begin{equation}
\mathbf{\bm{\Omega}}:=\left(\begin{array}{rc}
0 & 1\\
-1 & 0
\end{array}\right)\label{eq:Omegadef}
\end{equation}
for planar 90-degree rotations. In the absence of sources and sinks
($b(\mathbf{x}_{0},t)\equiv0)$, eq.~(\ref{eq:local_general-1-1})
simplifies to 
\begin{align}
\mathbf{x}_{0}^{\prime} & =\mathbf{\bm{\Omega}}\mathbf{\bar{T}}_{t_{0}}^{t_{1}}\left(\mathbf{x}_{0}\right)\mathbf{\bm{\nabla}}_{0}c_{0}\left(\mathbf{x}_{0}\right),\label{eq:local_nosource-1}
\end{align}
which leads to the following result.
\begin{prop}
\label{prop:nonexistence 3}In a two-dimensional flow, consider the
time $t_{0}$ position of a closed, perfect constrained diffusion
enhancer along which $\mathbf{\bm{\nabla}}_{0}c_{0}$ is not identically
zero. Then this closed enhancer cannot be fully contained in a domain
where the symmetric tensor $\mathbf{\bm{\Omega}}\mathbf{\bar{T}}_{t_{0}}^{t_{1}}\left(\mathbf{x}_{0}\right)-\mathbf{\bar{T}}_{t_{0}}^{t_{1}}\left(\mathbf{x}_{0}\right)\mathbf{\bm{\Omega}}$
is definite.
\end{prop}
\begin{proof}
As in Proposition \ref{prop: nonexistence2}, we use the function
$V(\mathbf{x}_{0})=c_{0}(\mathbf{x}_{0})$ to obtain
\begin{align*}
\frac{d}{ds}V(\mathbf{x}_{0}(s)) & =\mathbf{\bm{\nabla}}_{0}c_{0}\cdot\mathbf{x}_{0}^{\prime}=\left\langle \mathbf{\bm{\nabla}}_{0}c_{0},\mathbf{\bm{\Omega}}\mathbf{\bar{T}}_{t_{0}}^{t_{1}}\mathbf{\bm{\nabla}}_{0}c_{0}\right\rangle \\
 & =\left\langle \mathbf{\bm{\nabla}}_{0}c_{0},\left[\mathbf{\bm{\Omega}}\mathbf{\bar{T}}_{t_{0}}^{t_{1}}-\mathbf{\bar{T}}_{t_{0}}^{t_{1}}\mathbf{\bm{\Omega}}\right]\mathbf{\bm{\nabla}}_{0}c_{0}\right\rangle 
\end{align*}
along trajectories of (\ref{eq:local_nosource-1}). Under the assumptions
of the proposition, $V(\mathbf{x}_{0}(s))$ strictly increases or
decreases on domains where $\mathbf{\bm{\Omega}}\mathbf{\bar{T}}_{t_{0}}^{t_{1}}\left(\mathbf{x}_{0}\right)-\mathbf{\bar{T}}_{t_{0}}^{t_{1}}\left(\mathbf{x}_{0}\right)\mathbf{\bm{\Omega}}$
is definite, which excludes the existence of any closed trajectory
for eq.~(\ref{eq:local_nosource-1}).
\end{proof}

\subsection{Constrained diffusion barriers in two-dimensional flows}

The expression (\ref{eq:uniform barriers}) is a PDE in three and
more dimensions. In two dimensions, however, it is equivalent to two
ODEs, as we spell out in the following result.
\begin{thm}
\label{thm:2D_flows}In two-dimensional flows, time-$t_{0}$ positions
of constrained material diffusion barriers with uniform, pointwise
transport density $\mathcal{T}_{0}$ satisfy the following necessary
conditions:
\begin{description}
\item [{(i)}] Constrained uniform transport maximizers $\mathcal{M}_{0}$
are necessarily solutions of the differential equation family 
\begin{equation}
\mathbf{x}_{0}^{\prime}=\frac{\sqrt{\left|\mathbf{\bar{q}}_{t_{0}}^{t_{1}}\left(\mathbf{x}_{0}\right)\right|^{2}-\mathcal{T}_{0}^{2}}}{\left|\mathbf{\bar{q}}_{t_{0}}^{t_{1}}\left(\mathbf{x}_{0}\right)\right|^{2}}\mathbf{\bar{q}}_{t_{0}}^{t_{1}}\left(\mathbf{x}_{0}\right)\pm\frac{\mathcal{T}_{0}}{\left|\mathbf{\bar{q}}_{t_{0}}^{t_{1}}\left(\mathbf{x}_{0}\right)\right|^{2}}\mathbf{\bm{\Omega}}\mathbf{\bar{q}}_{t_{0}}^{t_{1}}\left(\mathbf{x}_{0}\right),\qquad\mathcal{T}_{0}>0.\label{eq:2ODEs1}
\end{equation}
\item [{(ii)}] If such a uniform transport maximizer $\mathcal{M}_{0}$
is a closed orbit of (\ref{eq:2ODEs1}) or a homoclinic or heteroclinic
orbit connecting a zero of the $\mathbf{\bar{q}}_{t_{0}}^{t_{1}}\left(\mathbf{x}_{0}\right)$
vector field to itself, then the symmetric matrix 
\begin{equation}
\mathbf{L}=\int_{\mathcal{M}_{0}}\mathrm{sign}\,\left\langle \mathbf{\bar{q}}_{t_{0}}^{t_{1}}\left(\mathbf{x}_{0}(s)\right),\mathbf{\bm{\Omega}}\mathbf{x}_{0}^{\prime}(s)\right\rangle \partial_{\mathbf{x}_{0}\mathbf{x}_{0}}^{2}\left\langle \mathbf{\bar{q}}_{t_{0}}^{t_{1}}\left(\mathbf{x}_{0}(s)\right),\mathbf{\bm{\Omega}}\mathbf{x}_{0}^{\prime}(s)\right\rangle ds\label{eq:Lformula}
\end{equation}
 must be negative semidefinite.
\item [{(iii)}] If a uniform transport maximizer $\mathcal{M}_{0}$ satisfies
\[
\left|\mathbf{\bar{q}}_{t_{0}}^{t_{1}}\left(\mathbf{x}_{0}(s_{i})\right)\right|=\mathcal{T}_{0},\,\,\,i=1,2,\quad\mathbf{\bar{q}}_{t_{0}}^{t_{1}}\left(\mathbf{x}_{0}(s_{1})\right)\parallel\mathbf{\bar{q}}_{t_{0}}^{t_{1}}\left(\mathbf{x}_{0}(s_{2})\right),
\]
 $\left|\mathbf{\bar{q}}_{t_{0}}^{t_{1}}\left(\mathbf{x}_{0}(s_{i})\right)\right|=\mathcal{T}_{0}$
at its endpoints, $\mathcal{M}_{0}$ 
\[
\mathbf{\bar{q}}_{t_{0}}^{t_{1}}\left(\mathbf{x}_{0}(s_{i})\right)\parallel\mathbf{x}_{0}^{\prime}(s_{i})\parallel\mathbf{x}_{0}^{\prime}(s_{j}),\quad i,j=1,2,\quad i\neq j,
\]
then 
\begin{equation}
\left\langle \mathbf{L}\mathbf{\bm{\Omega}}\mathbf{x}_{0}^{\prime},\mathbf{\bm{\Omega}}\mathbf{x}_{0}^{\prime}\right\rangle \leq0\label{eq:restricted cond}
\end{equation}
must hold along $\mathcal{M}_{0}$. 
\end{description}
\textbf{(iv)} Constrained uniform transport minimizers must necessarily
be perfect barriers, i.e., satisfy the differential equation
\[
\mathbf{x}_{0}^{\prime}=\mathbf{\bar{q}}_{t_{0}}^{t_{1}}\left(\mathbf{x}_{0}\right).
\]

\end{thm}
\begin{proof}
See \secref{Appendix-C}.
\end{proof}
\begin{rem}
The argument in the proof of (i) of Theorem \ref{thm:2D_flows} is
not applicable to perfect diffusion barriers, as for such material
lines, the leading-order term in the second variation of $\mathcal{E}(\mathcal{M}_{\epsilon})$
vanishes.
\end{rem}
In the absence of sources or sinks (i.e., for $b\left(\mathbf{x}_{0},t\right)\equiv0$
in (\ref{eq:qdef})), eq.~(\ref{eq:2ODEs1}) simplifies to 
\begin{equation}
\mathbf{x}_{0}^{\prime}=\frac{1}{\left|\mathbf{\bar{T}}_{t_{0}}^{t_{1}}\mathbf{\bm{\nabla}}_{0}c_{0}\right|^{2}}\mathbf{A}^{\pm}\left(\mathbf{x}_{0};\mathcal{T}_{0}\right)\mathbf{\bm{\nabla}}_{0}c_{0}\left(\mathbf{x}_{0}\right),\label{eq:direction field-1}
\end{equation}
with the tensor $\mathbf{A}^{\pm}\in\mathbb{R}^{2}$ defined as 
\begin{equation}
\mathbf{A}^{\pm}\left(\mathbf{x}_{0};\mathcal{T}_{0}\right)=\left(\mathcal{T}_{0}\mathbf{\bm{\Omega}}\pm\sqrt{\left|\mathbf{\bar{T}}_{t_{0}}^{t_{1}}\mathbf{\bm{\nabla}}_{0}c_{0}\right|^{2}-\mathcal{T}_{0}^{2}}\,\mathbf{I}\right)\mathbf{\bar{T}}_{t_{0}}^{t_{1}}.\label{eq:Apm}
\end{equation}
In this case, the following result is helpful in the numerical identification
of closed diffusion barriers as limit cycles of (\ref{eq:direction field-1}).
\begin{prop}
\label{prop: nonexistence 2D}Eq.~(\ref{eq:direction field-1}) will
have no closed (periodic or homoclinic) orbits contained entirely
in spatial domains where the symmetric part of the matrix $\mathbf{A}^{\pm}\in\mathbb{R}^{2}$
is definite and $\mathbf{\bm{\nabla}}_{0}c_{0}$ is not identically
zero.
\end{prop}
\begin{proof}
For the Lyapunov function 
\[
V(\mathbf{x}_{0})=c_{0}(\mathbf{x}_{0}),
\]
we obtain that 
\begin{align*}
\frac{d}{ds}V(\mathbf{x}_{0}(s)) & =\mathbf{\bm{\nabla}}_{0}c_{0}\cdot\mathbf{x}_{0}^{\prime}\\
 & =\frac{1}{\left|\mathbf{\bar{T}}_{t_{0}}^{t_{1}}\mathbf{\bm{\nabla}}_{0}c_{0}\right|^{2}}\left\langle \mathbf{\bm{\nabla}}_{0}c_{0},\mathbf{A}^{\pm}\mathbf{\bm{\nabla}}_{0}c_{0}\right\rangle .
\end{align*}
Therefore, $V(\mathbf{x}_{0}(s))$ is strictly monotonically increasing
or decreasing at least at one point of any orbit of (\ref{eq:direction field-1})
that lies entirely in a domain in which $\mathbf{A}^{\pm}\left(\mathbf{x}_{0};\mathcal{T}_{0}\right)$
is definite and $\mathbf{\bm{\nabla}}_{0}c_{0}$ is not identically
vanishing. This implies the statement of the proposition. 
\end{proof}
Note that $\mathbf{A}_{t_{0}}^{t_{1}}$ is certainly definite for
$\mathcal{T}_{0}=0$ at any point where $\mathbf{\bm{\nabla}}_{0}c_{0}$
is nonzero, and hence (\ref{eq:direction field-1}) has no closed
orbits for $\mathcal{T}_{0}=0$, except possibly for ones along which
the initial concentration gradient vanishes (curves of critical points,
which is non-generic, yet abundant in areas of constant initial concentration).
This statement remains valid for small enough $\mathcal{T}_{0}$ on
compact domains by the continuous dependence of the eigenvalues of
$\mathbf{A}^{\pm}\left(\mathbf{x}_{0};\mathcal{T}_{0}\right)$ on
the parameter $\mathcal{T}_{0}$.

\section{Particle transport barriers in stochastic velocity fields}

We showed in \cite{Haller2018} how our results on barriers to diffusive
scalar transport carry over to probabilistic transport barriers to
fluid particle motion with uncertainties, modeled by diffusive Itô
processes. Our derivation, however, specifically exploited the incompressibility
of the deterministic part of the velocity field. Here we show how
similar results continue to hold for compressible Itô processes of
the form
\begin{equation}
d\mathbf{x}(t)=\mathbf{v}_{0}(\mathbf{x}(t),t)dt+\sqrt{\nu}\mathbf{B}(\mathbf{x}(t),t)d\mathbf{W}(t).\label{eq:Ito}
\end{equation}
Here $\mathbf{x}(t)\in\mathbb{R}^{n}$ is the random position vector
of a particle at time $t$; $\mathbf{v}_{0}(\mathbf{x},t)$ denotes
the deterministic, generally compressible drift component in the velocity
of the particle motion; and $\mathbf{W}(t)$ in an $m$-dimensional
Wiener process with diffusion matrix $\sqrt{\nu}\mathbf{B}(\mathbf{x},t)\in\mathbb{R}^{n\times m}$.
Here the dimensionless, nonsingular diffusion structure matrix $\mathbf{B}$
is $\mathcal{O}(1)$ with respect to the small parameter $\nu>0$.

We let $p(\mathbf{x},t;\mathbf{x}_{0},t_{0})$ denote the probability
density function (PDF) for the current particle position $\mathbf{x}(t)$
with initial condition $\mathbf{x}_{0}(t_{0})=\mathbf{x}_{0}$. This
PDF satisfies the classic Fokker--Planck equation (see, e.g., Risken
\cite{Risken1984})
\begin{equation}
p_{t}+\mathbf{\mathbf{\bm{\nabla}}\cdot}\left(p\mathbf{v}_{0}\right)=\nu\tfrac{1}{2}\mathbf{\mathbf{\bm{\nabla}}}\cdot\left[\mathbf{\mathbf{\bm{\nabla}}}\cdot\left(\mathbf{B}\mathbf{B}^{\top}p\right)\right],\label{eq:FP equation}
\end{equation}
or, alternatively, 
\begin{equation}
p_{t}+\mathbf{\mathbf{\bm{\nabla}}\cdot}\left(p\mathbf{\tilde{v}}_{0}\right)=\nu\mathbf{\bm{\nabla}}\cdot\left(\tfrac{1}{2}\mathbf{B}\mathbf{B}^{\top}\mathbf{\bm{\nabla}}p\right),\quad\tilde{\mathbf{v}}_{0}=\mathbf{v}_{0}-\tfrac{\nu}{2}\mathbf{\bm{\nabla}}\cdot\left(\mathbf{B}\mathbf{B}^{\top}\right).\label{eq:FP1}
\end{equation}
This latter equation is of the advection--diffusion-form (\ref{eq:adv-diff-3})
if we select 
\begin{equation}
c\coloneqq\frac{p}{\rho},\quad\mathbf{D}\coloneqq\tfrac{1}{2}\mathbf{B}\mathbf{B}^{\top},\quad\mathbf{w}\coloneqq\tilde{\mathbf{v}}_{0}=\mathbf{v}_{0}-\nu\mathbf{\bm{\nabla}}\cdot\mathbf{D},\quad k(t)=f(\mathbf{x},t)\equiv0.\label{eq:density condition}
\end{equation}
Consequently, the Fokker--Planck equation (\ref{eq:FP1}) is equivalent
to the advection--diffusion equation (\ref{eq:adv-diff}) with the
velocity field
\begin{equation}
\mathbf{v}\coloneqq\mathbf{w}-\frac{\nu}{\rho}\mathbf{D}\mathbf{\bm{\nabla}}\rho=\mathbf{v}_{0}-\nu\mathbf{\bm{\nabla}}\cdot\mathbf{D}-\frac{\nu}{\rho}\mathbf{D}\mathbf{\bm{\nabla}}\rho.\label{eq:vdef}
\end{equation}

Since the equation of continuity,
\begin{equation}
\partial_{t}\rho+\mathbf{\bm{\nabla\cdot}}\left(\rho\mathbf{v}\right)=0,\label{eq:continuity-1}
\end{equation}
must hold for the velocity field $\mathbf{v}$ for our formulation
to apply, substitution of the definition of $\mathbf{v}$ from (\ref{eq:vdef})
into (\ref{eq:continuity-1}) gives
\[
\partial_{t}\rho+\mathbf{\bm{\nabla\cdot}}\left[\rho\left(\mathbf{v}_{0}-\frac{\nu}{\rho}\mathbf{D}\mathbf{\bm{\nabla}}\rho-\nu\mathbf{\bm{\nabla}}\cdot\mathbf{D}\right)\right]=0,
\]
or, equivalently, 
\begin{equation}
\partial_{t}\rho+\mathbf{\bm{\nabla\cdot}}\left(\rho\mathbf{v}_{0}\right)=\nu\mathbf{\bm{\nabla\cdot}}\left(\mathbf{\mathbf{\bm{\nabla}}\cdot\left(D\rho\right)}\right),\label{eq:density PDE}
\end{equation}
which is the same PDE (\ref{eq:FP1}) that the probability-density
$p$ satisfies. 

With the above choice of $\mathbf{v}$ and $\rho$ in (\ref{eq:vdef})
and (\ref{eq:density PDE}), all results in the earlier sections on
material diffusion extremizers in compressible flows carry over to
material diffusion barriers of the density-weighted probability-density
function $c=p/\rho$ with respect to the velocity field $\mathbf{v}$
in (\ref{eq:vdef}) if we re-define the transport tensor $\mathbf{T}_{t_{0}}^{t}(\mathbf{x}_{0})$
as
\begin{align}
\mathbf{T}_{t_{0}}^{t}(\mathbf{x}_{0}) & \coloneqq\tfrac{1}{2}\rho_{0}(\mathbf{x}_{0})\left[\mathbf{\bm{\nabla}}_{0}\mathbf{F}_{t_{0}}^{t}(\mathbf{x}_{0})\right]^{-1}\mathbf{B}(\mathbf{F}_{t_{0}}^{t}(\mathbf{x}_{0}),t)\mathbf{B}^{\top}(\mathbf{F}_{t_{0}}^{t}(\mathbf{x}_{0}),t)\left[\mathbf{\bm{\nabla}}_{0}\mathbf{F}_{t_{0}}^{t}(\mathbf{x}_{0})\right]^{-T}.\label{eq:probabilistic T}
\end{align}
Furthermore, let us denote the initial probability-density function
by $p_{0}(\mathbf{x}):=p(\mathbf{x},t_{0};\mathbf{x}_{0},t_{0})$
and assume the initial carrier fluid density $\rho_{0}(\mathbf{x})$
as given. Then, with the help of the vector field (cf.~(\ref{eq:qdef}))
\begin{equation}
\mathbf{\bar{q}}_{t_{0}}^{t_{1}}(\mathbf{x}_{0})=\int_{t_{0}}^{t_{1}}\left[\mathbf{T}_{t_{0}}^{t}(\mathbf{x}_{0})\mathbf{\bm{\nabla}}_{0}\frac{p_{0}(\mathbf{x}_{0})}{\rho_{0}(\mathbf{x}_{0})}\right]dt,\label{eq:probabilistic q}
\end{equation}
we collect the related results in the following theorem.
\begin{thm}
\label{thm: stoschastic case}(i) Unconstrained, uniform barriers
of transport for the mass-based PDF, $p/\rho,$ of particle positions
satisfy Theorems \ref{thm:uncnstrained barriers general}--\ref{thm:2D_flows}
with the transport tensor field $\mathbf{T}_{t_{0}}^{t}(\mathbf{x}_{0})$
defined as in (\ref{eq:probabilistic T}).

(ii) Constrained, uniform barriers of transport for the mass-based
PDF, $p/\rho,$ satisfy Theorems \ref{thm:first_integral}--\ref{thm:unconstrained barriers 2D}
with the transport vector field $\mathbf{\bar{q}}_{t_{0}}^{t_{1}}(\mathbf{x}_{0})$
defined as in (\ref{eq:probabilistic q}).
\end{thm}
\begin{proof}
Indeed, for the case of an unspecified initial density $c_{0}(\mathbf{x})$,
we obtain the normalized total transport of $\mu(\mathbf{x},t)$ in
the form (cf.~(\ref{eq:total transport}))

\begin{align}
\tilde{\Sigma}_{t_{0}}^{t_{1}}(\mathcal{M}_{0}) & =\frac{\Sigma_{t_{0}}^{t_{1}}(\mathcal{M}_{0})}{\nu K\left(t_{1}-t_{0}\right)A_{0}(\mathcal{M}_{0})}=\frac{\int_{\mathcal{M}_{0}}\left\langle \mathbf{n}_{0},\mathbf{\bar{T}}_{t_{0}}^{t_{1}}\mathbf{n}_{0}\right\rangle dA_{0}}{\int_{\mathcal{M}_{0}}dA_{0}}+o(\nu^{\alpha}),\quad\nu\in(0,1]\label{eq:total transport-1}
\end{align}
with the transport tensor (\ref{eq:probabilistic T}), where $\mathbf{F}_{t_{0}}^{t}(\mathbf{x}_{0})$
is the flow map associated with the velocity field $\mathbf{v}_{0}(\mathbf{x},t)$
and $\rho_{0}(\mathbf{x}_{0})$ is the initial density field of the
carrier fluid, serving as initial condition for the density evolution
equation (\ref{eq:density PDE}). The formulas (\ref{eq:total transport-1})-(\ref{eq:probabilistic T})
follow because the flow map of the full velocity field $\mathbf{v}$
defined in (\ref{eq:vdef}) is at least $\mathcal{O}(\nu)$ $C^{0}$-close
to the flow map $\mathbf{F}_{t_{0}}^{t}(\mathbf{x}_{0})$ of $\mathbf{v}_{0}(\mathbf{x},t)$
over the finite time interval $[t_{0},t_{1}].$ As a consequence,
only the leading order term, $\mathbf{F}_{t_{0}}^{t}(\mathbf{x}_{0})$,
of the flow map generated by (\ref{eq:vdef}) appears in the transport
tensor (\ref{eq:probabilistic T}). Higher-order corrections to the
the full flow map can be subsumed into the $o(\nu^{\alpha})$ term
in (\ref{eq:total transport-1}). With this observation, statements
(i) and (ii) can be deduced in the same fashion as Theorems \ref{thm:uncnstrained barriers general}--\ref{thm:unconstrained barriers 2D}
and Theorems \ref{thm:first_integral}--\ref{thm:2D_flows}.
\end{proof}
Based on Theorem \ref{thm: stoschastic case}, the arguments leading
to the diffusion barrier strength indicator in eqs.~(\ref{eq:DBS unconstrained})-(\ref{eq:DBSdef})
continue to apply, with the DBS field simplified to 
\[
DBS_{t_{0}}^{t_{1}}(\mathbf{x}_{0})=\left|\mathbf{\bar{T}}_{t_{0}}^{t_{1}}\left(\mathbf{x}_{0}\right)\mathbf{\bm{\nabla}}_{0}\frac{p_{0}(\mathbf{x})}{\rho_{0}(\mathbf{x})}\right|.
\]

\section{Examples}

\subsection{Two-dimensional channel flow}

\begin{figure}[H]
\begin{centering}
\includegraphics[width=0.5\textwidth]{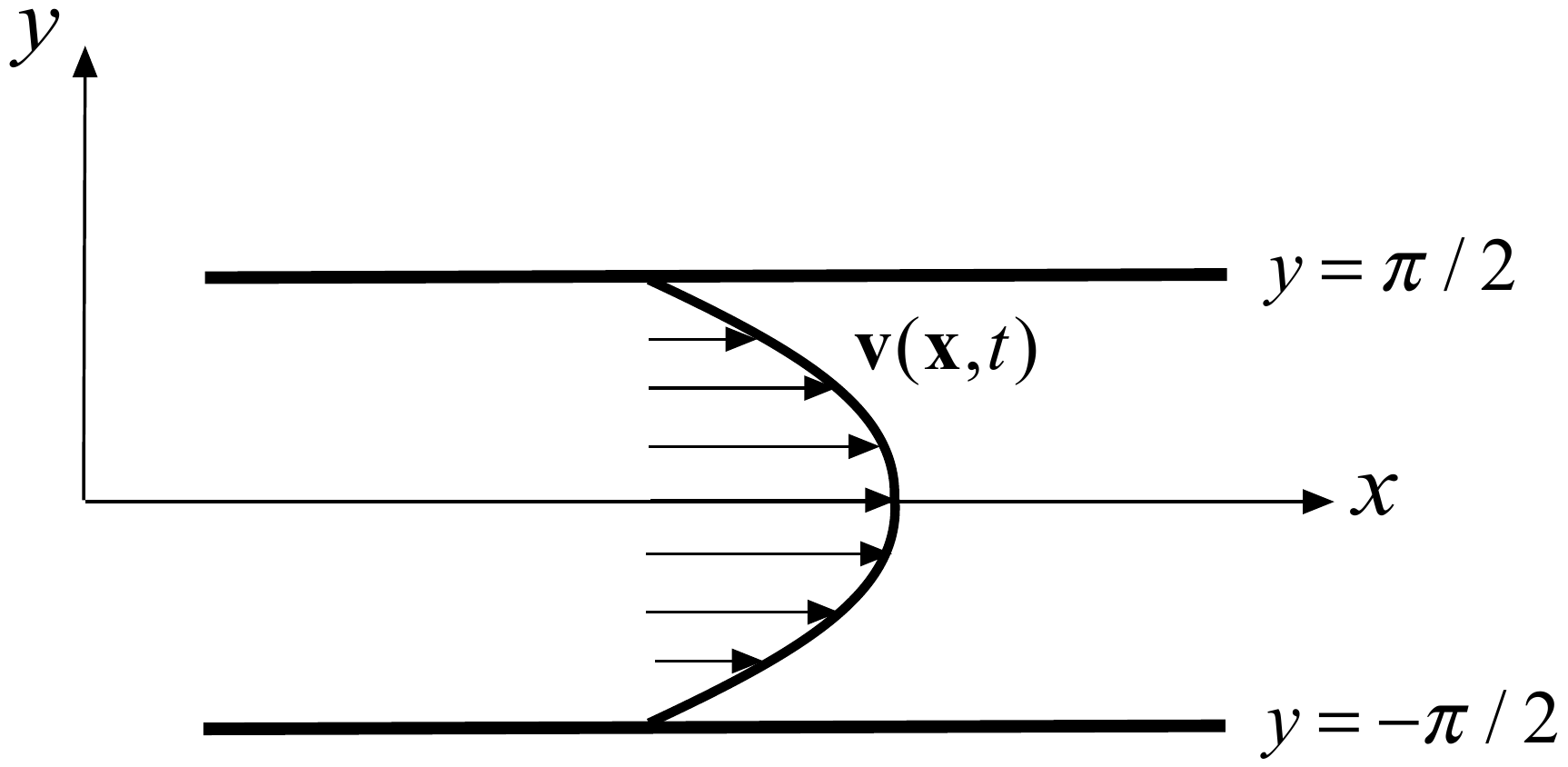}
\par\end{centering}
\caption{Unsteady, horizontal jet with a jet core at $y=0$.\label{fig:shear jet} }
\end{figure}

An unsteady solution of the 2D, unforced Navier-Stokes equations is
given by the decaying channel flow 
\begin{equation}
\mathbf{v}(\mathbf{x},t)=e^{-\nu t}\left(\begin{array}{c}
a\cos y\\
0
\end{array}\right),\label{eq:horizontal shear jet}
\end{equation}
whose vorticity field 
\[
\omega(\mathbf{x},t)=ae^{-\nu t}\sin y
\]
satisfies the advection--diffusion equation 
\begin{equation}
\partial_{t}\omega+\nabla\omega\cdot\mathbf{v}=\nu\Delta\omega,\label{eq:viscid-1}
\end{equation}
i.e., the two-dimensional vorticity-transport equation with viscosity
$\nu$. The simplest member of a more general Navier-Stokes solution
family (see, e.g., Majda and Bertozzi \cite{Majda2002}), the velocity
field (\ref{eq:horizontal shear jet}) describes a decaying horizontal
shear-jet between two no-slip boundaries at $y=\pm\frac{\pi}{2}$
(see Fig.~\ref{fig:shear jet}). 

The jet core is given by the horizontal line $y=0$. The constant
$a\in\mathbb{R}^{+}$ governs the strength of shear within the jet.
If we define the variable $x$ to be spatially periodic, the flow
becomes a model of a perfectly circular vortical flow in an annulus
with no-slip walls. 

All horizontal lines are invariant material lines in (\ref{eq:horizontal shear jet}).
Out of these invariant lines, the jet core at $y=0$ is the most often
noted barrier to the diffusion of vorticity, keeping positive and
negative vorticity values apart for all times. Indeed, for large values
of $a$, the norm of the vorticity gradient \emph{
\[
\nabla\omega(\mathbf{x},t)=e^{-\nu t}\left(\begin{array}{c}
0\\
a\cos y
\end{array}\right)
\]
}maintains its global maximum along the jet core for all times. The
upper and lower channel boundaries at $y=\pm\pi/2$ technically also
block the diffusion of vorticity into the wall, but vorticity tapers
off to zero anyway as one approaches these boundaries in the vertical
direction. 

As the initial distribution of $\omega(\mathbf{x},t)$ is constrained
by the velocity field, our theory of constrained diffusion barriers
is applicable to barriers to the transport of vorticity. To see the
predictions of this theory, we first note that the flow map $\mathbf{F}_{t_{0}}^{t}\left(\mathbf{x}_{0}\right)$
in this example is
\[
\mathbf{F}_{t_{0}}^{t}\left(\mathbf{x}_{0}\right)=\left(\begin{array}{c}
x_{0}-\frac{a}{\nu}\left(e^{-\nu t}-e^{-\nu t_{0}}\right)\cos y_{0}\\
y_{0}
\end{array}\right),
\]
which gives 
\begin{align*}
\mathbf{\bm{\nabla}}_{0}\left[\omega\left(\mathbf{F}_{t_{0}}^{t}\left(\mathbf{x}_{0}\right),t\right)\right] & =ae^{-\nu t}\left(\begin{array}{c}
0\\
\cos y_{0}
\end{array}\right),\\
\mathbf{T}_{t_{0}}^{t}(\mathbf{x}_{0}) & =\left(\begin{array}{cc}
1+\frac{a^{2}}{\nu^{2}}\left(e^{-\nu t}-e^{-\nu t_{0}}\right)^{2}\sin^{2}y_{0} & -\frac{a}{\nu}\left(e^{-\nu t}-e^{-\nu t_{0}}\right)\sin y_{0}\\
-\frac{a}{\nu}\left(e^{-\nu t}-e^{-\nu t_{0}}\right)\sin y_{0} & 1
\end{array}\right).
\end{align*}
Therefore,
\begin{align*}
\bar{\mathbf{q}}_{t_{0}}^{t_{1}}(\mathbf{x}_{0}) & =\frac{1}{t_{1}-t_{0}}\int_{t_{0}}^{t_{1}}\mathbf{T}_{t_{0}}^{t}(\mathbf{x}_{0})\mathbf{\bm{\nabla}}_{0}\left[\omega\left(\mathbf{F}_{t_{0}}^{t}\left(\mathbf{x}_{0}\right),t\right)\right]dt\\
 & =\frac{1}{2\nu\left(t_{1}-t_{0}\right)}\left(\begin{array}{c}
A\sin2y_{0}\\
B\cos y_{0}
\end{array}\right),
\end{align*}
where
\[
A=\frac{a^{2}}{\nu}\sin2y_{0}\left[\frac{1}{2}e^{-2\nu t_{1}}+\frac{1}{2}e^{-2\nu t_{0}}-e^{-\nu\left(t_{1}+t_{0}\right)}\right],\qquad B=a\left(e^{-\nu t_{0}}-e^{-\nu t_{1}}\right).
\]
Consequently, the ODE family describing the time $t_{0}$ position
of uniform constrained barriers is given by
\begin{equation}
\mathbf{x}_{0}^{\prime}=\frac{1}{2\nu\left(t_{1}-t_{0}\right)}\left\{ \frac{\sqrt{\left|\mathbf{\bar{q}}_{t_{0}}^{t_{1}}\left(\mathbf{x}_{0}\right)\right|^{2}-\mathcal{T}_{0}^{2}}}{\left|\mathbf{\bar{q}}_{t_{0}}^{t_{1}}\left(\mathbf{x}_{0}\right)\right|^{2}}\left(\begin{array}{c}
A\sin2y_{0}\\
B\cos y_{0}
\end{array}\right)+\frac{\mathcal{T}_{0}}{\left|\mathbf{\bar{q}}_{t_{0}}^{t_{1}}\left(\mathbf{x}_{0}\right)\right|^{2}}\left(\begin{array}{c}
B\cos y_{0}\\
-A\sin2y_{0}
\end{array}\right)\right\} \label{eq:2ODE}
\end{equation}
for some value of the transport density $\mathcal{T}_{0}\in\mathbb{R}$.
For the choice
\begin{equation}
\mathcal{T}_{0}=\left|\mathbf{\bar{q}}_{t_{0}}^{t_{1}}\left(\mathbf{x}_{0}\right)\right|_{y_{0}=0}=\frac{B}{2\nu\left(t_{1}-t_{0}\right)},\label{eq:T_0 for jet}
\end{equation}
 the ODE (\ref{eq:2ODE}) becomes
\begin{align}
\mathbf{x}_{0}^{\prime} & \vert_{y_{0}=0}=\frac{B}{2\nu\left(t_{1}-t_{0}\right)}\left(\begin{array}{c}
B\\
0
\end{array}\right)\parallel\mathbf{\bm{\Omega}}\mathbf{\bar{q}}_{t_{0}}^{t_{1}}\left(\mathbf{x}_{0}\right)\vert_{y_{0}=0}.\label{eq:jet parallel}
\end{align}
Therefore, $y_{0}=0$ is an invariant line for equation (\ref{eq:2ODE})
for the parameter value $\mathcal{T}_{0}$ selected as in (\ref{eq:T_0 for jet}).
Consequently, the jet core at $y_{0}=0$ is a uniform, constrained
barrier to vorticity diffusion along which the pointwise diffusive
transport of vorticity is equal to (\ref{eq:2ODE}). As noted earlier,
a barrier (as a stationary surface of the transport functional) is
not necessary a minimizer of transport. Indeed, any other horizontal
material curve admits a strictly lower transport density than the
jet core. 

In contrast, choosing the constant 
\[
\mathcal{T}_{0}=0
\]
in (\ref{eq:2ODE}) gives the ODE
\[
\mathbf{x}_{0}^{\prime}=\frac{1}{2\nu a\left(t_{1}-t_{0}\right)\left|\mathbf{\bar{q}}_{t_{0}}^{t_{1}}\left(\mathbf{x}_{0}\right)\right|}\left(\begin{array}{c}
A\sin2y_{0}\\
B\cos y_{0}
\end{array}\right),
\]
for which $y_{0}=\pm\pi/2$ are invariant lines. Along those invariant
lines, we have\emph{
\[
\mathbf{x}_{0}^{\prime}\vert_{y_{0}=\pm\pi/2}\,\parallel\,\mathbf{\bar{q}}_{t_{0}}^{t_{1}}\left(\mathbf{x}_{0}\right)\vert_{y_{0}=\pm\pi/2}.
\]
}Therefore, the channel walls at $y_{0}=\pm\pi/2$ are uniform, constrained
minimizers to vorticity diffusion along which the pointwise diffusive
transport of vorticity is equal to zero. In particular, the channel
walls are perfect constrained barriers to diffusive transport.

To evaluate the second necessary condition we need to check the definiteness
of the matrix 
\[
\mathbf{L}=\int_{\mathcal{M}_{0}}\mathrm{sign}\,\left\langle \mathbf{\bar{q}}_{t_{0}}^{t_{1}}\left(\mathbf{x}_{0}(s)\right),\mathbf{\bm{\Omega}}\mathbf{x}_{0}^{\prime}(s)\right\rangle \partial_{\mathbf{x}_{0}\mathbf{x}_{0}}^{2}\left\langle \mathbf{\bar{q}}_{t_{0}}^{t_{1}}\left(\mathbf{x}_{0}(s)\right),\mathbf{\bm{\Omega}}\mathbf{x}_{0}^{\prime}(s)\right\rangle ds.
\]
 Note that 
\begin{align*}
\partial_{\mathbf{x}_{0}\mathbf{x}_{0}}^{2}\left\langle \mathbf{\bar{q}}_{t_{0}}^{t_{1}}\left(\mathbf{x}_{0}(s)\right),\mathbf{\bm{\Omega}}\mathbf{x}_{0}^{\prime}(s)\right\rangle \vert_{y_{0}=0} & =\partial_{\mathbf{x}_{0}\mathbf{x}_{0}}^{2}\left[\frac{1}{2\nu\left(t_{1}-t_{0}\right)}\left(\begin{array}{c}
A\sin2y_{0}\\
B\cos y_{0}
\end{array}\right)\right]\vert_{y_{0}=0}\cdot\mathbf{\bm{\Omega}}\mathbf{x}_{0}^{\prime}(s)\vert_{y_{0}=0}\\
 & =\partial_{\mathbf{x}_{0}\mathbf{x}_{0}}^{2}\left[\frac{1}{2\nu\left(t_{1}-t_{0}\right)}\left(\begin{array}{c}
A\sin2y_{0}\\
B\cos y_{0}
\end{array}\right)\cdot\frac{B}{2\nu\left(t_{1}-t_{0}\right)}\left(\begin{array}{c}
0\\
-B
\end{array}\right)\right]\vert_{y_{0}=0}\\
 & =\frac{B^{2}}{4\nu^{2}\left(t_{1}-t_{0}\right)^{2}}\partial_{\mathbf{x}_{0}\mathbf{x}_{0}}^{2}\left[\left(\begin{array}{c}
A\sin2y_{0}\\
B\cos y_{0}
\end{array}\right)\cdot\left(\begin{array}{c}
0\\
-1
\end{array}\right)\right]\vert_{y_{0}=0}\\
 & =\frac{-B^{3}}{4\nu^{2}\left(t_{1}-t_{0}\right)^{2}}\partial_{\mathbf{x}_{0}\mathbf{x}_{0}}^{2}\left[\cos y_{0}\right]\vert_{y_{0}=0}\\
 & =\frac{-B^{3}}{4\nu^{2}\left(t_{1}-t_{0}\right)^{2}}\left(\begin{array}{cc}
0 & 0\\
0 & -\cos y_{0}
\end{array}\right)\vert_{y_{0}=0}\\
 & =\frac{B^{3}}{4\nu^{2}\left(t_{1}-t_{0}\right)^{2}}\left(\begin{array}{cc}
0 & 0\\
0 & 1
\end{array}\right),
\end{align*}
and 
\[
\mathrm{sign}\,\left\langle \mathbf{\bar{q}}_{t_{0}}^{t_{1}}\left(\mathbf{x}_{0}(s)\right),\mathbf{\bm{\Omega}}\mathbf{x}_{0}^{\prime}(s)\right\rangle \vert_{y_{0}=0}=\frac{-B^{3}}{4\nu^{2}\left(t_{1}-t_{0}\right)^{2}}\left[\cos y_{0}\right]\vert_{y_{0}=0}=-1.
\]
As a consequence, we have
\begin{align*}
\mathbf{L} & =\int_{\mathcal{M}_{0}}\mathrm{sign}\,\left\langle \mathbf{\bar{q}}_{t_{0}}^{t_{1}}\left(\mathbf{x}_{0}(s)\right),\mathbf{\bm{\Omega}}\mathbf{x}_{0}^{\prime}(s)\right\rangle \partial_{\mathbf{x}_{0}\mathbf{x}_{0}}^{2}\left\langle \mathbf{\bar{q}}_{t_{0}}^{t_{1}}\left(\mathbf{x}_{0}(s)\right),\mathbf{\bm{\Omega}}\mathbf{x}_{0}^{\prime}(s)\right\rangle ds\\
 & =-\frac{B^{6}\mathrm{length}\left(\mathcal{M}_{0}\right)}{16\nu^{4}\left(t_{1}-t_{0}\right)^{4}}\left(\begin{array}{cc}
0 & 0\\
0 & 1
\end{array}\right),
\end{align*}
implying 
\[
\left\langle \mathbf{L}\mathbf{\bm{\Omega}}\mathbf{x}_{0}^{\prime},\mathbf{\bm{\Omega}}\mathbf{x}_{0}^{\prime}\right\rangle \leq0
\]
satisfying the necessary condition (\ref{eq:restricted cond}) for
a maximizer.

In summary, our theory correctly identifies the walls and the jet
core as noteworthy features of this channel flow. While these features
are all considered as inhibitors of transport in informal descriptions
of jet-type flows, our approach reveals that in strict mathematical
terms, only the walls act as diffusion minimizers. The jet core, in
contrast, is a diffusion maximizer with respect to any localized perturbation
and with respect to parallel translations.

\subsection{Spatially periodic recirculation cells}

A spatially periodic, unsteady solution of the 2D Navier--Stokes
equations is given by (cf.~Majda and Bertozzi \cite{Majda2002})
\begin{equation}
\mathbf{v}(\mathbf{x},t)=ae^{-4\pi^{2}\nu t}\left(\begin{array}{c}
\sin(2\pi x)\sin(2\pi y)\\
\cos(2\pi x)\cos(2\pi y)
\end{array}\right),\label{eq:vorte array}
\end{equation}
whose vorticity field and Jacobian are given by
\begin{align}
\omega(\mathbf{x},t) & =-4a\pi e^{-4\pi^{2}\nu t}\sin(2\pi x)\cos(2\pi y),\nonumber \\
\mathbf{\bm{\nabla}}\mathbf{v}(\mathbf{x},t) & =2\pi ae^{-4\pi^{2}\nu t}\left(\begin{array}{cc}
\cos(2\pi x)\sin(2\pi y) & \sin(2\pi x)\cos(2\pi y)\\
-\sin(2\pi x)\cos(2\pi y) & -\cos(2\pi x)\sin(2\pi y)
\end{array}\right),\label{eq:Jacobian}
\end{align}
satisfies the advection--diffusion equation (\ref{eq:viscid-1})
with viscosity $\nu$ and a real parameter $a$ that controls the
overall strength of the vorticity field. 

\begin{figure}[H]
\begin{centering}
\includegraphics[width=0.5\textwidth]{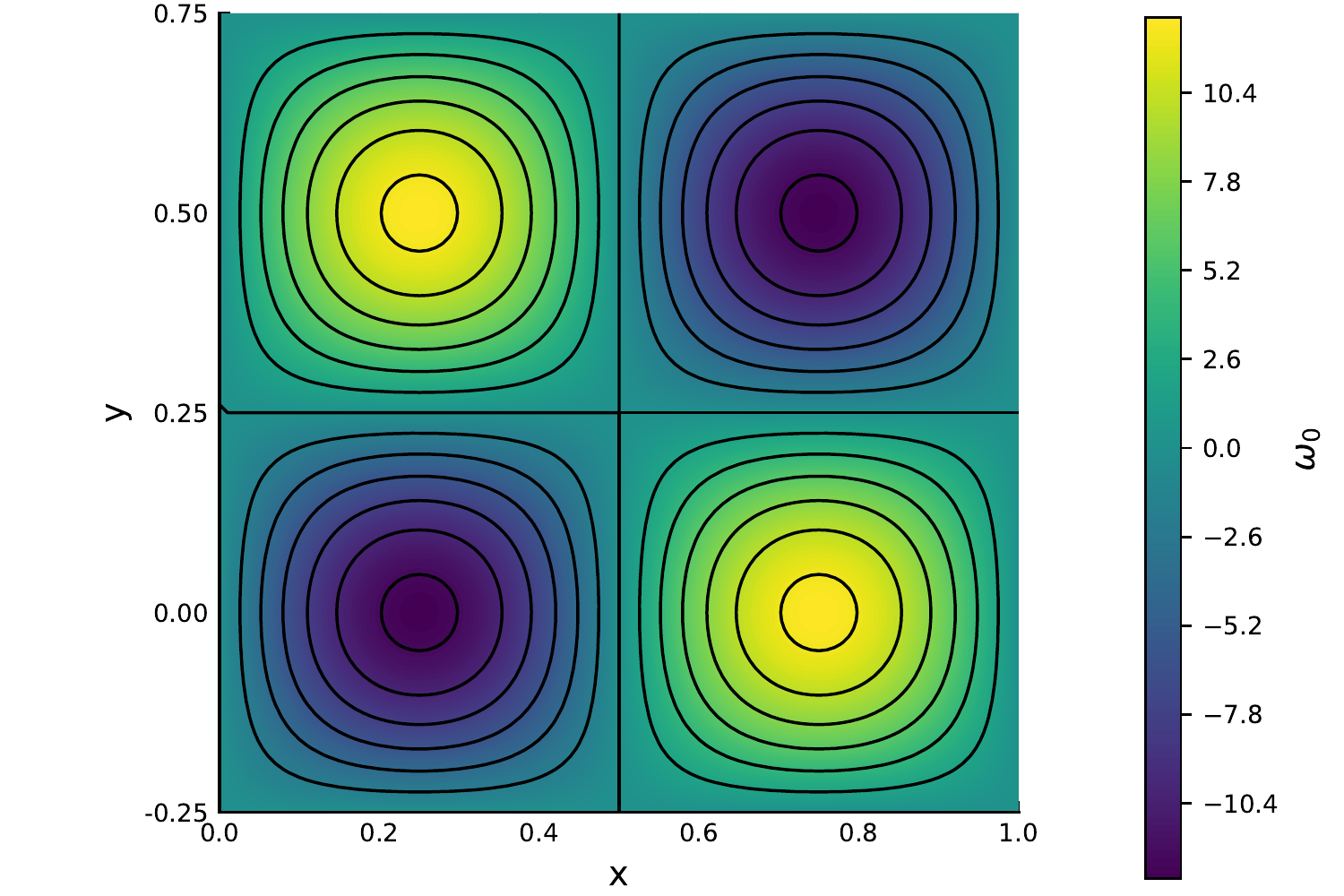}
\par\end{centering}
\caption{Unsteady vortex array with $a=1$ at time $t=0$. \label{fig:vortex array} }
\end{figure}

The vorticity gradient is 

\[
\mathbf{\bm{\nabla}}\omega\left(\mathbf{x},t\right)=-8a\pi^{2}e^{-4\pi^{2}\nu t}\left(\begin{array}{c}
\cos(2\pi x)\cos(2\pi y)\\
-\sin(2\pi x)\sin(2\pi y)
\end{array}\right)=-8a\pi^{2}\mathbf{\bm{\Omega}\mathbf{v}}(\mathbf{\mathbf{x}},t),
\]
 whose squared norm satisfies
\begin{align*}
\frac{\left|\mathbf{\bm{\nabla}}\omega\left(\mathbf{x},t\right)\right|^{2}}{\left(8\pi^{2}e^{-4\pi^{2}\nu t}\right)^{2}} & =a^{2}\left[\cos^{2}(2\pi x)\cos^{2}(2\pi y)+\sin^{2}(2\pi x)\sin^{2}(2\pi y)\right]\\
 & \leq a^{2}.
\end{align*}
The flow has horizontal and vertical heteroclinic orbits connecting
the array of saddle-type fixed points at $\left(x,y\right)=\left(\frac{j}{4},\frac{k}{4}\right)$
for arbitrary integers $j$ and $k$. This heteroclinic network surrounds
an array of vortical recirculation regions. Even though the velocity
field is unsteady, its streamline geometry consists of steady material
lines (cf. Fig. \ref{fig:vortex array}). Only the value of the vorticity
changes in time by the same factor along these material lines, just
as in our previous example.

The main observed features in the diffusive transport of vorticity
in this flow are the cell boundaries formed by the heteroclinic orbits.
Along these orbits, $\left|\mathbf{\bm{\nabla}}\omega\left(\mathbf{x},t\right)\right|^{2}$
admits maximum ridges that decay slowly in time by a uniform factor.
The ridges contain global minima with $\left|\mathbf{\bm{\nabla}}\omega\right|^{2}=0$
at the hyperbolic equilibria and global maxima with $\left|\mathbf{\bm{\nabla}}\omega\right|^{2}=a^{2}\left(8\pi^{2}e^{-4\pi^{2}\nu t}\right)^{2}$halfway
between them. Inside the cells, $\left|\mathbf{\bm{\nabla}}\omega\right|^{2}$
decays away from the ridge boundaries and reaches the global minimum
$\left|\mathbf{\bm{\nabla}}\omega\right|^{2}=0$ again at the elliptic
equilibria. All closed, periodic streamlines in the vortical region
are also perceived as features hindering the spread of high vorticity
from the centers of the vortical regions.

We now examine how our theory of constrained diffusion extremizers
bears on the vorticity field features identified above from observations.
Along, say, the $y=0.25$ horizontal heteroclinic streamline, the
velocity Jacobian (\ref{eq:Jacobian}) becomes
\[
\mathbf{\bm{\nabla}}\mathbf{v}(\left(x,0.25\right),t)=2\pi a\cos(2\pi x)e^{-4\pi^{2}\nu t}\left(\begin{array}{cc}
1 & 0\\
0 & -1
\end{array}\right),
\]
 implying 
\begin{align*}
\left[\mathbf{\bm{\nabla}}_{0}\mathbf{F}_{t_{0}}^{t}((x_{0},0.25))\right]^{-1} & =\left(\begin{array}{cc}
\exp\left[-\int_{t_{0}}^{t}2\pi a\cos(2\pi x(s))e^{-4\pi^{2}\nu s}ds\right] & 0\\
0 & \exp\left[\int_{t_{0}}^{t}2\pi a\cos(2\pi x(s))e^{-4\pi^{2}\nu s}ds\right]
\end{array}\right),\\
\mathbf{\bm{\nabla}}\omega\left(\mathbf{F}_{t_{0}}^{t}\left((x_{0},0.25)\right),t\right) & =8a\pi^{2}\sin(2\pi x(t))e^{-4\pi^{2}\nu t}\left(\begin{array}{c}
0\\
1
\end{array}\right).
\end{align*}
This gives 
\[
\left[\mathbf{\bm{\nabla}}_{0}\mathbf{F}_{t_{0}}^{t}(\mathbf{x}_{0})\right]^{-1}\mathbf{\bm{\nabla}}\omega\left(\mathbf{F}_{t_{0}}^{t}\left(\mathbf{x}_{0}\right),t\right)\parallel\left(\begin{array}{c}
0\\
1
\end{array}\right).
\]
and hence, for any $t_{1}>t_{0}$ and for any initial point $\mathbf{x}_{0}=\left(x_{0},0.25\right)$,
we have (cf.~Remark \ref{rem:simplified q}) 
\begin{align*}
\bar{\mathbf{q}}_{t_{0}}^{t_{1}}(\mathbf{x}_{0}) & =\frac{1}{t_{1}-t_{0}}\int_{t_{0}}^{t_{1}}\left[\mathbf{\bm{\nabla}}_{0}\mathbf{F}_{t_{0}}^{t}(\mathbf{x}_{0})\right]^{-1}\mathbf{\bm{\nabla}}\omega\left(\mathbf{F}_{t_{0}}^{t}\left(\mathbf{x}_{0}\right),t\right)dt\parallel\left(\begin{array}{c}
0\\
1
\end{array}\right).
\end{align*}
We conclude that $y=0.25$ horizontal heteroclinic streamline with
unit normal $\mathbf{n}_{0}(\mathbf{x}_{0})=\left(0,1\right)$ is
a perfect transport enhancer in the sense of formula (\ref{eq:zero inner product-1}).
An identical conclusion holds for all other heteroclinic connections.

In contrast, along closed, vortical streamlines, we find the integrand
of $\bar{\mathbf{q}}_{t_{0}}^{t_{1}}(\mathbf{x}_{0})$ to align with
these streamlines due to the shearing effect of the inverse flow map
$\left[\mathbf{\bm{\nabla}}_{0}\mathbf{F}_{t_{0}}^{t}(\mathbf{x}_{0})\right]^{-1}$
on the streamline-normal vorticity gradient $\mathbf{\bm{\nabla}}\omega\left(\mathbf{F}_{t_{0}}^{t}\left(\mathbf{x}_{0}\right),t\right)$.
\emph{This implies that in the $t_{1}\to\infty$ limit, all closed
streamlines become asymptotically perfect transport barriers, with
their normals asymptotically aligning with $\bar{\mathbf{q}}_{t_{0}}^{t_{1}}(\mathbf{x}_{0})$
at the same rate at all of their points.} 

\begin{figure}
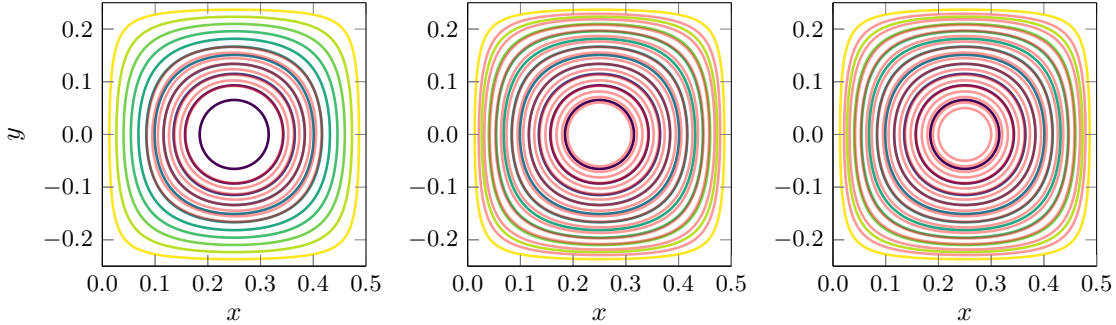

\input{cell_int_time_3.tex}\input{cell_int_time_13.tex}\input{cell_int_time_23.tex}\caption{Closed material vorticity transport barriers (in light red) in the
lower-left recirculation cell of Fig. \ref{fig:vortex array} for
integration times $T=3$, $T=13$, and $T=23$ (from left to right),
on top of vorticity level sets (yellow-green-blue).}

\label{fig:recirculation_closed_orbits}
\end{figure}

For finite times, the exact closed transport barriers in this flow
can be identified numerically by computing the ODEs appearing in Theorem
\ref{thm:2D_flows} for the velocity field (\ref{thm:2D_flows}).
For this computation, we set $a=1$ and $\nu=0.001$. The results
shown in Fig.~\ref{fig:recirculation_closed_orbits} show a close
match between an increasing number of detected barriers and vorticity
level sets as the integration time is extended from $T=3$ (left)
via $T=13$ (middle) up to $T=23$ (right).

\section{Application to transport-barrier detection in ocean-surface dynamics}

We now illustrate our results on two different ocean surface velocity
data sets. The first one is HYCOM, a data-assimilating hybrid ocean
model, whose ocean-surface velocity output is generally not divergence-free
and hence represents a compressible 2D-flow. The second data set is
a two-dimensional unsteady velocity field obtained from AVISO satellite
altimetry measurements under the geostrophic approximation. This data
set is currently distributed by the Copernicus Marine and Environment
Monitoring Service (CMEMS). Due to the geostrophic approximation,
this velocity field is constructed as divergence-free.

All simulations in this section have been performed with the package
\href{https://github.com/CoherentStructures/CoherentStructures.jl}{CoherentStructures.jl},
a collection of implementations of objective coherent structure detection
methods written in the open-source programming language Julia. Our
computations rely crucially on the ODE integration codes provided
by the \href{https://github.com/JuliaDiffEq/DifferentialEquations.jl}{DifferentialEquations.jl}
package \cite{Rackauckas2017}.

\subsection{Unconstrained transport-barriers in the compressible HYCOM velocity
data set}

\begin{figure}
\centering\includegraphics[width=0.8\textwidth]{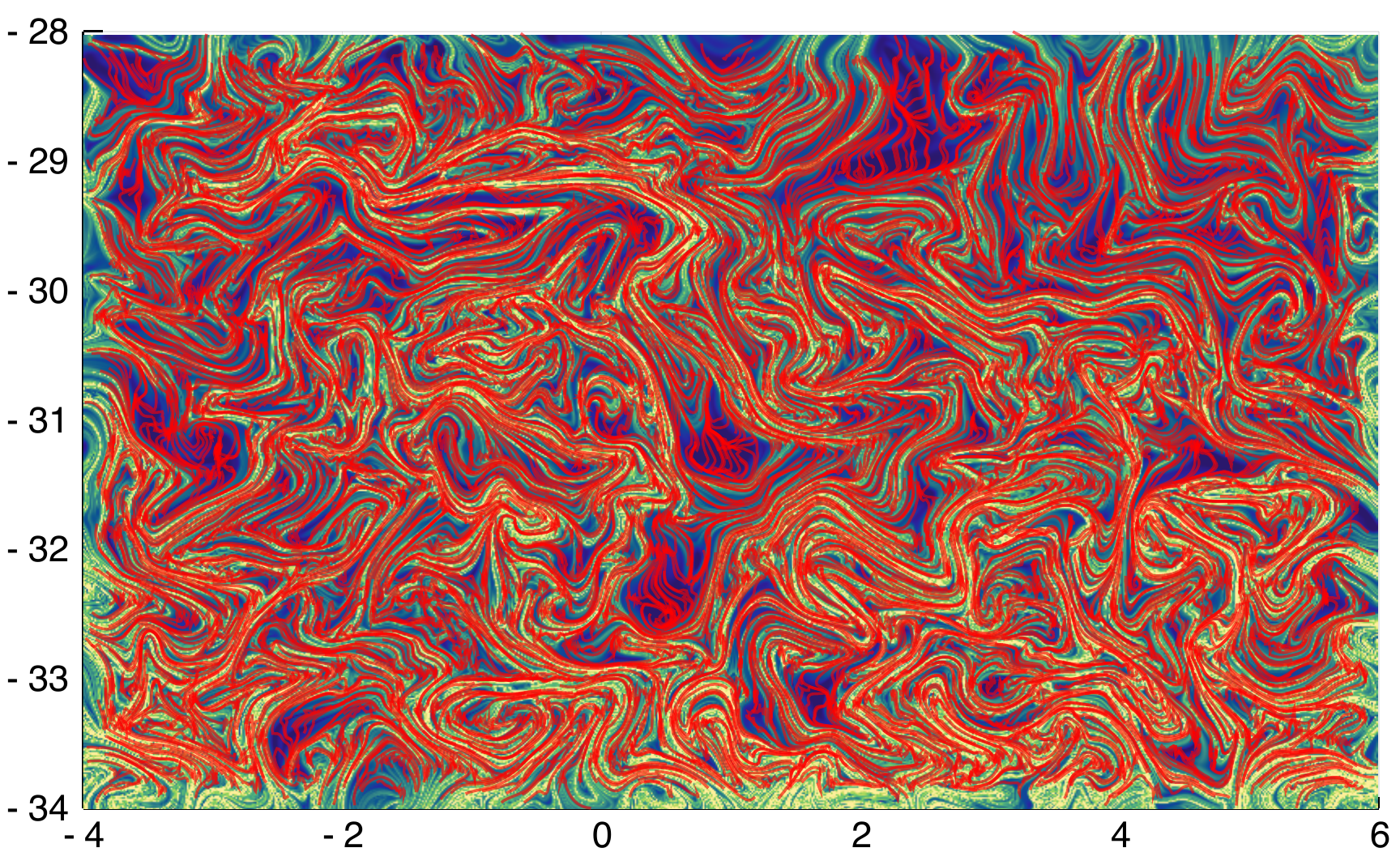}

\caption{Open transport barriers in the HYCOM ocean surface data set. Shown
in the background is the $DBS$ field (with tails cut for visualization
purposes). Overlaid are short integral curve segments of the $\bm{\eta}_{\mathcal{T}_{0}}^{+}$
field (\ref{eq:eta field}) for $\mathcal{T}_{0}=1$. Other values
of $\mathcal{T}_{0}$ produce similar results.}

\label{fig:HYCOM_DBS}
\end{figure}

We use ocean surface velocity data from 2013-12-15 to 2014-01-14,
i.e., 30 days, taken from the Agulhas leakage area at the southern
tip of Africa. In Fig.~(\ref{fig:HYCOM_DBS}), we show the diagnostic
$DBS$ field (\ref{eq:DBS unconstrained}), whose features align,
as expected, remarkably close with the unconstrained, uniform transport
barriers extracted as trajectories of the $\bm{\eta}_{\mathcal{T}_{0}}^{+}$
field (\ref{eq:eta field}) for $\mathcal{T}_{0}=1$.

\begin{figure}
\centering\includegraphics[width=0.49\textwidth]{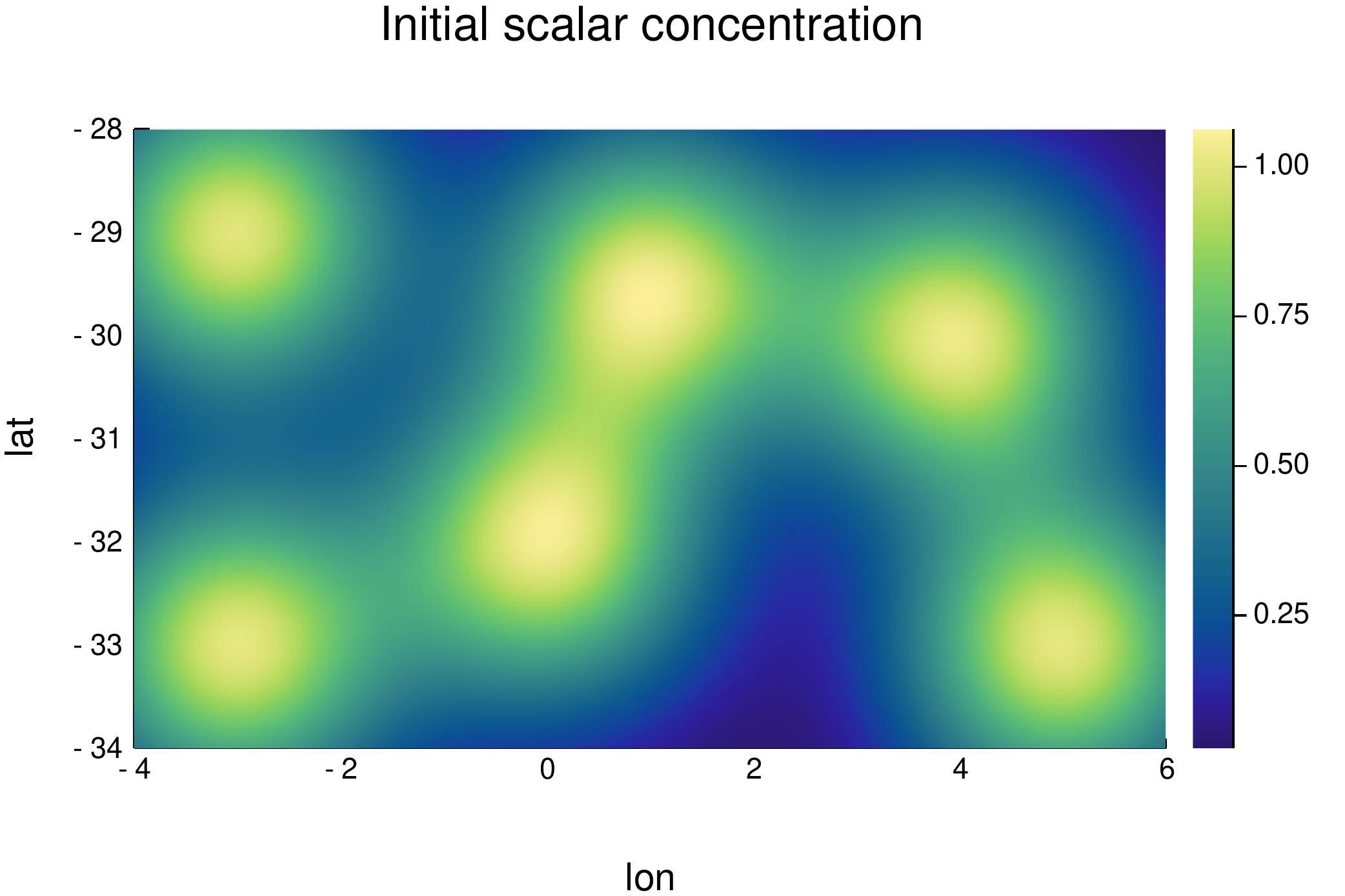}\includegraphics[width=0.49\textwidth]{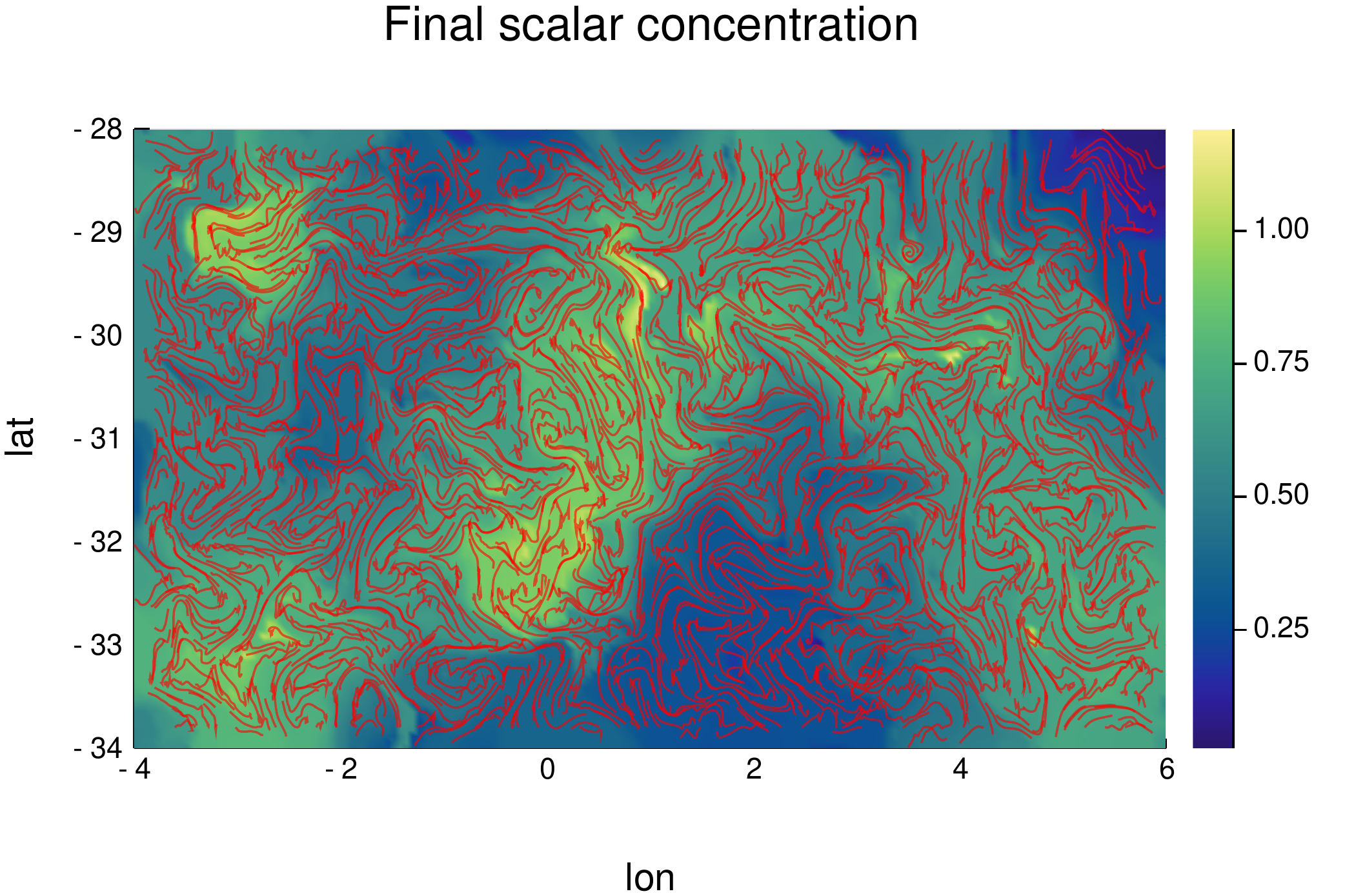}

\caption{Transport barriers in the HYCOM ocean surface data set. The scalar
field corresponds to the initial (left) and final (right) scalar density,
evolved under the advection--diffusion equation (\ref{eq:adv-diff})
in Lagrangian coordinates. The final scalar concentration field is
shown with short integral curve segments of the $\bm{\eta}_{\mathcal{T}_{0}}^{+}$
field (\ref{eq:eta field}) overlaid for a small value $\mathcal{T}_{0}=1$
of the nondimensionalized, uniform transport density.}

\label{fig:HYCOM_concentration}
\end{figure}

As a second step, we now verify if these material curves (obtained
from purely advective calculations) indeed act as observed transport
barriers for a diffusive scalar field. To this end, we solve the advection--diffusion
equation (\ref{eq:adv-diff}), with $k(t)=f(\mathbf{x},t)\equiv0$
and $\mathbf{D}(\mathbf{x},t)\equiv\mathbf{I}$, in Lagrangian coordinates
for the initial concentration shown in Fig.~\ref{fig:HYCOM_concentration}
(left). The final density obtained from this computation is then shown
in Fig.~\ref{fig:HYCOM_concentration} (right), with the same uniform
transport barriers overlaid as in Fig.~(\ref{fig:HYCOM_DBS}). Note
how the predicted barriers indeed capture detailed features in the
evolving concentration field.

\subsection{Constrained diffusion barriers in the AVISO velocity data}

We now illustrate our results on two-dimensional unsteady velocity
data obtained from AVISO satellite altimetry measurements. The domain
of the dataset is the Agulhas leakage in the Southern Ocean. Under
the assumption of a geostrophic flow, the sea surface height $h$
serves as a streamfunction for the surface velocity field. In longitude--latitude
coordinates $\left(\varphi,\theta\right)$, particle trajectories
are then solutions of the system
\begin{equation}
\dot{\varphi}=-\frac{g}{R^{2}f(\theta)\cos\theta}\partial_{\theta}h(\varphi,\theta,t),\qquad\dot{\theta}=\frac{g}{R^{2}f(\theta)\cos\theta}\partial_{\varphi}h(\varphi,\theta,t),\label{eq:aviso flow}
\end{equation}
where $g$ is the constant of gravity, $R$ is the mean radius of
the Earth and $f(\theta)\coloneqq2\Omega\sin\theta$ is the Coriolis
parameter with $\Omega$ denoting the mean angular velocity of the
Earth. The computational domain is chosen as in several other studies
before (see, e.g., \cite{Haller2013a,Karrasch2015,Haller2018}), with
integration time $T$ equal to 90 days.

In this two-dimensional flow, we wish to determine material transport
barriers for the vorticity $\omega(\mathbf{x},t)$, i.e., the single
nonzero component of $\nabla\times\mathbf{v}$ normal to the plane
of the flow (\ref{eq:aviso flow}). Following Remark \ref{rem:simplified q},
we use the exact transport vector field $\mathbf{\bar{q}}_{t_{0}}^{t_{1}}$,
(\ref{eq:q redefined}). For closed-orbit detection, we employ a numerical
scheme analogous to \cite{Karrasch2015}, with
\begin{enumerate}
\item an index-theory based preselection of elliptic-type subdomains;
\item placement of Poincaré sections in regions with appropriate index; 
\item launch of integral curves from seed points, solving for the transport
parameter which yields a closed orbit at the respective seed point.
\end{enumerate}
\begin{figure}
\centering\includegraphics[width=0.49\textwidth]{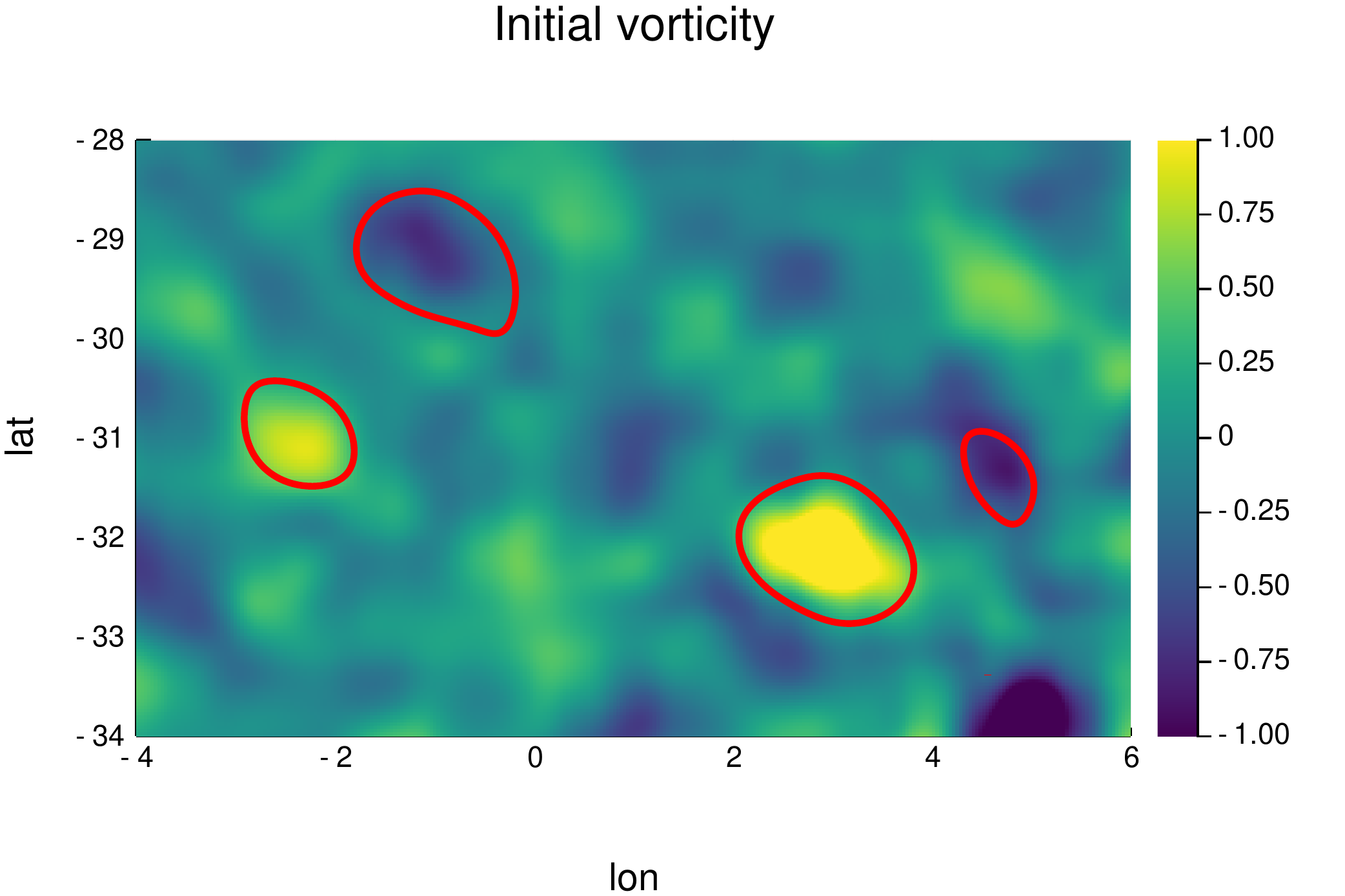}\includegraphics[width=0.49\textwidth]{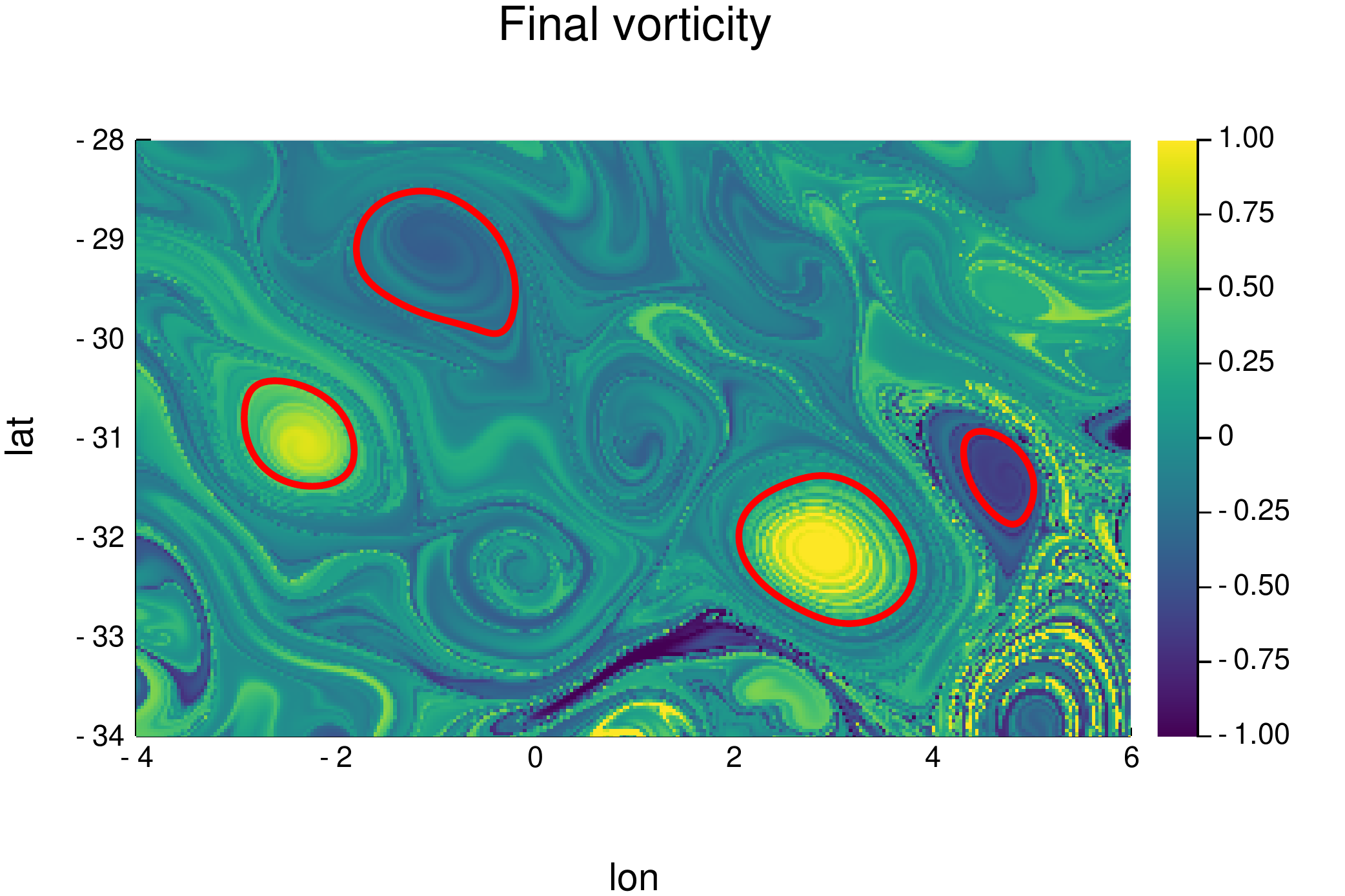}

\caption{Closed, constrained vorticity transport barriers in the AVISO ocean
surface data set. The scalar field corresponds to vorticity (truncated
at $\pm1$ for visualization purposes) at the initial (left) and the
final (right) time instances in Lagrangian coordinates.}

\label{fig:AVISO_vorticity}
\end{figure}

The closed material vorticity transport barriers obtained from this
procedure are shown in Fig.~\ref{fig:AVISO_vorticity}, on top of
the initial (left) and final (right) vorticity fields in Lagrangian
coordinates. We note that closed diffusion barriers arise around all
four vortical regions identified by previous studies (see, e.g., \cite{Haller2013a,Karrasch2015,Hadjighasem2016a,Serra2017,Haller2018})
as materially coherent. The closed region boundaries here are optimized
to be extremizers of the vorticity transport, as opposed to be outermost
coherent material curves.

\section{Conclusions}

We have shown how recent results on barriers to diffusive transport
extend from the incompressible case treated in \cite{Haller2018}
to compressible flows. In our present setting, we have also allowed
for the presence of sources and sinks in the concentration field.
In addition, we have distinguished between the case of an unknown
initial concentration (unconstrained extremizers) and the case of
a specifically known initial concentration field (constrained extremizers)
with respect to which diffusive transport is to be extremized over
material surfaces.

For unconstrained barriers, we have obtained results that formally
coincide with those in \cite{Haller2018}, except that the flow map
here is compressible and the initial density of the fluid appears
in the transport tensor. Despite the similarity with the results in
\cite{Haller2018}, the present results have required a different
derivation and additional assumptions on the selection of the most
diffusion-prone initial concentration field near each material surface
in the flow. In this formulation, concentration sinks and sources
turn out to play no explicit role in the leading-order transport extremization
problem. As in the incompressible case, we have obtained explicit
direction fields defining the barriers for two-dimensional flows.
For higher-dimensional flows, the barriers continue to be null-surfaces
of a tensor field, but satisfy partial differential equations. In
any dimension, however, the diffusion barriers strength (DBS) field
can directly be computed from the velocity field and serves as a diagnostic
to map out the global barrier distribution.

For constrained barriers, we have sought material surfaces that block
transport more than their neighbors do under a specific initial concentration
field. In this case, the equations defining the diffusion extremizers
depend explicitly on the sink or source distribution for the concentration,
as well as on the (possibly time-dependent) spontaneous concentration
decay rate. Constrained barrier surfaces also satisfy explicit differential
equations in two dimensions and partial differential equations in
higher dimensions. A DBS scalar field is again available in any dimensions
for diagnostic purposes. We have found in canonical examples that
some classically documented transport barriers (such as jet cores
and unstable manifolds) are, in fact, perfect enhancers of diffusive
transport. Barriers that are strict local minimizers of diffusive
transport, by contrast, are rare and must completely block transport
at leading order, such as the walls of a channel flow. In two dimensions,
with the exception of perfect barriers, constrained uniform barriers
are local maximizers of diffusive transport with respect to all localized
perturbations. 

Finally, we have shown how the above results extend to barriers to
the transport of probability densities for particle motion in compressible,
stochastic velocity fields modeled by Itô processes. This extension
enables one to locate barriers to stochastic transport from purely
deterministic calculations, as long as the diffusivities involved
are small, which is generally the case for geophysical flows.

As we illustrated on two explicitly known Navier-Stokes flows, the
present results on constrained barriers enable the detection of barriers
to the diffusion of vorticity in two-dimensional flows. This follows
because planar vorticity transport is governed by an (incompressible)
advection--diffusion equation whose initial condition (the initial
vorticity) cannot be considered unknown or uncertain once the velocity
field is known. Barriers to vorticity diffusion in three dimensions,
however, cannot be treated by the present results, given that vorticity
is an active vector field, rather than a passively diffusing scalar
field, in three-dimensional flows. More generally, the construction
of barriers to the transport of active scalar- and vector field requires
new ideas relative to those in the present work.

\subsubsection*{Acknowledgment}

We would like to thank Ryan Abernathey, Francisco J. Beron--Vera,
Stergios Katsanoulis and Jean-Luc Thiffeault for useful discussions,
and Nate Schilling for valuable contributions to the code as part
of \href{https://github.com/CoherentStructures/CoherentStructures.jl}{CoherentStructures.jl}.
The 1/12 deg global HYCOM+NCODA Ocean Reanalysis was funded by the
U.S. Navy and the Modeling and Simulation Coordination Office. Computer
time was made available by the DoD High Performance Computing Modernization
Program. The output is publicly available at \href{http://hycom.org}{http://hycom.org}.
The Ssalto/Duacs altimeter products were produced and distributed
by the Copernicus Marine and Environment Monitoring Service (CMEMS)
(\href{http://www.marine.copernicus.eu}{http://www.marine.copernicus.eu}).
G.H.~and D.K.~acknowledge partial support from the Turbulent Superstructures
priority program of the German National Science Foundation (DFG). 

\appendix

\section{Proof of \thmref{leading_order_transport}\label{sec:Appendix-A}}

We start by establishing an expression for the instantaneous flux
vector associated with the transport of $\mu$ through a material
surface. We consider first the transport of $\mu$ out of an arbitrary,
closed material volume $V(t)$ to obtain an expression for the flux
of $\mu$ through $\partial V(t)$ via the divergence theorem. The
same flux expression is then applicable to a general material surface
$\mathcal{M}(t)$, given that $V(t)$ is arbitrary. Indeed, for any
$\mathcal{M}(t)$, the set $V(t)$ can be selected as an $\mathcal{O}(\epsilon)$
volume whose boundary is the union of $\mathcal{M}(t)$, a parallel
translate of $\mathcal{M}(t)$ by a distance $\mathcal{O}(\epsilon)$,
and a cylindrical surface of area $\mathcal{O}(\epsilon^{2}).$ Taking
the $\epsilon\to0$ limit, one then obtains the same flux vector for
$\mathcal{M}(t)$ as for $\partial V(t)$.

The rate of change in the mass-based concentration of $\mu$ in a
closed material volume $V(t)=\mathbf{F}_{t_{0}}^{t}\left(V(t)\right)$
is, by definition, 

\begin{align}
\frac{d}{dt}\int_{V(t)}\mu(\mathbf{x},t)\rho(\mathbf{x},t)dV & =\frac{d}{dt}\int_{V(t_{0})}\mu(\mathbf{F}_{t_{0}}^{t}(\mathbf{x}_{0}),t)\rho_{0}(\mathbf{x}_{0})\,dV_{0}=\int_{V(t_{0})}\partial_{t}\hat{\mu}(\mathbf{x}_{0},t)\rho_{0}(\mathbf{x}_{0})\,dV_{0}\nonumber \\
 & =\nu\int_{V(t_{0})}\mathbf{\bm{\nabla}}_{0}\cdot\left[\mathbf{T}_{t_{0}}^{t}(\mathbf{x}_{0})\mathbf{\bm{\nabla}}_{0}\left(\hat{\mu}(\mathbf{x}_{0},t)+b(\mathbf{x}_{0},t)\right)\right]\,dV_{0}\nonumber \\
 & =\nu\int_{\partial V(t_{0})}\left\langle \mathbf{T}_{t_{0}}^{t}\mathbf{\bm{\nabla}}_{0}\left(\hat{\mu}(\mathbf{x}_{0},t)+b(\mathbf{x}_{0},t)\right),\mathbf{n}_{0}\right\rangle \,dA_{0},\label{eq:mass integral}
\end{align}
with the last integral denoting the surface integral over the $\left(n-1\right)$-dimensional
boundary $\partial V(t_{0})$ of the $n$-dimensional volume $V(t_{0}).$
As discussed above, the calculation implies that in Lagrangian coordinates,
the flux vector for the field $\mu(\mathbf{x},t)$ through an arbitrary
material surface $\mathcal{M}(t)$ is also $\left\langle \mathbf{T}_{t_{0}}^{t}\mathbf{\bm{\nabla}}_{0}\left(\hat{\mu}(\mathbf{x}_{0},t)+b(\mathbf{x}_{0},t)\right),\mathbf{n}_{0}\right\rangle ,$
with $\mathbf{n}_{0}\left(\mathbf{x}_{0}\right)$ denoting an oriented
unit normal field to $\mathcal{M}(t_{0})$. 

To proceed, we take the $\mathbf{\bm{\nabla}}_{0}$-gradient of both
sides of (\ref{eq:mu-adv-diff}) and integrate in time to obtain
\begin{equation}
\mathbf{\bm{\nabla}}_{0}\hat{\mu}(\mathbf{x}_{0},t)=\mathbf{\bm{\nabla}}_{0}c_{0}(\mathbf{x}_{0})+\nu\int_{t_{0}}^{t}\mathbf{\bm{\nabla}}_{0}\left[\frac{1}{\rho_{0}(\mathbf{x}_{0})}\mathbf{\bm{\nabla}}_{0}\cdot\left(\mathbf{T}_{t_{0}}^{s}(\mathbf{x}_{0})\mathbf{\bm{\nabla}}_{0}\left[\hat{\mu}(\mathbf{x}_{0},s)+b(\mathbf{x}_{0},s)\right]\right)\right]ds,\label{eq:mu hat gradient}
\end{equation}
where we have used the relation $\mathbf{\bm{\nabla}}_{0}\hat{\mu}(\mathbf{x}_{0},t_{0})=\mathbf{\bm{\nabla}}_{0}c(\mathbf{x}_{0})$,
which follows from (\ref{eq:mudef}). Then with the flux vector obtained
in the last eq. (\ref{eq:mass integral}) and with expression (\ref{eq:mu hat gradient})
for $\mathbf{\bm{\nabla}}_{0}\hat{\mu}(\mathbf{x}_{0},t)$ and hand,
the total transport of $\mu$ through $\mathcal{M}(t)$ can be written
as
\begin{align}
\Sigma_{t_{0}}^{t_{1}}(\mathcal{M}_{0}) & =\nu\int_{t_{0}}^{t_{1}}\int_{\mathcal{M}_{0}}\left\langle \mathbf{T}_{t_{0}}^{t}(\mathbf{x}_{0})\mathbf{\bm{\nabla}}_{0}\left(\hat{\mu}(\mathbf{x}_{0},t)+b(\mathbf{x}_{0},t)\right),\mathbf{n}_{0}(\mathbf{x}_{0})\right\rangle \,dA_{0}\,dt\label{eq:sigma first eq}\\
 & =\nu\int_{t_{0}}^{t_{1}}\int_{\mathcal{M}_{0}}\left\langle \mathbf{T}_{t_{0}}^{t}\left(\mathbf{\bm{\nabla}}_{0}c_{0}(\mathbf{x}_{0})+\mathbf{\bm{\nabla}}_{0}b(\mathbf{x}_{0},t)\right),\mathbf{n}_{0}\right\rangle \,dA_{0}\,dt+\nonumber \\
 & \phantom{=}+\nu^{2}\int_{t_{0}}^{t_{1}}\int_{\mathcal{M}_{0}}\int_{t_{0}}^{t}\mathbf{\bm{\nabla}}_{0}\left[\frac{1}{\rho_{0}(\mathbf{x}_{0})}\mathbf{\bm{\nabla}}_{0}\cdot\left(\mathbf{T}_{t_{0}}^{s}(\mathbf{x}_{0})\mathbf{\bm{\nabla}}_{0}\left[\hat{\mu}(\mathbf{x}_{0},s)+b(\mathbf{x}_{0},s)\right]\right)\right]^{T}\mathbf{T}_{t_{0}}^{s}\mathbf{n}_{0}ds\,dA_{0}\,dt.\nonumber 
\end{align}
The statement of the theorem, therefore, follows if the last term
in (\ref{eq:sigma first eq}) is of order $o(\nu)$, i.e., if

{\small{}
\begin{equation}
\lim_{\nu\to0}\nu\int_{t_{0}}^{t_{1}}\int_{\mathcal{M}_{0}}\int_{t_{0}}^{t}\mathbf{\bm{\nabla}}_{0}\left[\frac{1}{\rho_{0}(\mathbf{x}_{0})}\mathbf{\bm{\nabla}}_{0}\cdot\left(\mathbf{T}_{t_{0}}^{s}(\mathbf{x}_{0})\mathbf{\bm{\nabla}}_{0}\left[\hat{\mu}(\mathbf{x}_{0},s)+b(\mathbf{x}_{0},s)\right]\right)\right]^{T}\mathbf{T}_{t_{0}}^{s}\mathbf{n}_{0}ds\,dA_{0}\,dt=0.\label{eq:ansatz}
\end{equation}
}{\small\par}

To prove (\ref{eq:ansatz}), we need estimates on the solution of
the initial value problem 
\begin{align}
\partial_{t}\hat{\mu}(\mathbf{x}_{0},t) & =\nu\frac{1}{\rho_{0}(\mathbf{x}_{0})}\mathbf{\bm{\nabla}}_{0}\cdot\left(\mathbf{T}_{t_{0}}^{t}(\mathbf{x}_{0})\mathbf{\bm{\nabla}}_{0}\left[\hat{\mu}(\mathbf{x}_{0},t)+b(\mathbf{x}_{0},t)\right]\right),\label{main}\\
\hat{\mu}(\mathbf{x}_{0},t_{0}) & =c_{0}(\mathbf{x}_{0}).\nonumber 
\end{align}
Based on our initial assumptions, we have the following bounds on
the entries $T_{ij}(\mathbf{x}_{0},t):=\left[\mathbf{T}_{t_{0}}^{t}(\mathbf{x}_{0})\right]_{ij}$
of the matrix representation of $\mathbf{T}_{t_{0}}^{t}$: 
\begin{equation}
\begin{split} & \left|\rho_{0}(\mathbf{x}_{0})^{-1}T_{ij}(\mathbf{x}_{0},t)-\rho_{0}(\mathbf{y}_{0})^{-1}T_{ij}(\mathbf{y}_{0},s)\right|\leq(C_{1}\left|\mathbf{x}_{0}-\mathbf{y}_{0}\right|^{\alpha}+C_{2}\left|t-s\right|^{\frac{\alpha}{2}}),\\
 & \left|\rho_{0}(\mathbf{x}_{0})^{-1}\bm{\nabla}_{0}T_{ij}(\mathbf{x}_{0},t)-\rho_{0}(\mathbf{y}_{0})^{-1}\bm{\nabla}_{0}T_{ij}(\mathbf{y}_{0},t)\right|\leq C_{3}|\mathbf{x}_{0}-\mathbf{y}_{0}|^{\alpha},
\end{split}
\label{Hold}
\end{equation}
for some constant $0<\alpha\leq1$ and for all $\mathbf{x}_{0},\mathbf{y}_{0}\in U$
and $t,s\in[t_{0},t_{1}]$. By the positive definiteness of $\mathbf{T}_{t_{0}}^{t}(\mathbf{x}_{0})$
and the positivity of $\rho_{0}$, we also have 
\begin{equation}
\lambda|\mathbf{u}|^{2}\leq\left\langle \mathbf{u},\frac{1}{\rho_{0}(\mathbf{x}_{0})}\mathbf{T}_{t_{0}}^{t}(\mathbf{x}_{0})\mathbf{u}\right\rangle \leq\Lambda|\mathbf{u}|^{2},\quad\mathbf{u}\in\mathbb{R}^{n},\,\,\mathbf{x}_{0}\in U,\,\,t\in[t_{1},t_{2}],\label{ell}
\end{equation}
which implies the bounds 
\begin{equation}
\frac{|\mathbf{u}|^{2}}{\Lambda}\leq\left\langle \mathbf{u},\rho_{0}(\mathbf{x}_{0})\left[\mathbf{T}_{t_{0}}^{t}(\mathbf{x}_{0})\right]^{-1}\mathbf{u}\right\rangle \leq\frac{|\mathbf{u}|^{2}}{\lambda},\quad\lambda^{n}\leq\rho_{0}(\mathbf{x}_{0})^{-n}\det\mathbf{T}_{t_{0}}^{t}(\mathbf{x}_{0})\leq\Lambda^{n},\label{Ainv}
\end{equation}
for all $\mathbf{u}\in\mathbb{R}^{n},\,\,\mathbf{x}_{0}\in U$ and
$t\in[t_{1},t_{2}]$.

Next, we observe that (\ref{eq:ansatz}) is satisfied when 
\begin{equation}
\sup_{\mathbf{x}_{0}\in U,t\in[t_{0},t_{1}]}\left|\mathbf{\mathbf{\bm{\nabla}}}_{0}\hat{\mu}(\mathbf{x}_{0},t)-\mathbf{\mathbf{\bm{\nabla}}}_{0}c_{0}(\mathbf{x}_{0})\right|=\mathcal{O}(\nu^{q}),\label{sup}
\end{equation}
holds for some $q>0$, as one obtains using (\ref{eq:mu hat gradient})
and estimating the supremum norm in $\mathbf{x}_{0}$ and $t$ using
(\ref{Hold}). Using the assumption that $c_{0}\in C^{2}(U),$ we
will now show that (\ref{sup}) holds, and hence (\ref{eq:ansatz})
is indeed satisfied. In our presentation, we will utilize a scaling
approach described by Friedman \cite{Friedman2013}.

Introducing the rescaled time variable $\tau:=\nu(t-t_{0})$ as well
as the shifted and rescaled concentration $w(\mathbf{x}_{0},\tau):=\hat{\mu}(\mathbf{x}_{0},t_{0}+\frac{\tau}{\nu})-c_{0}(\mathbf{x}_{0})$,
then setting $\mathbf{T}_{\nu}(\mathbf{x}_{0},\tau):=\mathbf{T}_{t_{0}}^{t_{0}+\frac{\tau}{\nu}}(\mathbf{x}_{0})$,
we can rewrite (\ref{main}) as 
\begin{equation}
\begin{cases}
w_{\tau}=\frac{1}{\rho_{0}}\mathbf{{\bm{\nabla}}}_{0}\cdot(\mathbf{T}_{\nu}\mathbf{\bm{\nabla}}_{0}w)+\frac{1}{\rho_{0}}\mathbf{\mathbf{\bm{\nabla}}}_{0}\cdot\left(\mathbf{T}_{\nu}\bm{\nabla}_{0}(c_{0}+b)\right),\\
w(\mathbf{x}_{0},0)=0,\quad(\mathbf{x}_{0},\tau)\in U\times[0,\nu(t_{1}-t_{0})].
\end{cases}\label{mainw}
\end{equation}
Condition (\ref{sup}) is then equivalent to 
\begin{equation}
\sup_{\mathbf{x}_{0}\in\Omega,t\in[0,\tau_{1}]}|\mathbf{\mathbf{\bm{\nabla}}}_{0}w(\mathbf{x}_{0},\tau)|=\mathcal{O}(\nu^{q}),\qquad\tau_{1}:=\nu(t_{1}-t_{0}),\label{supw}
\end{equation}
for some $q>0$. In non-divergence form, equation (\ref{mainw}) takes
the form 
\begin{equation}
w_{\tau}=\sum_{i,j=1}^{n}\frac{T_{\nu}^{ij}}{\rho_{0}}\frac{\partial^{2}w}{\partial x_{0}^{i}\partial x_{0}^{j}}+\sum_{i=1}^{n}\frac{1}{\rho_{0}}\left(\sum_{j=1}^{n}\frac{\partial T_{\nu}^{ij}}{\partial x_{0}^{j}}\right)\frac{\partial w}{\partial x_{0}^{i}}+f_{\nu},
\end{equation}
where we have defined 
\begin{equation}
f_{\nu}(\mathbf{x}_{0},\tau)\coloneqq\frac{1}{\rho_{0}(\mathbf{x}_{0})}\mathbf{\mathbf{\bm{\nabla}}}_{0}\cdot\left(\mathbf{T}_{\nu}(\mathbf{x}_{0},\tau)\bm{\nabla}_{0}(c_{0}(\mathbf{x}_{0})+b(\mathbf{x}_{0},\tau)\right).
\end{equation}

Let 
\begin{align}
Z(\mathbf{x}_{0},\tau;\bm{\xi},s) & \coloneqq\frac{\exp\left[-\frac{\left\langle \mathbf{x}_{0}-\bm{\xi},\rho_{0}\mathbf{T}_{\nu}^{-1}(\bm{\xi},s)(\mathbf{x}_{0}-\bm{\xi})\right\rangle }{4(\tau-s)}\right]}{(2\sqrt{\pi})^{n}\left[\rho_{0}^{-n}\det\mathbf{T}_{\nu}(\bm{\xi},s)\right]{}^{\frac{1}{2}}(\tau-s)^{\frac{n}{2}}},\label{defZ}\\
Z_{\tau} & =\rho_{0}^{-1}\mathbf{T}_{\nu}\mathbf{\mathbf{\bm{\nabla}}}_{0}^{2}Z,\nonumber 
\end{align}
for $\mathbf{x}_{0},\bm{\xi}\in\Omega$ and $\tau,s\in[0,\tau_{1}]$,
denote the fundamental solution of the homogeneous, second-order part
of (\ref{mainw}). For later computations, we note that with the $n$-dimensional
volume element $d\bm{\xi}=d\xi_{1}...d\xi_{n}$, we have the estimate{\tiny{}
\begin{align}
\int_{\Omega}Z(\mathbf{x}_{0},\tau;\mathbf{\bm{\xi}},s)\,d\bm{\xi} & =\int_{\Omega}(2\sqrt{\pi})^{-n}\left[\rho_{0}^{n}\det\mathbf{T}_{\nu}^{-1}\right]^{-\frac{1}{2}}(\tau-s)^{-\frac{n}{2}}e^{-\frac{\left\langle \mathbf{x}_{0}-\bm{\xi},\rho_{0}\mathbf{T}_{\nu}^{-1}(\mathbf{x}_{0}-\bm{\xi})\right\rangle }{4(\tau-s)}}\,d\bm{\xi}\nonumber \\
 & \leq\int_{\Omega}(2\sqrt{\pi})^{-n}\lambda^{-\frac{n}{2}}(\tau-s)^{-\frac{n}{2}}e^{-\frac{|\mathbf{x}_{0}-\bm{\mathbf{\xi}}|^{2}}{4\Lambda(\tau-s)}}\,d\mathbf{\bm{\xi}},\label{estZ}
\end{align}
}where we have used the inequalities in (\ref{Ainv}). With the rescaled
spatial variable $\mathbf{y}$ and the rescaled volume form $d\mathbf{y}$
defined as 
\begin{equation}
\begin{split}\mathbf{y} & =(2\Lambda)^{-\frac{1}{2}}(\tau-s)^{-1/2}(\mathbf{x}-\bm{\xi}),\quad d\mathbf{y}=(2\Lambda)^{-\frac{n}{2}}(\tau-s)^{-\frac{n}{2}}d\mathbf{\bm{\xi}},\end{split}
\label{scaling}
\end{equation}
we define the set $\Omega_{\mathbf{x}_{0},\tau,s}:=(2\Lambda)^{-\frac{1}{2}}(\tau-s)^{-1/2}(\mathbf{x}_{0}-\Omega)$
to obtain from (\ref{estZ}) the estimate 
\begin{equation}
\begin{split}\int_{\Omega}Z(\mathbf{x}_{0},\tau;\mathbf{\bm{\xi}},s)\,d\bm{\xi} & \leq\pi^{-\frac{n}{2}}\left(\frac{\Lambda}{\lambda}\right)^{\frac{n}{2}}\int_{\Omega_{\mathbf{x},\tau,s}}e^{-|\mathbf{y}|^{2}}d\mathbf{y}\\
 & \leq\pi^{-\frac{n}{2}}\left(\frac{\Lambda}{\lambda}\right)^{\frac{n}{2}}\int_{\mathbb{R}^{n}}e^{-|\mathbf{y}|^{2}}d\mathbf{y}=\left(\frac{\Lambda}{\lambda}\right)^{\frac{n}{2}},
\end{split}
\label{estZ2}
\end{equation}
where we have used that $\int_{-\infty}^{\infty}e^{-r^{2}}\,dr=\sqrt{\pi}$.
We also recall from \cite[Thm. 3, p. 8]{Friedman2013}, that for any
continuous function $f:\Omega\times[0,\tau_{1}]\to\mathbb{R}$, the
integral 
\begin{equation}
V(\mathbf{x}_{0},\tau):=\int_{0}^{\tau}\int_{\Omega}Z(\mathbf{x}_{0},\tau;\mathbf{\bm{\xi}},s)f(\bm{\xi},s)\,d\bm{\xi}ds\label{volpot}
\end{equation}
is continuously-differentiable with respect to $\mathbf{x}_{0}$ and
satisfies 
\begin{equation}
\mathbf{\mathbf{\bm{\nabla}}}_{0}V(\mathbf{x}_{0},\tau)=\int_{0}^{\tau}\int_{\Omega}\mathbf{\mathbf{\bm{\nabla}}}_{0}Z(\mathbf{x}_{0},\tau;\bm{\xi},s)f(\mathbf{x}_{0},s)\,d\bm{\xi}ds.\label{derpot}
\end{equation}

As shown in \cite[Thm. 9, p.21]{Friedman2013}, the variation of constants
formula applied to (\ref{mainw}) gives its solution in the form {\footnotesize{}
\begin{equation}
\begin{split}w(\mathbf{x}_{0},\tau) & =\int_{0}^{\tau}\int_{\Omega}Z\mathbf{\mathbf{\bm{\nabla}}}_{0}\cdot\left(\rho_{0}^{-1}\mathbf{T}_{\nu}\bm{\nabla}_{0}c_{0}\right)\,d\bm{\xi}\,ds+\\
 & \phantom{=}\int_{0}^{\tau}\int_{\Omega}Z(\mathbf{x}_{0},\tau;\mathbf{\bm{\xi}},s)\times\\
 & \,\,\,\,\,\,\,\times\left(\int_{0}^{s}\int_{\Omega}\Phi\left(\bm{\xi},s;\bm{\eta},\sigma\right)\left(\rho_{0}^{-1}\mathbf{T}_{\nu}(\bm{\eta},\sigma)\bm{\nabla}_{0}c_{0}(\bm{\eta})\right)\,d\bm{\eta}\,d\sigma\right)\,d\bm{\xi}\,ds\\
 & \eqqcolon W_{1}(\mathbf{x}_{0},\tau)+W_{2}(\mathbf{x}_{0},\tau),
\end{split}
\label{Fund}
\end{equation}
}for some (not explicitly known) function $\Phi$ that satisfies the
estimate 
\begin{equation}
\left|\Phi\left(\bm{\xi},s;\bm{\eta},\sigma\right)\right|\leq C_{4}\frac{1}{|s-\sigma|^{h_{0}}}\frac{1}{|\xi-\eta|^{n+2-2h_{0}-\alpha}},\label{phi}
\end{equation}
for any constant $h_{0}\in\left(1-\frac{\alpha}{2},1\right)$, where
$\alpha$ is the Hölder exponent in (\ref{Hold}).

To estimate the spatial gradient of $W_{1}$, we use the formula for
the $\mathbf{x}_{0}$-derivative of (\ref{Fund}) in (\ref{derpot})
to obtain{\footnotesize{}
\begin{equation}
\begin{split}\left|\bm{\nabla}_{0}W_{1}\right| & =\left|\bm{\nabla}_{0}\int_{0}^{\tau}\int_{\Omega}Z\mathbf{\mathbf{\bm{\nabla}}}_{0}\cdot\left(\rho_{0}^{-1}\mathbf{T}_{\nu}\bm{\nabla}_{0}c_{0}\right)\,d\bm{\xi}\,ds\right|\\
 & =\left|\int_{0}^{\tau}\int_{\Omega}\left(\bm{\nabla}_{0}Z\right)\mathbf{\mathbf{\bm{\nabla}}}_{0}\cdot\left(\rho_{0}^{-1}\mathbf{T}_{\nu}\bm{\nabla}_{0}c_{0}\right)\,d\bm{\xi}\,ds\right|\\
 & \leq\int_{0}^{\tau}\int_{\Omega}\frac{1}{2|\tau-s|}\left|\rho_{0}\mathbf{T}_{\nu}^{-1}(\bm{\xi},s)(\mathbf{x}_{0}-\bm{\xi})\right|\left|Z\right|\left|\mathbf{\mathbf{\bm{\nabla}}}_{0}\cdot\left(\rho_{0}^{-1}\mathbf{T}_{\nu}\bm{\nabla}_{0}c_{0}\right)\right|\,d\bm{\xi}\,ds,
\end{split}
\label{essW1}
\end{equation}
}where we also used the definition (\ref{defZ}) in evaluating $\bm{\nabla}_{0}Z$.
From (\ref{Ainv}), we obtain $\|\rho_{0}\mathbf{T}_{\nu}^{-1}\|=\lambda^{-1}$,
and hence we can further write (\ref{essW1}) as {\footnotesize{}
\begin{equation}
\begin{split}\left|\bm{\nabla}_{0}W_{1}\right| & \leq\frac{1}{\lambda}\int_{0}^{\tau}\int_{\Omega}\frac{\left|Z\right|}{2\left|\tau-s\right|}\left|\mathbf{\mathbf{\bm{\nabla}}}_{0}\cdot\left(\rho_{0}^{-1}\mathbf{T}_{\nu}\bm{\nabla}_{0}c_{0}\right)\right|\,d\bm{\xi}\,ds\\
 & \leq\frac{\|\mathbf{\mathbf{\bm{\nabla}}}_{0}\cdot\left(\rho_{0}^{-1}\mathbf{T}_{\nu}\bm{\nabla}_{0}c_{0}\right)\|_{C^{0}(\Omega)}}{\lambda}\int_{0}^{\tau}\int_{\Omega}\frac{1}{2|\tau-s|}\left|\mathbf{x}_{0}-\bm{\xi}\right|\left|Z\right|\,d\bm{\xi}\,ds\\
 & \leq C_{5}\frac{\|u_{0}\|_{C^{2}(\Omega)}}{\lambda}\int_{0}^{\tau}\int_{\Omega}\frac{1}{2|\tau-s|}\left|\mathbf{x}_{0}-\bm{\xi}\right|\left|Z\right|\,d\bm{\xi}\,ds.
\end{split}
\label{eq:w1est1}
\end{equation}
}Next, as in the calculation of the integral in (\ref{estZ}), we
use the scaling (\ref{scaling}) in (\ref{eq:w1est1}) to obtain 
\begin{equation}
\begin{split}\left|\bm{\nabla}_{0}W_{1}\right| & \leq C_{5}\frac{\Lambda\|u_{0}\|_{C^{2}(\Omega)}}{\lambda}\int_{0}^{\tau}\frac{1}{\sqrt{\tau-s}}\left(\int_{\mathbb{R}^{n}}|\bm{y}|e^{-|\bm{y}|^{2}}\,d\bm{y}\right)\,ds\\
 & \leq C_{6}\frac{\Lambda\|u_{0}\|_{C^{2}(\Omega)}}{\lambda}\int_{0}^{\tau}\frac{1}{\sqrt{\tau-s}}\,ds\\
 & \leq C_{7}\sqrt{\tau}=\mathcal{O}\left(\nu^{\frac{1}{2}}\right).
\end{split}
\label{eq:w1estfinal}
\end{equation}

To estimate the spatial gradient of $W_{2}$ in (\ref{Fund}), we
proceed similarly by using the growth condition (\ref{phi}) to obtain
{\tiny{}
\begin{equation}
\begin{split}\left|\bm{\nabla}_{0}W_{2}\right| & \leq\int_{0}^{\tau}\int_{\Omega}\frac{1}{2\left|\tau-s\right|}\left|\rho_{0}\mathbf{T}_{\nu}^{-1}(\mathbf{x}_{0}-\bm{\xi})\right|\left|Z\right|\times\\
 & \,\,\,\,\,\,\,\,\,\,\,\,\,\,\,\left(\int_{0}^{s}\int_{\Omega}\left|\Phi\right|\left|\mathbf{\mathbf{\bm{\nabla}}}_{0}\cdot\left(\rho_{0}^{-1}\mathbf{T}_{\nu}\bm{\nabla}_{0}c_{0}\right)\right|\,d\bm{\eta}\,d\sigma\right)\,d\bm{\xi}\,ds\\
 & \leq C_{8}\frac{\|\mathbf{\mathbf{\bm{\nabla}}}_{0}\cdot\left(\rho_{0}^{-1}\mathbf{T}_{\nu}\bm{\nabla}_{0}c_{0}\right)\|_{C^{0}(\Omega)}}{\lambda}\times\\
 & \,\,\,\,\,\times\int_{0}^{\tau}\int_{\Omega}\frac{1}{2|\tau-s|}\left|\mathbf{x}_{0}-\bm{\xi}\right|\left|Z\right|\times\\
 & \,\,\,\,\,\,\,\,\,\,\,\,\,\,\,\left(\int_{0}^{s}\frac{d\sigma}{|s-\sigma|^{h_{0}}}\int_{\Omega}\frac{d\bm{\eta}}{|\bm{\xi}-\bm{\eta}|^{n+2-2h_{0}-\alpha}}\right)\,d\bm{\xi}\,ds.
\end{split}
\label{eq:w2est1}
\end{equation}
}Since $\Omega$ is bounded, there exists a ball of radius $R$ such
that $\Omega+\Omega\subset B_{R}$ and therefore, noticing that $2-2h_{0}-\alpha>0$
by $1-\frac{\alpha}{2}<h_{0}<1$, we find that 
\begin{equation}
\int_{\Omega}\frac{d\bm{\eta}}{|\bm{\xi}-\bm{\eta}|^{n+2-2h_{0}-\alpha}}\leq C_{9}\left.r^{2-2h_{0}-\alpha}\right|_{r=0}^{r=R}=C_{9}R^{2-2h_{0}-\alpha}.
\end{equation}
As in (\ref{essW1}), we can estimate the integral of $\left|\mathbf{x}_{0}-\bm{\xi}\right|\left|Z\right|$
to obtain {\scriptsize{}
\begin{equation}
\begin{split}\left|\bm{\nabla}_{0}W_{2}\right| & \leq C_{9}\frac{R^{2-2h_{0}-\alpha}\|u_{0}\|_{C^{2}(\Omega)}}{\lambda}\times\\
 & \,\,\,\,\,\,\,\times\int_{0}^{\tau}\int_{\Omega}\frac{1}{2|\tau-s|}\left|\mathbf{x}_{0}-\bm{\xi}\right|\left|Z\right|\left(\int_{0}^{s}\frac{d\sigma}{|s-\sigma|^{h_{0}}}\right)\,d\bm{\xi}\,ds\\
 & \leq C_{10}\frac{R^{2-2h_{0}-\alpha}\|u_{0}\|_{C^{2}(\Omega)}}{\lambda}\int_{0}^{\tau}\frac{1}{\sqrt{\tau-s}}\left(\int_{0}^{s}\frac{1}{|s-\sigma|^{h_{0}}}\,d\sigma\right)\,ds\\
 & \leq C_{11}\int_{0}^{\tau}\frac{|\tau-s|^{1-h_{0}}}{\sqrt{\tau-s}}\,ds\\
 & \leq C_{12}|\tau|^{\frac{3}{2}-h_{0}}=\mathcal{O}\left(\nu^{\frac{\alpha+1}{2}}\right).
\end{split}
\label{eq:w2estfinal}
\end{equation}
}The estimates (\ref{eq:w1estfinal})-(\ref{eq:w2estfinal}) together
prove (\ref{supw}), which then implies (\ref{sup}), which in turn
implies (\ref{eq:ansatz}), as claimed. 

\section{Proof of \thmref{first_integral}\label{sec:Appendix-B}}

Applying the classic result on the variational derivative of quotient
functional (see, e.g., \cite{Castillo2008}), we obtain that eq. (\ref{eq:variproblem_known_IC})
is equivalent to
\begin{equation}
\delta\mathcal{\tilde{\mathcal{E}}}(\mathcal{M}_{0})=\frac{1}{\int_{\mathcal{M}_{0}}dA_{0}}\delta\int_{\mathcal{M}_{0}}\left[\left|\left\langle \mathbf{\bar{q}}_{t_{0}}^{t_{1}},\mathbf{n}_{0}\right\rangle \right|-\mathcal{T}_{0}\right]\,dA_{0}=0,\qquad\mathcal{T}_{0}:=\frac{\int_{\mathcal{M}_{0}}\left|\left\langle \mathbf{\bar{q}}_{t_{0}}^{t_{1}},\mathbf{n}_{0}\right\rangle \right|\,dA_{0}}{\int_{\mathcal{M}_{0}}dA_{0}}=const..\label{eq:delta=00003D00003D0}
\end{equation}
 Therefore, extrema of the functional 
\begin{equation}
\mathcal{E}(\mathcal{M}_{0}):=\int_{\mathcal{M}_{0}}\left[\left|\left\langle \mathbf{\bar{q}}_{t_{0}}^{t_{1}},\mathbf{n}_{0}\right\rangle \right|-\mathcal{T}_{0}\right]\,dA_{0}\label{eq:epsilon of M}
\end{equation}
coincide with those of $\tilde{\mathcal{E}}$. 

We proceed by introducing a parametrization $\mathbf{x}_{0}(\mathbf{s}):=\mathbf{x}_{0}(s_{1},\ldots s_{n-1})$
of $\mathcal{M}_{0}$ under which $\mathcal{E}(\mathcal{M}_{0})$
becomes 
\begin{equation}
\mathcal{E}(\mathcal{M}_{0})=\int_{\mathcal{M}_{0}}L\left(\mathbf{x}_{0}(\mathbf{s}),\partial_{\mathbf{s}}\mathbf{x}_{0}(\mathbf{s})\right)\,ds_{1}\ldots ds_{n-1},\label{eq:calE}
\end{equation}
with the Lagrangian 
\begin{equation}
L\left(\mathbf{x}_{0},\partial_{\mathbf{s}}\mathbf{x}_{0}\right):=\left[\left|\left\langle \mathbf{\bar{q}}_{t_{0}}^{t_{1}}\left(\mathbf{x}_{0}\right),\mathbf{n}_{0}\left(\partial_{\mathbf{s}}\mathbf{x}_{0}\right)\right\rangle \right|-\mathcal{T}_{0}\right]\sqrt{\det\mathbf{G}\left(\partial_{\mathbf{s}}\mathbf{x}_{0}\right)}.\label{eq:Lagrangian}
\end{equation}
Here $G_{ij}=\left\langle \frac{\partial\mathbf{x}_{0}}{\partial s_{i}},\frac{\partial\mathbf{x}_{0}}{\partial s_{j}}\right\rangle $
denotes the $(i,j)$ entry of the Gramian matrix $\mathbf{G}$ of
the parametrization, which therefore satisfies, for any real number
$c>0$, the identity
\[
\sqrt{\det\mathbf{G}\left(\partial_{s_{1}}\mathbf{x}_{0},\ldots,c\partial_{s_{i}}\mathbf{x}_{0},\ldots,\partial_{s_{n-1}}\mathbf{x}_{0}\right)}=c\sqrt{\det\mathbf{G}\left(\partial_{\mathbf{s}}\mathbf{x}_{0}\right)}.
\]
Thus, by definition, $\sqrt{\det\mathbf{G}}$ is a positively homogenous
function of $\partial_{s_{i}}\mathbf{x}_{0}$ with order $1$, and
hence, by Euler's theorem \cite{Lewis1969}, satisfies 
\begin{equation}
\partial_{s_{i}}\mathbf{x}_{0}(\mathbf{s})\cdot\frac{\partial\sqrt{\det\mathbf{G}}}{\partial\left(\partial_{s_{i}}\mathbf{x}_{0}(\mathbf{s})\right)}=1\cdot\sqrt{\det\mathbf{G}}=\sqrt{\det\mathbf{G}.}\label{eq:Euler1}
\end{equation}

Furthermore, once an orientation is fixed, the unit normal $\mathbf{n}_{0}$
is a unique smooth function of the $n-1$ tangent vectors $\partial_{s_{i}}\mathbf{x}_{0},$
even though the lengths of these vectors are arbitrary and depend
on the parametrization. Consequently, for any real number $c>0$,
we have 
\[
\mathbf{n}_{0}\left(\partial_{s_{1}}\mathbf{x}_{0},\ldots,c\partial_{s_{i}}\mathbf{x}_{0},\ldots,\partial_{s_{n-1}}\mathbf{x}_{0}\right)=\mathbf{n}_{0}\left(\partial_{\mathbf{s}}\mathbf{x}_{0}\right)=c^{k}\cdot\mathbf{n}_{0}\left(\partial_{\mathbf{s}}\mathbf{x}_{0}(\mathbf{s})\right),\quad k=0,
\]
which, by definition, implies that $\mathbf{n}_{0}\left(\partial_{\mathbf{s}}\mathbf{x}_{0}(\mathbf{s})\right)$
is a positively homogeneous function of order zero. Then, again by
Euler's theorem, we conclude that
\begin{equation}
\partial_{s_{i}}\mathbf{x}_{0}(\mathbf{s})\cdot\frac{\partial\mathbf{n}_{0}\left(\partial_{\mathbf{s}}\mathbf{x}_{0}(\mathbf{s})\right)}{\partial\left(\partial_{s_{i}}\mathbf{x}_{0}(\mathbf{s})\right)}=0.\label{eq:Euler2}
\end{equation}
As the Lagrangian $L\left(\mathbf{x}_{0},\partial_{\mathbf{s}}\mathbf{x}_{0}\right)$
has no explicit dependence on \textbf{$\mathbf{s}$}, Noether's theorem
provides partial conservation laws (cf., Logan \cite[Ch. 4, Example 4.2]{Logan1977}
for the associated Euler--Lagrange equation in the form 
\begin{equation}
\frac{\partial H_{j}^{i}}{\partial s_{k}}=0,\quad H_{j}^{i}:=\partial_{s_{j}}\mathbf{x}_{0}\cdot\frac{\partial L}{\partial\left(\partial_{s_{i}}\mathbf{x}_{0}\right)}-\delta_{ij}L,\quad i,j,k=1,\ldots,n-1,\label{eq:Hamiltonians}
\end{equation}
with $\delta_{ij}$ referring to the Kronecker delta. A direct calculation
using (\ref{eq:Euler1})-(\ref{eq:Euler2}), however, gives
\begin{align*}
H_{i}^{i} & =\partial_{s_{i}}\mathbf{x}_{0}(\mathbf{s})\cdot\frac{\partial L}{\partial\left(\partial_{s_{i}}\mathbf{x}_{0}(\mathbf{s})\right)}-L\\
 & =\left[\left|\left\langle \mathbf{\bar{q}}_{t_{0}}^{t_{1}},\mathbf{n}_{0}\right\rangle \right|-\mathcal{T}_{0}\right]\partial_{s_{i}}\mathbf{x}_{0}(\mathbf{s})\cdot\frac{\partial\sqrt{\det\mathbf{G}}}{\partial\left(\partial_{s_{i}}\mathbf{x}_{0}(\mathbf{s})\right)}-L\\
 & =\left[\left|\left\langle \mathbf{\bar{q}}_{t_{0}}^{t_{1}}\left(\mathbf{x}_{0}\right),\mathbf{n}_{0}\left(\partial_{\mathbf{s}}\mathbf{x}_{0}\right)\right\rangle \right|-\mathcal{T}_{0}\right]\sqrt{\det\mathbf{G}\left(\partial_{\mathbf{s}}\mathbf{x}_{0}\right)}\\
 & =0,
\end{align*}
and hence no nontrivial first integral arises from the relations (\ref{eq:Hamiltonians}). 

Nevertheless, for the modified variational problem
\begin{equation}
\hat{\mathcal{E}}(\mathcal{M}_{0})=\int_{\mathcal{M}_{0}}L^{2}\left(\mathbf{x}_{0}(\mathbf{s}),\partial_{\mathbf{s}}\mathbf{x}_{0}(\mathbf{s})\right)\,ds_{1}\ldots ds_{n-1},\label{eq:Ehat}
\end{equation}
 Noether's theorem gives 
\begin{equation}
H_{i}^{i}:=\partial_{s_{i}}\mathbf{x}_{0}\cdot\frac{\partial L^{2}}{\partial\left(\partial_{s_{i}}\mathbf{x}_{0}\right)}-L^{2}=2L^{2}-L^{2}=L^{2},\quad i=1,\ldots,n-1,\label{eq:Hii}
\end{equation}
by Euler's theorem, given that $L^{2}$ is a positively homogeneous
function of order $2$ in the variables $\partial_{s_{i}}\mathbf{x}_{0}$.
By the definition of $H_{i}^{i}$ in (\ref{eq:Hamiltonians}), eq.
(\ref{eq:Hii}) implies that $L$ is a first a first integral for
solutions of the variational problem (\ref{eq:Ehat}). Then, by a
generalization of the Maupertuis principle for PDEs (see \cite[Appendix S4]{Haller2018}),
$L$ is also a first integral for the Euler--Lagrange equation of
the original variational problem on all nonzero level sets of $L.$
This in turn implies the invariance of the $\left\{ L=0\right\} $
level set as well. We conclude that barrier surfaces must satisfy
$\left[\left|\left\langle \mathbf{\bar{q}}_{t_{0}}^{t_{1}}\left(\mathbf{x}_{0}\right),\mathbf{n}_{0}\left(\partial_{\mathbf{s}}\mathbf{x}_{0}\right)\right\rangle \right|-\mathcal{T}_{0}\right]\sqrt{\det\mathbf{G}\left(\partial_{\mathbf{s}}\mathbf{x}_{0}\right)}=C$
in any dimension.

\section{Proof of \thmref{2D_flows}\label{sec:Appendix-C}}

For $n=2,$ a diffusion extremizer surface $\mathcal{M}_{0}$ is a
one-dimensional curve parametrized by the single variable $s.$ The
unit normal to this curve at a point $\mathbf{x}_{0}\in\mathcal{M}_{0}$
can be written as
\[
\mathbf{n}_{0}(\mathbf{x}_{0})=\mathbf{\bm{\Omega}}\frac{\mathbf{x}_{0}^{\prime}}{\left|\mathbf{x}_{0}^{\prime}\right|},
\]
with the rotation matrix $\mathbf{\bm{\Omega}}$ defined as in (\ref{eq:Omegadef}).
The Lagrangian $L$ defined in (\ref{eq:Lagrangian}) then takes the
specific form
\begin{align}
L\left(\mathbf{x}_{0},\mathbf{x}_{0}^{\prime}\right) & =\left(\sqrt{\left\langle \mathbf{\bar{q}}_{t_{0}}^{t_{1}}\left(\mathbf{x}_{0}\right),\mathbf{\bm{\Omega}}\frac{\mathbf{x}_{0}^{\prime}}{\left|\mathbf{x}_{0}^{\prime}\right|}\right\rangle ^{2}}-\mathcal{T}_{0}\right)\sqrt{\left\langle \mathbf{x}_{0}^{\prime},\mathbf{x}_{0}^{\prime}\right\rangle }\nonumber \\
 & =\sqrt{\left\langle \mathbf{\bar{q}}_{t_{0}}^{t_{1}}\left(\mathbf{x}_{0}\right),\mathbf{\bm{\Omega}}\mathbf{x}_{0}^{\prime}\right\rangle ^{2}}-\mathcal{T}_{0}\left|\mathbf{x}_{0}^{\prime}\right|.\label{eq:2D Ldef}
\end{align}
The general equation (\ref{eq:uniform barriers}), therefore, simplifies
in two-dimensions to the implicit ODE
\begin{equation}
\left|\left\langle \mathbf{\bm{\Omega}}\mathbf{\bar{q}}_{t_{0}}^{t_{1}}\left(\mathbf{x}_{0}\right),\frac{\mathbf{x}_{0}^{\prime}}{\left|\mathbf{x}_{0}^{\prime}\right|}\right\rangle \right|=\mathcal{T}_{0}.\label{eq:cons law}
\end{equation}
Note from (\ref{eq:cons law}) that within the $\left\{ L=0\right\} $
level set, arbitrary re-parametrizations of the solutions (which are
also solutions, by the rescaling invariance of the variational principle)
also happen to preserve the value of the first integral $L$. We are,
therefore, free to select the arc-length parametrization for diffusion
barriers by letting $\mathbf{x}_{0}^{\prime}=\alpha\mathbf{\bar{q}}_{t_{0}}^{t_{1}}\left(\mathbf{x}_{0}\right)+\beta\mathbf{\bm{\Omega}}\mathbf{\bar{q}}_{t_{0}}^{t_{1}}\left(\mathbf{x}_{0}\right)$
with $\left(\alpha^{2}+\beta^{2}\right)=1/\left|\mathbf{\bar{q}}_{t_{0}}^{t_{1}}\left(\mathbf{x}_{0}\right)\right|^{2}.$
Substituting this form of $\mathbf{x}_{0}^{\prime}$ into (\ref{eq:cons law})
gives the vector field family
\begin{equation}
\mathbf{x}_{0}^{\prime}=\frac{\sqrt{\left|\mathbf{\bar{q}}_{t_{0}}^{t_{1}}\left(\mathbf{x}_{0}\right)\right|^{2}-\mathcal{T}_{0}^{2}}}{\left|\mathbf{\bar{q}}_{t_{0}}^{t_{1}}\left(\mathbf{x}_{0}\right)\right|^{2}}\mathbf{\bar{q}}_{t_{0}}^{t_{1}}\left(\mathbf{x}_{0}\right)\pm\frac{\mathcal{T}_{0}}{\left|\mathbf{\bar{q}}_{t_{0}}^{t_{1}}\left(\mathbf{x}_{0}\right)\right|^{2}}\mathbf{\bm{\Omega}}\mathbf{\bar{q}}_{t_{0}}^{t_{1}}\left(\mathbf{x}_{0}\right),\label{eq:2ODES1-2}
\end{equation}
 as stated in (\ref{eq:2ODEs1}), where we have used the equality
$\left|\mathbf{\bm{\Omega}}\mathbf{\bar{q}}_{t_{0}}^{t_{1}}\right|=\left|\mathbf{\bar{q}}_{t_{0}}^{t_{1}}\right|$.
Trajectories of (\ref{eq:2ODEs1}) are, therefore, stationary curves
of $\mathcal{E}$. 

It is yet unclear, however, whether trajectories of (\ref{eq:2ODEs1})
are minimizers or maximizers of the functional $\tilde{\mathcal{E}}$.
As we have seen, stationary curves of $\tilde{\mathcal{E}}$ coincide
with those of
\[
\mathcal{E}(\mathcal{M}_{0})=\int_{\mathcal{M}_{0}}L\left(\mathbf{x}_{0}(s),\mathbf{x}_{0}^{\prime}(s)\right)\,ds,
\]
with $L$ defined in (\ref{eq:2D Ldef}). As in any classic calculus
of variations problem, the admissible variations $\mathbf{h}(s)$
of a $\mathcal{M}_{0}$ are those that make the boundary term arising
in the integration by parts vanish, i.e., 
\begin{equation}
\left[\partial_{\mathbf{x}_{0}^{\prime}}L\left(\mathbf{x}_{0}(s),\mathbf{x}_{0}^{\prime}(s)\right)\cdot\mathbf{h}(s)\right]_{s_{1}}^{s_{2}}=0.\label{eq:boundary conditions}
\end{equation}
Noting that
\begin{equation}
\partial_{\mathbf{x}_{0}^{\prime}}L\left(\mathbf{x}_{0},\mathbf{x}_{0}^{\prime}\right)=\mathrm{sign\,\left\langle \left\langle \mathbf{\bar{q}}_{t_{0}}^{t_{1}}\left(\mathbf{x}_{0}\right),\mathbf{\bm{\Omega}}\mathbf{x}_{0}^{\prime}\right\rangle \right\rangle }\mathbf{\bm{\Omega}}^{T}\mathbf{\bar{q}}_{t_{0}}^{t_{1}}\left(\mathbf{x}_{0}\right)-\mathcal{T}_{0}\frac{\mathbf{x}_{0}^{\prime}}{\sqrt{\left\langle \mathbf{x}_{0}^{\prime},\mathbf{x}_{0}^{\prime}\right\rangle }},\quad\left\langle \mathbf{\bar{q}}_{t_{0}}^{t_{1}}\left(\mathbf{x}_{0}\right),\mathbf{\bm{\Omega}}\mathbf{x}_{0}^{\prime}\right\rangle \neq0,\label{eq:Lgradient}
\end{equation}
 we can distinguish the following types of variations based on (\ref{eq:boundary conditions}):
\begin{description}
\item [{\emph{B1.}}] \emph{Vanishing endpoint variations}: $\mathbf{h}(s_{1})=\mathbf{h}(s_{2})=\mathbf{0}.$
\item [{\emph{B2.}}] \emph{Orthogonal endpoint variations}: The variation
$\mathbf{h}(s)$ is nonzero but orthogonal to $\partial_{\mathbf{x}_{0}^{\prime}}L\left(\mathbf{x}_{0}(s),\mathbf{x}_{0}^{\prime}(s)\right)$
at the endpoints, i.e.,
\begin{equation}
\partial_{\mathbf{x}_{0}^{\prime}}L\left(\mathbf{x}_{0}(s_{i}),\mathbf{x}_{0}^{\prime}(s_{i})\right)\perp\mathbf{h}(s_{i}),\quad i=1,2.\label{eq:normalcond}
\end{equation}
Restricting to normal variations ($\mathbf{h}(s)\perp\mathbf{x}_{0}^{\prime}(s)$)
and using (\ref{eq:Lgradient}), we find that (\ref{eq:normalcond})
is equivalent to
\[
\mathbf{\bm{\Omega}}\mathbf{\bar{q}}_{t_{0}}^{t_{1}}\left(\mathbf{x}_{0}(s_{i})\right)\cdot\mathbf{h}(s_{i})=0,\quad i=1,2,
\]
as longs as $\left\langle \mathbf{\bar{q}}_{t_{0}}^{t_{1}}\left(\mathbf{x}_{0}\right),\mathbf{\bm{\Omega}}\mathbf{x}_{0}^{\prime}\right\rangle \neq0,$
i.e., as long as $L$ is differentiable. We conclude that arbitrary
normal variations are admissible to the endpoints of $\mathcal{M}_{0}$
whenever $\mathbf{\bar{q}}_{t_{0}}^{t_{1}}\left(\mathbf{x}_{0}(s_{i})\right)\parallel\mathbf{x}_{0}^{\prime}(s_{i}),$
$i=1,2$ holds at those endpoints. 
\item [{\emph{B3.}}] \emph{Free endpoint variations}: The factor $\partial_{\mathbf{x}_{0}^{\prime}}L\left(\mathbf{x}_{0},\mathbf{x}_{0}^{\prime}\right)$
vanishes at the endpoints, i.e., 
\begin{equation}
\partial_{\mathbf{x}_{0}^{\prime}}L\left(\mathbf{x}_{0}(s_{i}),\mathbf{x}_{0}^{\prime}(s_{i})\right)=\mathbf{0},\quad i=1,2.\label{eq:zeroderivative}
\end{equation}
Using (\ref{eq:Lgradient}) and the expression (\ref{eq:2ODES1-2})
for $\mathbf{x}_{0}^{\prime}$ along $\mathcal{M}_{0}$, we find that
condition (\ref{eq:zeroderivative}) is satisfied whenever
\[
\partial_{\mathbf{x}_{0}^{\prime}}L\vert_{\mathcal{M}_{0}}=-\left[\mathrm{sign}\,\left\langle \left\langle \mathbf{\bar{q}}_{t_{0}}^{t_{1}}\left(\mathbf{x}_{0}\right),\mathbf{\bm{\Omega}}\mathbf{x}_{0}^{\prime}\right\rangle \right\rangle \pm\mathcal{T}_{0}\right]\mathbf{\bm{\Omega}}\mathbf{\bar{q}}_{t_{0}}^{t_{1}}\left(\mathbf{x}_{0}\right)-\mathcal{T}_{0}\frac{\sqrt{\left|\mathbf{\bar{q}}_{t_{0}}^{t_{1}}\left(\mathbf{x}_{0}\right)\right|^{2}-\mathcal{T}_{0}^{2}}}{\left|\mathbf{\bar{q}}_{t_{0}}^{t_{1}}\left(\mathbf{x}_{0}\right)\right|^{2}}\mathbf{\bar{q}}_{t_{0}}^{t_{1}}\left(\mathbf{x}_{0}\right).
\]
 Therefore, condition (\ref{eq:zeroderivative}) is satisfied if either
\[
\mathbf{\bar{q}}_{t_{0}}^{t_{1}}\left(\mathbf{x}_{0}(s_{i})\right)=\mathbf{0},\quad i=1,2
\]
 or 
\[
\mathcal{T}_{0}=\left|\mathbf{\bar{q}}_{t_{0}}^{t_{1}}\left(\mathbf{x}_{0}(s_{i})\right)\right|=1,\quad i=1,2.
\]
\item [{\emph{B4.}}] \emph{Periodic variations}: $\mathcal{M}_{0}$ is
a closed curve and the variation $\mathbf{h}(s)$ is periodic with
the same period in $s$, i.e., 
\[
\partial_{\mathbf{x}_{0}^{\prime}}L\left(\mathbf{x}_{0}(s_{1}),\mathbf{x}_{0}^{\prime}(s_{1})\right)\cdot\mathbf{h}(s_{1})=\partial_{\mathbf{x}_{0}^{\prime}}L\left(\mathbf{x}_{0}(s_{2}),\mathbf{x}_{0}^{\prime}(s_{2})\right)\cdot\mathbf{h}(s_{2})).
\]
\end{description}
A simple calculation gives, furthermore, that the second variation
of $\tilde{\mathcal{E}}$ along any of its stationary curves, $\mathcal{M}_{0}$,
satisfies
\[
\delta^{2}\tilde{\mathcal{E}}(\mathcal{M}_{0})=\frac{\delta^{2}\mathcal{E}(\mathcal{M}_{0})}{\int_{\mathcal{M}_{0}}\sqrt{\left\langle \mathbf{x}_{0}^{\prime}(s),\mathbf{x}_{0}^{\prime}(s)\right\rangle }\,ds}.
\]
Therefore, $\tilde{\mathcal{E}}$ and $\mathcal{E}$ also share the
type of their stationary curves (minimizers, maximizers or saddle-type
stationary curves). To obtain a necessary condition for a stationary
curve of $\tilde{\mathcal{E}}$ to be an extremizer, we can therefore
apply the classic Legendre condition to the functional $\mathcal{E}$,
which is based on the definiteness of the Hessian $\partial_{\mathbf{x}_{0}^{\prime}\mathbf{x}_{0}^{\prime}}^{2}L\left(\mathbf{x}_{0},\mathbf{x}_{0}^{\prime}\right).$ 

Note that whenever $\partial_{\mathbf{x}_{0}^{\prime}}L\left(\mathbf{x}_{0},\mathbf{x}_{0}^{\prime}\right)$
is well-defined in (\ref{eq:Lgradient}), ie., $\left\langle \mathbf{\bar{q}}_{t_{0}}^{t_{1}}\left(\mathbf{x}_{0}\right),\mathbf{\bm{\Omega}}\mathbf{x}_{0}^{\prime}\right\rangle \neq0$
holds, the function $\mathrm{sign}\,\left\langle \left\langle \mathbf{\bar{q}}_{t_{0}}^{t_{1}}\left(\mathbf{x}_{0}\right),\mathbf{\bm{\Omega}}\mathbf{x}_{0}^{\prime}\right\rangle \right\rangle $
is locally constant in $\mathbf{x}_{0}^{\prime}$. Therefore, whenever
$\partial_{\mathbf{x}_{0}^{\prime}}L\left(\mathbf{x}_{0},\mathbf{x}_{0}^{\prime}\right)$
is differentiable, its Jacobian equals 
\begin{align}
\partial_{\mathbf{x}_{0}^{\prime}\mathbf{x}_{0}^{\prime}}^{2}L\left(\mathbf{x}_{0},\mathbf{x}_{0}^{\prime}\right) & =-\mathcal{T}_{0}\partial_{\mathbf{x}_{0}^{\prime}}\frac{\mathbf{x}_{0}^{\prime}}{\sqrt{\left\langle \mathbf{x}_{0}^{\prime},\mathbf{x}_{0}^{\prime}\right\rangle }}\label{eq:secondvari}\\
 & =-\frac{\mathcal{T}_{0}}{\sqrt{\left\langle \mathbf{x}_{0}^{\prime},\mathbf{x}_{0}^{\prime}\right\rangle }}\left[\mathbf{I}-\frac{1}{\left\langle \mathbf{x}_{0}^{\prime},\mathbf{x}_{0}^{\prime}\right\rangle }\mathbf{x}_{0}^{\prime}\left(\mathbf{x}_{0}^{\prime}\right)^{T}\right],\quad\left\langle \mathbf{\bar{q}}_{t_{0}}^{t_{1}}\left(\mathbf{x}_{0}\right),\mathbf{\bm{\Omega}}\mathbf{x}_{0}^{\prime}\right\rangle \neq0,\nonumber 
\end{align}
Recall that any two-dimensional dyadic product matrix $\mathbf{x}_{0}^{\prime}\left(\mathbf{x}_{0}^{\prime}\right)^{T}$
has an eigenvalue equal to $\left\langle \mathbf{x}_{0}^{\prime},\mathbf{x}_{0}^{\prime}\right\rangle $
(corresponding to the eigenvector $\mathbf{x}_{0}^{\prime})$ and
another eigenvalue equal to zero (corresponding to $\mathbf{\bm{\Omega}}\mathbf{x}_{0}^{\prime}$).
As a consequence, the eigenvalues of $\partial_{\mathbf{x}_{0}^{\prime}\mathbf{x}_{0}^{\prime}}^{2}L\left(\mathbf{x}_{0},\mathbf{x}_{0}^{\prime}\right)$
are 
\[
\rho_{1}=0,\quad\rho_{2}=-\frac{\mathcal{T}_{0}}{\sqrt{\left\langle \mathbf{x}_{0}^{\prime},\mathbf{x}_{0}^{\prime}\right\rangle }}<0,
\]
which implies that the Hessian $\partial_{\mathbf{x}_{0}^{\prime}\mathbf{x}_{0}^{\prime}}^{2}L\left(\mathbf{x}_{0},\mathbf{x}_{0}^{\prime}\right)$
is negative semidefinite on any stationary curve of $\tilde{\mathcal{E}}$
satisfying $\left\langle \mathbf{\bar{q}}_{t_{0}}^{t_{1}}\left(\mathbf{x}_{0}\right),\mathbf{\bm{\Omega}}\mathbf{x}_{0}^{\prime}\right\rangle \neq0.$
(The kernel of $\partial_{\mathbf{x}_{0}^{\prime}\mathbf{x}_{0}^{\prime}}^{2}L\left(\mathbf{x}_{0},\mathbf{x}_{0}^{\prime}\right)$
is spanned by $\mathbf{x}_{0}^{\prime}$, confirming that tangential
perturbations to the stationary curve at its endpoints result in no
second-order change in the transport functional due to our length-normalized
definition of transport.) Consequently, each stationary curve of $\tilde{\mathcal{E}}$
satisfies the Legendre necessary condition for a strict maximum. 

For stationary curves $\mathcal{M}_{0}$ that are either closed or
connect zeros of the $\mathbf{\bar{q}}_{t_{0}}^{t_{1}}\left(\mathbf{x}_{0}\right)$
vector field (cf. the boundary condition types B3. and B4. above),
we now derive another necessary condition under which a closed stationary
curve $\mathcal{M}_{0}$ is a local maximum of $\mathcal{E}$. We
consider parallel translations of $\mathcal{M}_{0}$ described by
a constant vector field $\mathbf{h}(s)\equiv\mathbf{h}_{0}$. Since
we have $\mathbf{h}^{\prime}(s)\equiv\mathbf{0}$ along such perturbations,
we obtain
\[
\delta^{2}\mathcal{E}[\mathbf{h}_{0}]=\frac{1}{2}\int_{\mathcal{M}_{0}}\left\langle \mathbf{h}_{0},\partial_{\mathbf{x}_{0}\mathbf{x}_{0}}^{2}L(\mathbf{x}_{0},\mathbf{x}_{0}^{\prime})\mathbf{h}_{0}\right\rangle ds=\frac{1}{2}\left\langle \mathbf{h}_{0},\mathbf{L}\mathbf{h_{0}}\right\rangle ,\qquad\mathbf{L}:=\int_{\mathcal{M}_{0}}\partial_{\mathbf{x}_{0}\mathbf{x}_{0}}^{2}L\left(\mathbf{x}_{0}(s),\mathbf{x}_{0}^{\prime}(s)\right)ds.
\]
Therefore, $\delta^{2}\mathcal{E}[\mathbf{h}_{0}]\leq0$ holds for
arbitrary $\mathbf{h}_{0}$ if $\mathbf{L}$ is negative semidefinite.
Note that wherever $L$ is differentiable, its Hessian with respect
to $\mathbf{x}_{0}$ satisfies
\[
\partial_{\mathbf{x}_{0}\mathbf{x}_{0}}^{2}L\left(\mathbf{x}_{0},\mathbf{x}_{0}^{\prime}\right)=\mathrm{sign}\,\left\langle \mathbf{\bar{q}}_{t_{0}}^{t_{1}}\left(\mathbf{x}_{0}\right),\mathbf{\bm{\Omega}}\mathbf{x}_{0}^{\prime}\right\rangle \partial_{\mathbf{x}_{0}\mathbf{x}_{0}}^{2}\left\langle \mathbf{\bar{q}}_{t_{0}}^{t_{1}}\left(\mathbf{x}_{0}\right),\mathbf{\bm{\Omega}}\mathbf{x}_{0}^{\prime}\right\rangle ,
\]
which then implies formula (\ref{eq:Lformula}) of the theorem. The
same argument is also valid for barriers with endpoints satisfying
\[
\mathbf{\bar{q}}_{t_{0}}^{t_{1}}\left(\mathbf{x}_{0}(s_{i})\right)\parallel\mathbf{x}_{0}^{\prime}(s_{i})\parallel\mathbf{x}_{0}^{\prime}(s_{j}),\quad i,j=1,2,\quad i\neq j,
\]
 (cf. the boundary conditions B2), except that in that case
\[
\left\langle \mathbf{L}\mathbf{\bm{\Omega}}\mathbf{x}_{0}^{\prime},\mathbf{\bm{\Omega}}\mathbf{x}_{0}^{\prime}\right\rangle \leq0
\]
must hold for the stationary curve to be a minimizer, given that the
admissible parallel translations are restricted to those normal translations
satisfying $\mathbf{h}(s)\equiv\mathbf{h}_{0}\perp\mathbf{x}_{0}^{\prime}(s_{i}),$
$i=1,2.$

In summary, no trajectory of (\ref{eq:2ODEs1}) can be a minimizer
of the constrained transport functional $\mathcal{E}$ for $\mathcal{T}_{0}$.
For $\mathcal{T}_{0}=0$, however, one recovers the case of perfect
barriers that are clearly global minimizers of $\mathcal{E}$. Thus,
only perfect barriers can be transport-minimizing material surfaces,
as stated in Theorem \ref{thm:2D_flows}.


\end{document}